\documentclass[11pt,preprint,amssymb,showpacs,nofootinbib,eqsecnum,floatfix,byrevtex]{revtex4}
\usepackage{eucal}
\usepackage{color}
\usepackage{graphics}
\def\be{\begin{equation}}
\def\ee{\end{equation}}
\def\bea{\begin{eqnarray}}
\def\eea{\end{eqnarray}}
\def\nn{\nonumber}
\def\mz{\mathbb Z}

\def\ms{{M_{\rm string}}}
\newcommand{\lgp}[2]{{\rm {#1}_{#2}}}
\def\ps{\rm SU_{4C}\times SU_{2L}\times SU_{2R}}
\def\lat{{\lgp{G}{2}\oplus\lgp{SU}{3}\oplus\lgp{SO}{4}}}
\def\Eee{\lgp{E}{8}\times\lgp{E}{8}}

\newcommand{\bm}[1]{\mbox{\boldmath{$#1$}}}

\begin{document}
\preprint{\sf KUNS-1929, OHSTPY-HEP-T-04-008, MCTP-04-13}

\title{Searching for realistic 4d string models with a
Pati-Salam symmetry\\ -- {\sl Orbifold grand unified theories from
heterotic string compactification on a $\mz_6$ orbifold}}

\author{{\sc Tatsuo
Kobayashi}}
\affiliation{Department of Physics, Kyoto University, Kyoto
606-8502, Japan.}
\author{{\sc Stuart Raby}}
\affiliation{Department of Physics, The Ohio
State University, Columbus, OH 43210 USA.}
\author{\sc Ren-Jie Zhang}
\affiliation{Michigan
Center for Theoretical Physics, Randall Laboratory, University of
Michigan, Ann Arbor, MI 48109 USA.}
\altaffiliation{ Also at Institute of
Theoretical Physics, The Chinese Academy of Sciences, Beijing
 100080, China.}

\begin{abstract}
Motivated by orbifold grand unified theories, we construct a class of three-family Pati-Salam models in a
${\mathbb Z}_6$ abelian symmetric orbifold with two discrete Wilson lines. These models have marked differences
from previously-constructed three-family models in prime-order orbifolds. In the limit where one of the six
compactified dimensions (which lies in a ${\mathbb Z}_2$ sub-orbifold) is large compared to the string length
scale, our models reproduce the supersymmetry and gauge symmetry breaking pattern of 5d orbifold grand unified
theories on an ${\rm S^1}/{\mathbb Z}_2$ orbicircle. We find a horizontal $2+1$ splitting in the chiral matter
spectra -- 2 families of matter are localized on the ${\mathbb Z}_2$ orbifold fixed points, and 1 family
propagates in the 5d bulk -- and identify them as the first-two and third families. Remarkably, the first two
families enjoy a non-abelian dihedral $\lgp{D}{4}$ family symmetry, due to the geometric setup of the
compactified space. In all our models there are always some color triplets, i.e. $({\bf 6,1,1})$ representations
of the Pati-Salam group, survive orbifold projections. They could be utilized to spontaneously break the
Pati-Salam symmetry to that of the Standard Model. One model, with a 5d E$_6$ symmetry, may give rise to
interesting low energy phenomenology. We study gauge coupling unification, allowed Yukawa couplings and some of
their phenomenological consequences. The $\lgp{E}{6}$ model has a renormalizable Yukawa coupling only for the
third family. It predicts a gauge-Yukawa unification relation at the 5d compactification scale, and is capable of
generating reasonable quark/lepton masses and mixings. Potential problems are also addressed, they may point to
the direction for refining our models.
\end{abstract}
\date{September 8, 2004}

\maketitle

\section{Introduction and motivation}

The Standard Model (SM) has been a cornerstone of modern-day
particle physics. Although during the past three decades it has
passed all experimental tests, nevertheless there are many open
questions remaining to be answered. We have yet to understand ---
(i) the mechanism of electro-weak symmetry breaking, and find the
Higgs boson which might be responsible for this breaking; (ii) the
quantized fermion charges (why the up and down quarks have charges
$2/3$ and $-1/3$ respectively) and the weak mixing angle (why it is
0.23); (iii) the 3 replicas of quarks and leptons, the observed
fermion mass hierarchy, and the Cabibbo-Kobayashi-Maskawa (CKM)
mixing matrix and its leptonic cousin. We ought to understand these
problems from more fundamental principles, rather than simply take
various charges, masses, mixing angles and CP phases as input
parameters as in the SM. Addressing these questions may eventually
lead us to a more fundamental theory such as string theory at high
energy scales.

String theory \cite{GSW} is a leading candidate for a consistent
theory of quantum gravity. It has a rich structure and many believe
it can easily accommodate the SM as a subset. Moreover there have
been many attempts in the past to construct supersymmetric
generalizations of the SM (which will be loosely referred to as the
minimal supersymmetric standard model, or the MSSM) or grand unified
theories (GUTs) from the heterotic string
\cite{het,CY,Dixon,Schellekens,BL} and superstrings \cite{os}.
Partially successful results have been obtained. For example, many
string theoretical models can explain the existence of three chiral
families at low energy scales
\cite{greene,IMNQ,IMNQ2,FIQS,ff,sgut,ALR,os}, and in principle can
also provide a natural framework for understanding the fermion
masses and mixings \cite{fm}. In this paper, we construct a new
class of three-family models in the heterotic string theory.

Before presenting our models, it is important to note some caveats
common to all known string models. They are due to two main
difficulties facing the string theoretical model constructions. The
first difficulty concerns the compactification of the string itself,
i.e. the mechanism by which the desirable string vacuum is selected.
The vacua of string theory compactifications are parameterized by
many scalar fields with flat potentials. These fields are the
modulus fields. They characterize the sizes and shapes of the
compactified spaces and the strengths of the string interactions;
none of them can be fixed in perturbation theory
\cite{ds}.\footnote{Stabilizing moduli by fluxes in the context of
heterotic string theory has been discussed recently in
\cite{keshav}.} The modulus problem and the related issue of
supersymmetry breaking will not be dealt with in this article,
instead we will simply assume that the moduli are fixed by some
unknown mechanism at the string scale.

The second difficulty concerns our ignorance of the physics between the electro-weak and unification scales.
Except for some indication that the SM gauge couplings may unify at about $10^{16}$ GeV in certain supersymmetric
extensions of the SM with minimal matter content \cite{unif}, we can hardly have any confidence in extrapolating
the low energy data by some $14$ to $16$ orders of magnitude to the unification scale and infer what the gauge
symmetries, matter spectra and physical parameters are at that scale. Hence, we are far from having a clear-cut
field theoretical model at the unification scale to which a string-derived model is
supposed to match.  Any string construction must combine both bottom-up and top-down analyses.

The new class of string models in this paper are mainly motivated by
the recent discussions on orbifold GUT models \cite{kawamura,OGUT}.
These GUT models utilize properties of higher-dimensional field
theories, and have some advantages over conventional 4d GUTs. For
example, GUT symmetry breaking can be accomplished by an orbifold
parity, instead of by a complicated Higgs sector. The
doublet-triplet splitting problem, which plagues conventional GUTs,
can also be solved by assigning appropriate orbifold parities to the
doublet and triplet Higgs bosons. Note, however, that like all field
theoretical models in higher dimensions, these GUT models are not
renormalizable quantum field theories. They can only make sense as
low-energy effective theories of some more fundamental theory with
better ultra-violet (UV) behavior. Our string models provide exactly
such kind of UV completions, in the sense that they reproduce many
interesting features of the orbifold GUTs in certain low energy
limits. (Connections between orbifold GUTs and an $\lgp{SO}{10}$
string model have already been established in ref.~\cite{KRZ0}. The
present paper contains more detailed discussions on these
connections. See also the recent paper~\cite{Forste:2004ie}.)

To make the connections between string and field theoretical models
more concrete, we consider some examples, in particular, the 5d
$\lgp{SO}{10}$ model of ref.~\cite{OGUT} and a generalization with
bulk gauge group E$_6$. In these models, the extra dimension is
taken to be an orbicircle ${\rm S^1}/{\mathbb Z}_2$ and the 4d
effective theory has a Pati-Salam (PS) symmetry, $\ps$ \cite{PS}.
The technical apparatus we adopt to build string models is the
simplest abelian symmetric orbifold compactification
\cite{Dixon,BL,IMNQ,IMNQ2,FIQS,katsuki} of the heterotic string
\cite{het}. More specifically, we consider a non-prime-order
${\mathbb Z}_6$ orbifold (or equivalently, ${\mathbb
Z}_2\times{\mathbb Z}_3$) model with the orbifold twist vector ${\bf
v}_6=\frac{1}{6}(1,2,-3)$. To achieve three chiral PS families at
low energies, we also introduce several (in fact, two) discrete
Wilson lines \cite{wl}. \footnote{Prime-order orbifold models (such
as the $\mz_3$ orbifold models) with Wilson lines
\cite{IMNQ,IMNQ2,FIQS} and non-prime-order orbifold models without
Wilson lines \cite{katsuki} have been extensively studied in the
literature. Non-prime-order orbifold models with Wilson lines, on
the other hand, possess a number of complications, and to our
knowledge they have not been studied to the same extent. Our work
can be regarded as the first serious attempt at constructing
three-family models from non-prime-order orbifolds.} It is obvious
that the third compactified complex dimension has a ${\mathbb Z}_2$
symmetry in the ${\mathbb Z}_6$ model, hence it can consistently be
taken to be the root lattice of the $\lgp{SO}{4}$ Lie algebra. The
string models are effectively 5d when the length of one of the
$\lgp{SO}{4}$ simple roots is large compared to the string scale,
while all other dimensions are kept comparable to the string scale
(i.e. the geometry of the compactified space is equivalent to that
of the orbifold GUTs, ${\rm S^1}/{\mathbb Z}_2$). In this limit, the
${\mathbb Z}_6$ heterotic models are similar to the orbifold GUT
models in the following respects:
\begin{itemize}
\item The 5d ${\mathcal N}=2$ supersymmetry\footnote{By ${\mathcal N}=2$ supersymmetry in 5 or
6d, we mean the minimal number of supersymmetries in these
dimensions, (i.e. the fermions satisfy the pseudo-reality
condition). It reduces to ${\mathcal N}=2$ in 4d by dimensional
reduction and is sometimes called ${\mathcal N}=1$ supersymmetry in
the literature.} is broken to that of ${\mathcal N}=1$ in 4d by the
${\mathbb Z}_2$ orbifold twist and the ``bulk" gauge group is broken
to two different regular subgroups  at the two inequivalent fixed
points by degree-2 non-trivial gauge embedding and Wilson line. The
surviving gauge group in the 4d effective theory is the intersection
of groups at the fixed points. It is the PS group in our models.
More specifically, we find two types of models. In the first type we
have an $\lgp{E}{6}$ symmetry in the 5d bulk which is broken to
$\lgp{SO}{10}$ and $\lgp{SU}{6}\times\lgp{SU}{2}$ respectively. In
the second type we have an $\lgp{SO}{10}\times\lgp{SU}{2}$ in the
bulk, broken to PS at one of the two fixed points.

\item Untwisted-sector and twisted-sector states that are not localized on the
${\mathbb Z}_2$ fixed points of the $\lgp{SO}{4}$ lattice can be identified with the ``bulk" states of the
orbifold GUT. Interpretation of the Kaluza-Klein (KK) towers of the bulk gauge and matter fields agree in the
string-based and orbifold GUT models.

\item Twisted-sector states that are localized on the ${\mathbb Z}_2$ fixed points of the $\lgp{SO}{4}$ lattice
have no field theoretical counterparts, although they can correctly be identified with the ``brane'' states of
the orbifold GUT. In the orbifold GUT models, these states are only constrained by the requirement of (chiral)
anomaly cancellation.

\end{itemize}

Of course, string theoretical models are more intricate than the corresponding field theoretic orbifold GUT
models. They need to satisfy more stringent consistency conditions and thus they are physically more constrained.
We find it is highly non-trivial (or impossible) to implement all the features of the orbifold GUTs.  For
example, we cannot arbitrarily place the three families of quarks and leptons in the bulk or on either brane.
Moreover, the very act of obtaining three families, along with their respective locations, is fixed by the
requirement that the gauge embeddings and Wilson lines have to satisfy the modular invariance conditions
\cite{Dixon,vafa}. In addition, we cannot utilize the orbifold projections to remove all the $({\bf 6},{\bf
1},{\bf 1})$ color-triplet states as in the $\lgp{SO}{10}$ orbifold GUTs \cite{OGUT} and at the same time obtain
three families. We also find many massless states carrying unconventional representations under the SM gauge
group. These exotic states are commonplace in almost all known three-family models. Whether these models can give
rise to satisfactory phenomenology needs more detailed knowledge of the low-energy effective actions.   The
present status of our analysis is contained in this paper.

The paper is organized as follows.  In sect.~\ref{sec:ft} we briefly
review 5d field theories on the orbicircle ${\rm S^1}/{\mathbb Z}_2$
and present two orbifold GUT models with bulk gauge groups
$\lgp{SO}{10}$ and $\lgp{E}{6}$. The latter (model A1\footnote{This
model is denoted A1 since it corresponds to the first of several
string models discussed in the paper.}) is a novel 5d model with
many nice phenomenological features.  Then in sect.~\ref{z2z3} we
discuss the heterotic string construction of model A1. Using this
model as a guide we compare the heterotic string construction with
generic orbifold GUT models by restricting the compactified space to
a specific type (which is referred to as the orbifold GUT limit). We
show the equivalence between the matter states (in the untwisted and
some twisted sectors) in string-based models and the bulk states in
orbifold GUTs, as well as their KK excitations. We interpret
orbifold parities (for the bulk states) in the orbifold GUTs in
string theory language, and explain why the gauge embeddings and
Wilson lines cannot project away all the $({\bf 6,1,1})$
color-triplet states. These states may be needed to break the PS
group to that of the SM, as in the field theoretical model of
sect.~\ref{sec:E6}. In sect.~\ref{pheno} we focus on more of the
phenomenological aspects of model A1. In sect.~\ref{unif} we discuss
gauge coupling unification and the determination of the
compactification and string scales.  In sect.~\ref{sec:yukawa} we
examine the allowed Yukawa couplings (at both the renormalizable and
non-renormalizable levels) and their phenomenological consequences,
concentrating on the possibility of breaking the PS symmetry, mass
generation for the color-triplet fields and SM fermions, and proton
stability. We conclude in sect.~\ref{con}, listing the pros and cons
of the present models. Hopefully one can learn from the problems to
design better models in the future.

We have made an effort to make the paper more accessible to field theory model builders. Many of the details of
string constructions are relegated to four appendices. In appendix~\ref{review} we review the construction of
non-prime-order orbifold models with Wilson lines, highlighting its differences with the prime-order orbifold
construction. In appendix~\ref{model} we present three three-family ${\mathbb Z}_6$ models with PS gauge
symmetry. The complete matter spectra are listed in appendix \ref{notation}, where we also explain the notation
for the twisted-sector states. In appendix \ref{sec:sr} we review the string selection rules necessary for
determining non-trivial Yukawa couplings in a 4d effective theory, and in appendix \ref{sec:ac} list some allowed
couplings involving operators of interest in model A1. Finally, in appendix \ref{sec:RG} we study gauge coupling
unification and derive the Georgi-Quinn-Weinberg (GQW) relations \cite{GQW} in the orbifold GUT limit. These
relations allow us to determine various mass scales in our models.

\section{5d orbifold GUT models on ${\rm S^1}/{\mathbb
Z}_2$}\label{sec:ft}

Let us briefly review the geometric picture of orbifold GUT models
compactified on an orbicircle ${\rm S}^1/{\mathbb Z}_2$. The space
group of ${\rm S}^1/\mz_2$ is composed of two actions, a
translation, ${\cal T}:\,x^5\rightarrow x^5+2\pi R$, and a space
reversal, ${\cal P}:\,x^5\rightarrow -x^5$. There are two
(conjugacy) classes of fixed points, $x^5=2n\pi R$ and $(2n+1)\pi
R$, where $n\in\mz$.

The space group multiplication rules imply ${\cal T}{\cal P}{\cal
T}={\cal P}$, so we can replace the translation by a composite
$\mz_2$ action ${\cal P}'={\cal P}{\cal T}: x^5\rightarrow -x^5+2\pi
R$. The orbicircle ${\rm S}^1/{\mathbb Z}_2$ is equivalent to an
${\mathbb R}/({\mathbb Z}_2\times{\mathbb Z}'_2)$ orbifold, whose
fundamental domain is the interval $[0,\,\pi R]$, and the two ends
$x^5=0$ and $x^5=\pi R$ are fixed points of the ${\mathbb Z}_2$ and
${\mathbb Z}_2'$ actions respectively.

A generic 5d field $\Phi$  has the following transformation
properties under the ${\mathbb Z}_2$ and ${\mathbb Z}'_2$
orbifoldings (the 4d space-time coordinates are suppressed),
\be {\cal P}:\,\Phi(x^5)\rightarrow\Phi(-x^5)=P\Phi(x^5)\,,\qquad
{\cal P}':\,\Phi(x^5)\rightarrow \Phi(-x^5+2{\pi R})=P'\Phi(x^5)\,,
\ee
where $P,\,P'=\pm$ are \textit{orbifold parities}. In general cases
$P'\neq P$; this corresponds to the translation ${\cal T}$ being
realized non-trivially by a degree-2 Wilson line (i.e., background
gauge field). The four combinations of orbifold parities give four
types of states, with wavefunctions
$\Phi_{++}(x^5)\sim\cos(mx^5/R)$,
$\Phi_{+-}(x^5)\sim\cos[(2m+1)x^5/2R]$,
$\Phi_{-+}(x^5)\sim\sin[(2m+1)x^5/2R]$ and
$\Phi_{--}(x^5)\sim\sin[(m+1)x^5/R]$, where $m\in\mathbb Z$. The
corresponding KK towers have masses
\bea M_{\rm KK}=\left\{
\begin{array}{ll}
m/R  &\,\,{\rm for}\,(PP')=(++)\,,\\
(2m+1)/2R &\,\,{\rm for}\,(PP')=(+-)\,\,{\rm and}\,\,(-+)\,,\\
(m+1)/R &\,\, {\rm for}\,(PP')=(--)\,.\label{kkmass}
\end{array}\right.
\eea
Note that only the $\Phi_{++}$ field possesses a massless zero mode.

\subsection{An $\lgp{SO}{10}$ orbifold GUT}\label{sec:so10}

Consider the 5d orbifold GUT model of ref.~\cite{OGUT}. The model
has an $\lgp{SO}{10}$ symmetry broken to the PS gauge group, $\ps$,
in 4d, by orbifold parities. The compactification scale $M_c = (\pi
R)^{-1}$ is assumed to be much less than the cutoff scale. (In
string theory the cutoff scale is given by the string scale $M_{\rm
string}$.)

The gauge field is a 5d vector multiplet ${\cal
V}=(A_M,\lambda,\lambda',\sigma)$, where $A_M,\,\sigma$ (and their
fermionic partners $\lambda, \ \lambda'$) are in the adjoint
representation (${\bf 45}$) of $\lgp{SO}{10}$. This multiplet
consists of one 4d ${\mathcal N}= 1$ supersymmetric vector multiplet
$V=(A_\mu,\lambda)$ and one 4d chiral multiplet
$\Sigma=((\sigma+iA_5)/\sqrt{2},\lambda')$. We also add a 5d
hypermultiplet ${\cal H}=(\phi,\phi^c,\psi,\psi^c)$ in the
${\bf 10}$ representation. It decomposes into two
4d chiral multiplets $H=(\phi,\psi)$ and $H^c=(\phi^c,\psi^c)$ in
complex conjugate representations. This model has an ${\mathcal N}=2$
extended supersymmetry. The 5d gravitino
$\Psi_M=(\psi^1_M,\psi_M^2)$ decomposes into two 4d gravitini
$\psi_\mu^1$, $\psi_\mu^2$ and two dilatini $\psi_5^1$, $\psi_5^2$.
To be consistent with the 5d supersymmetry transformations one can
assign positive parities to $\psi_\mu^1+\psi_\mu^2$,
$\psi_5^1-\psi_5^2$ and negative parities to
$\psi_\mu^1-\psi_\mu^2$, $\psi_5^1+\psi_5^2$; this assignment partially breaks
${\mathcal N}=2$ to ${\mathcal N}=1$ in 4d.

The orbifold parities for various states in the vector and hyper
multiplets are chosen as follows \cite{OGUT} (where we have
decomposed all the fields into PS irreducible representations)
\bea \setlength{\arraycolsep}{0.25in}
\begin{array}{llllll}
\hline
{\rm States} & P & P' &{\rm States} & P & P'\\
 \hline
V({\bf 15,1,1}) & + & + & \Sigma({\bf 15,1,1}) & - & -\\
V({\bf 1,3,1})  & + & + & \Sigma({\bf 1,3,1}) & - & -\\
V({\bf 1,1,3})  & + & + & \Sigma({\bf 1,1,3}) & - & -\\
V({\bf 6,2,2})  & + & - & \Sigma({\bf 6,2,2}) & - & +\\
H({\bf 6,1,1})  & + & - & H^c({\bf 6,1,1}) & - & + \\
H({\bf 1,2,2})  & + & + & H^c({\bf 1,2,2}) & - & -\\
\hline
\end{array}\,.
\eea
We see the fields supported at the orbifold fixed points $x^5=0$ and
$\pi R$ have parities $P=+$ and $P'=+$ respectively. They form complete
representations under the $\lgp{SO}{10}$ and PS groups; the
corresponding fixed points are called $\lgp{SO}{10}$ and PS
``branes.'' In a 4d effective theory one would integrate out all the
massive states, leaving only massless modes of the $P=P'=+$ states.
With the above choices of orbifold parities, the PS gauge fields and
the $H({\bf 1,2,2})$ chiral multiplet are the only surviving states
in 4d. The $H({\bf 6,1,1})$ and $H^c({\bf 6,1,1})$ color-triplet states
are projected out, solving the doublet-triplet splitting problem
that plagues conventional 4d GUTs.

\subsection{An $\lgp{E}{6}$ orbifold GUT}\label{sec:E6}

We now consider a novel  5d orbifold GUT with an $\lgp{E}{6}$ gauge
symmetry.
In analogy to the model in sect.~\ref{sec:so10} we take the 5d gauge
field, given by $(V,\Sigma)$, in the adjoint representation (${\bf
78}$) of $\lgp{E}{6}$. In addition to this we add a matter
hypermultiplet $H({\bf 27}) + H^c({\bf\overline{27}})$.

We define two orbifold parities \be P = \exp(\pi{\rm i} Q_Z/3) \times
P_F\,,\qquad P^\prime = \exp[3 \pi{\rm i} (B- L)/2] \times
P^\prime_F\,,\label{eq:pp'}
\ee
which break the $\lgp{E}{6}$ via $P$ to $\lgp{SO}{10}$ and then via
$P^\prime$ to PS. $Q_Z$ is the abelian charge in $\lgp{E}{6}$
commuting with $\lgp{SO}{10}$, normalized such that the {\bf 27}
decomposes to ${\bf 16}_1 + {\bf 10}_{-2} + {\bf 1}_4$, and $P_F,
P^\prime_F$ are appropriate discrete flavor charges. (For explicit
definition of the parities in the corresponding string model, see
sect. \ref{parities}.) It is easy to obtain the following projections
to $(++)$ modes, where the first step follows from $P$ alone and the
second follows from the subsequent action of $P^\prime$, \bea V =
{\bf 78} \rightarrow & {\bf 45}
\rightarrow & {\rm adjoint \; of \; PS}\,,\nn \\
\Sigma  = {\bf 78} \rightarrow &   {\bf 16} + {\bf \overline{16}}
\rightarrow & f_3^c + \overline\chi^c\,, \nn\\
{\bf 27}  \rightarrow & {\bf 16} \rightarrow  & f_3\,,  \nn \\
 {\bf \overline{27}} \rightarrow & {\bf 10} \rightarrow & h\,. \eea
In this equation, we have identified the third family of quarks and
leptons as well as the MSSM Higgs-doublet pair ($h=H_U+H_D$ where
$H_U$ and $H_D$ are the MSSM Higgs doublets responsible for the up-
and down-type quark/charged lepton masses), \be f_3^c = ({\bf
\overline 4, 1, 2}), \qquad f_3 = ({\bf 4, 2, 1}), \qquad h = ({\bf
1, 2, 2}) . \ee

As a consequence of the fact that the third family and Higgs doublet
come from the bulk gauge and {\bf 27} hypermultiplets we obtain a
gauge-Yukawa unification relation, \be \lambda_t = \lambda_b =
\lambda_\tau = g^{}_{\rm 4d} \equiv \sqrt{4 \pi \alpha_{\rm GUT}}\,,
\label{eq:yg} \ee where $g^{}_{\rm 4d}$ is the 4d gauge coupling
constant at the compactification scale. This relation can be seen by
inspecting the 5d bulk gauge interaction \be \int_0^{\pi R}
dx^5\,\biggl(g^{}_{\rm 5d} H^c \Sigma H\biggr) \rightarrow g^{}_{\rm
4d}\,h f_3^c f_3\,, \ee where $g^{}_{\rm 4d} = g^{}_{\rm
5d}\sqrt{M_c}$.

Of course, we then need to spontaneously break PS to the SM via the
standard Higgs mechanism.   This can be accomplished when the
``right-handed neutrino'' fields in \be \chi^c = ({\bf \overline 4,
1, 2}), \qquad  \overline \chi^c = ({\bf 4, 1, 2}) \ee obtain
non-vanishing vacuum expectation values (vevs) \be \langle \nu^c
\rangle_{\chi^c} = \langle \overline \nu^c \rangle_{\overline
\chi^c} = M_{\rm PS} . \label{eq:nuvev}\ee We already have one such
state but we need more (if only for anomaly cancellation). Consider
the addition of three more {\bf 27} hypermultiplets given by
$3\times( {\bf 27} + {\bf \overline{27}} )$. Upon applying the
orbifold parities we find \be 3 \times({\bf 27} + {\bf
\overline{27}} )
 \rightarrow 2 ( {\bf 16} ) + {\bf \overline{16}} + 3 ( {\bf 10} ) \rightarrow
2 ( \chi^c ) + { \overline \chi^c } + 3 ( C ), \ee  where $C= ({\bf 6, 1, 1})$.  We now have a total $2 (\chi^c +
\overline \chi^c)$ fields.  Note, with one $C$, one $\chi^c, \ \bar \chi^c$ pair and a superpotential given by
\be {\cal W} = \chi^c \chi^c C + \overline \chi^c \overline \chi^c  C, \label{eq:sp} \ee we can give mass to the
color triplets and also break PS to the SM along a D- and F-flat direction. (The D-flatness condition requires
$\langle\nu^c\rangle_{\chi^c} = \langle\overline\nu^c\rangle_{\overline \chi^c}$ and $\langle
\overline{D}_1\rangle_{\chi^c} = \langle\overline{D}^c_1\rangle_{\overline \chi^c}$, then the F-flatness
condition requires further that one of these vevs, say the second, be zero.)   In the end, however, we must
guarantee that the extra $\chi^c, \ \bar \chi^c, \ C$ states obtain mass above the PS breaking scale.

\begin{figure}[ht]
\scalebox{0.6}{\includegraphics{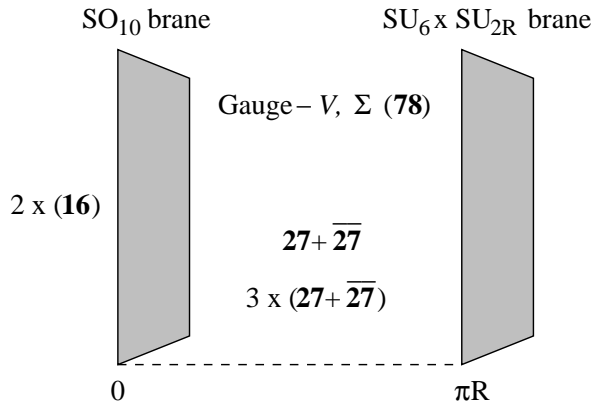}} \caption{5d $\lgp{E}{6}$ orbifold GUT model with bulk and brane
states. The bulk gauge symmetry is broken to $\lgp{SO}{10}$ on the end of world brane at $x^5=0$ and to
$\lgp{SU}{6}\times\lgp{SU}{2R}$ at $x^5=\pi R$. The massless sector of the 4d effective theory has a PS gauge
symmetry. In addition, the bulk contains four hypermultiplets, and the $\lgp{SO}{10}$ brane contains two spinor
representations, giving rise to the first two matter families.} \label{fig5d}
\end{figure}

But what about the first two families? When constructing an orbifold
GUT, one has the option of whether to place the first two families
in the bulk or on either brane.   One of the main considerations is
to avoid rapid proton decay due to gauge exchange and another is to
generate a hierarchy of fermion masses. If the compactification
scale is much smaller than the GUT scale, say $M_c \ll M_{\rm GUT}$,
then it is not possible to place the first two families on the
$\lgp{SO}{10}$ brane. It would however be fine to place them in the
bulk or on the $\lgp{SU}{6}\times\lgp{SU}{2R}$ brane, since in the
first case the families are in irreducible representations with
massive KK modes, while in the latter case one family is contained
in two irreducible representations $({\bf 15, 1}) + ({\bf \overline
6, 2})$, also with massive KK modes. In both cases, gauge exchange
takes massless quarks and leptons into massive states. Hence there
is \textit{no} problem with proton decay. If however $M_c \geq
M_{\rm GUT}$ then one can place the first two families on either
brane. Unfortunately, in string theory, we do not get to choose
\textit{easily} where to place the families. It is determined by the
choice of vacuum.   In the heterotic string version of the model
(model A1 in appendix \ref{app:models}) we find two families sitting
on the $\lgp{SO}{10}$ brane, as in fig. \ref{fig5d}.

\section{Heterotic string construction of effective orbifold GUTs \label{z2z3}}

In appendix \ref{review} we review the rules for constructing
heterotic string models compactified on an abelian symmetric
orbifold with discrete Wilson lines.   Then in appendix
\ref{app:models} we construct three three-family ${\mathbb Z}_6$
orbifold models with two Wilson lines, labelled models A1, A2 and B.
We have obtained the complete spectra of massless states (plus KK
excitations for these models in certain limits). As we now show,
model A1 is the string equivalent to the orbifold GUT in
sect.~\ref{sec:E6}.

The following discussion relies greatly on the notation and discussion in appendices \ref{review} and
\ref{app:models}. Briefly stated, the heterotic string combines a 10d superstring for right movers and a 26d
bosonic string for left movers. However 16 of the 26 left-moving dimensions are compactified on the $\Eee$ root
lattice. In order to obtain an effective 4d theory, we compactify six of the remaining ten dimensions on a
symmetric orbifold defined by a six torus modded by a point group ${\mathbb Z}_6$ with the twist vector \be {\bf
v}_6=\frac{1}{6}(1,2,-3),\label{eq:tv} \ee i.e. the three compactified complex coordinates transform as
$Z_i\rightarrow\exp(2\pi{\rm i}{\bf v}_6^i) Z_i$ under the twist. The embedding of orbifold twists in the gauge
degrees of freedom is realized by gauge twists, ${\bf V}$, and lattice translations by discrete Wilson lines,
${\bf W}$.  In abelian orbifolds these vectors simply shift the appropriate $\lgp{E}{8}\times\lgp{E}{8}$ roots.

\begin{figure}
\scalebox{0.65}{\includegraphics{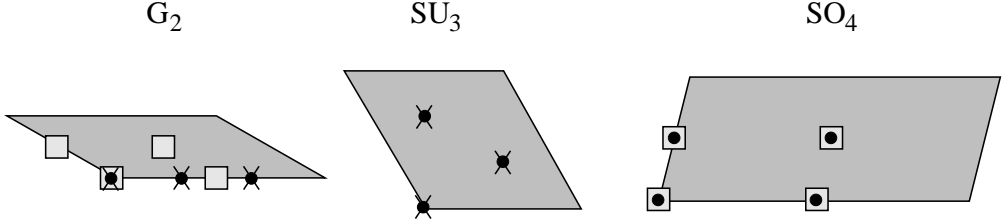} }\caption{Fundamental
region of the root lattice $\lat$. The filled circles, crosses and
squares represent fixed points in the $T_1$, $T_{2,4}$ and $T_3$
twisted sectors. See appendix~\ref{app:models} for further
details.\label{fig:lattice}}
\end{figure}

To be definite, we choose the six torus as the Lie algebra root
lattice $\lat$, as shown in fig.~\ref{fig:lattice}. Denoting the
basis of the lattice by ${\bf e}=\left(\begin{array}{c}
{\bf e}_1\\
{\bf e}_2
\end{array}\right)
\oplus
\left(\begin{array}{c}
{\bf e}_3\\
{\bf e}_4
\end{array}\right)
\oplus
\left(\begin{array}{c}
{\bf e}_5\\
{\bf e}_6
\end{array}\right)$, whose inner
product gives the Cartan matrix of the corresponding Lie algebra,
the ${\mathbb Z}_6$ discrete symmetry can be realized by the Coxeter
element,\footnote{The Coxeter element is an inner automorphism of the
lattice, composed of products of Weyl reflections of the
corresponding root lattice. For example, the Coxeter element of $\lgp{G}{2}$
is simply $s_1s_2$ where $s_1$, $s_2$ are the two reflections with respect to
planes orthogonal to the two simple roots.
A generalized Coxeter element may also
include outer automorphism of the lattice.}
\be
C=\left(
\begin{array}{rr}
2 & -1\\
3 & -1
\end{array}\right)\oplus
\left(
\begin{array}{rr}
0 & -1 \\
1 & -1
\end{array}\right)\oplus
\left(\begin{array}{rr}
-1 & 0\\
0 & -1
\end{array}\right)\,,
\ee
under which the basis is transformed to $C{\bf e}$.  The Coxeter
element has eigenvalues ${\rm e}^{\pm{{\rm i} \pi }/{3}}$,
${\rm e}^{\pm{2{\rm i} \pi}/{3}}$
and ${\rm e}^{\pm{{\rm i} \pi}}$, thus the three two-dimensional sub-lattices
have degree-6, 3, and 2 cyclic symmetries, and the corresponding
numbers of fixed points are 1, 3 and 4.

There are three K\"ahler class moduli (${\cal T}_{1,2,3}$), whose
real parts parameterize the sizes of the three tori, and one complex
structure modulus (${\cal U}_3$), which parameterizes the shape of
the third torus. Explicitly, ${\rm Re}{\cal T}_3= 2 R  R^\prime
\sin\phi$, and ${\cal U}_3=\frac{R}{R^\prime}\,{\rm e}^{{\rm i}\phi}$,
where $R,
R^\prime$ are the lengths of the two axes of the
$\lgp{SO}{4}$-lattice and $\phi$ their relative angle. These moduli
are arbitrary parameters. One may make the length of one axis (along
which one puts the degree-2 Wilson line, ${\bf W}_2$), say $R$,
large compared to the string length scale while keeping all other
dimensions small. In this limit (for length scales larger than the
string scale but smaller than the radius $R$), the low energy theory
is effectively five dimensional.\footnote{It should be obvious that our
construction can be generalized to 6d models, simply by taking both
$R$ and $R'$ large compared to the string length scale. These models
are related to 6d orbifold GUTs compactified on T$^2/{\mathbb
Z}_2$.} The $\lgp{SO}{4}$ lattice, on which only the ${\mathbb Z}_2$
sub-orbifold twist acts, has four fixed points. With only one
degree-2 Wilson line, the fixed points split into two inequivalent
classes, labelled by the winding number $n_2=0, 1$. Thus in our setup
the fifth dimension is equivalent to the orbicircle S$^1/{\mathbb
Z}_2$ where each of the two fixed points has a degree-2
degeneracy.

Note that we can reinterpret the ${\mathbb Z}_6$ models of appendix
\ref{app:models} in terms of the equivalent ${\mathbb Z}_2\times
{\mathbb Z}_3$ orbifold (where the ${\mathbb Z}_2$ (${\mathbb Z}_3$)
sub-orbifold twist acts on the $\lgp{G}{2}$ and $\lgp{SO}{4}$
($\lgp{G}{2}$ and $\lgp{SU}{3}$) sub-lattices). This point of view
is more useful for our comparisons with the orbifold GUTs in sect.
\ref{sec:E6}. Labelling a twisted sector in the ${\mathbb Z}_6$
model by $T_k$ where $k=1,2,\cdots,5$ and  in the ${\mathbb
Z}_2\times {\mathbb Z}_3$ model by $T_{(k,l)}$ where $k=0,1$ and
$l=0,1,2$, then the correspondence between the twisted sectors in
the ${\mathbb Z}_6$ and ${\mathbb Z}_2\times {\mathbb Z}_3$
orbifolds is the following: \be \setlength{\arraycolsep}{0.1in}
\begin{array}{l|lllll}
\hline
{\mathbb Z}_6\ {\text{orbifold}}& T_1 & T_2 & T_3 & T_4 & T_5\\\hline {\mathbb
Z}_2\times {\mathbb Z}_3\ {\text{orbifold}}
& T_{(1,2)} & T_{(0,1)} & T_{(1,0)} &
T_{(0,2)} & T_{(1,1)}\\\hline
\end{array}\,.\label{cor}
\ee
The $T_{2,4}$ sectors, which will shortly be identified with the bulk states in the language of orbifold GUTs,
have $k=0, \ l = 1,2$; therefore they are untwisted by the ${\mathbb Z}_2$ twist.

\subsection{Model A1 from the ${\mathbb Z}_6$ orbifold
compactification \label{ogut}}

We now examine model A1 of appendix \ref{app:models}. Consider first
the model with only the ${\mathbb Z}_3$ sub-orbifolding being
imposed (i.e., with twist vector ${\bf v}_3=2{\bf v}_6$, gauge twist
${\bf V}_3=2{\bf V}_6$ and a degree-3 Wilson line ${\bf W}_3$, where
${\bf v}_6$, ${\bf V}_6$ and ${\bf W}_3$ are given in
eqs.~\ref{eq:tv}, \ref{gt1} and \ref{wla1}), we find a 6d ${\mathcal
N}=2$ model with observable-sector gauge group E$_6$ (modulo abelian
factors). Matter fields of the observable sector consist of 6d
${\mathcal N}=2$ hypermultiplets in the following representations,
\be
U\,{\rm sectors}:\,{\bf 27}+{\bf\overline{27}} ,\qquad T\,{\rm
sectors}:\,3\times({\bf 27}+{\bf\overline{27}})\,.\label{bulkmatter}
\ee

The remaining ${\mathbb Z}_2$ twist acts as a space reversal on the third compactified complex dimension,
$Z_3\rightarrow -Z_3$. The ${\mathbb Z}_3$ models have two gravitini with the $\lgp{SO}{8}$ momentum vectors,
${\bf r}=\frac{1}{2}(1,1,1,1)$ and $\frac{1}{2}(1,-1,-1,1)$, in the Ramond sector of the right-moving superstring
(see appendix \ref{review} for notation). Only one of them, ${\bf r}=\frac{1}{2}(1,1,1,1)$, satisfies the
${\mathbb Z}_2$ projection, ${\bf r}\cdot {\bf v}_2={\mathbb Z}$. Hence the ${\cal N}=2$ supersymmetry is broken
to that of ${\mathcal N}=1$ in 4d.

Gauge symmetry breaking induced by the ${\mathbb Z}_2$ orbifolding is as follows. The twist vector ${\bf v}_2$ is
embedded in the gauge degrees of freedom in two different ways, with gauge twists ${\bf V}_2$ and ${\bf
V}'_2={\bf V}_2+{\bf W}_2$ where ${\bf V}_2=3{\bf V}_6$ and ${\bf W}_2$ is given in eq.~\ref{wla1}.
$\lgp{E}{6}$ generators in the Cartan-Weyl basis are transformed under the ${\mathbb Z}_2$ action as $E_{\bf
P}\rightarrow {\rm e}^{2\pi {\rm i} {\bf P}\cdot {\bf V}_2} E_{\bf P}$
and $E_{\bf P}\rightarrow {\rm e}^{2\pi {\rm i} {\bf P}\cdot {\bf
V}'_2} E_{\bf P}$, thus the linearly-realized gauge groups consist of
roots satisfying ${\bf P}\cdot{\bf
V}_2$ and ${\bf P}\cdot{\bf V}'_2={\mathbb Z}$ respectively.
The pattern of symmetry breaking in the
observable sector can be summarized as follows:
\be
\setlength{\unitlength}{0.4in}
\begin{picture}(6,3)
\put(0,1.25){$\lgp{E}{6}$}
\put(2.3,2.5){$\lgp{SO}{10}$}
\put(1.7,0){$\lgp{SU}{6}\times\lgp{SU}{2R}$}
\put(4.9,1.25){PS}
\put(0.5,1.25){\vector(1,-1){1}}
\put(0.5,1.55){\vector(1,1){1}}
\put(3.8,0.25){\vector(1,1){1}}
\put(3.8,2.55){\vector(1,-1){1}}
\end{picture}
\ee
At the final step we have the complete ${\mathbb Z}_6$ model
with two discrete Wilson lines being imposed simultaneously; this
gives the PS symmetry group in the 4d effective theory.

In these two inequivalent implementations of the ${\mathbb Z}_2$
twist the non-trivial matter fields of $\lgp{SO}{10}$ and
$\lgp{SU}{6}\times\lgp{SU}{2R}$ are:
\be
\setlength{\arraycolsep}{0.25in}
\begin{array}{lll}
\hline {\rm Sectors} & \lgp{SO}{10}
& \lgp{SU}{6}\times\lgp{SU}{2R}\\
\hline
U_1 & {\bf 16} & ({\bf 15,1})\\
U_2 & {\bf 10} & ({\bf 6,2}) \\
U_3 & {\bf 16}+{\bf\overline{16}} & ({\bf 20,2})\\
T_{(0,1)} & 2\times {\bf 16}_++{\bf 10}_- &
2 ({\bf\overline 6,2})_++({\bf 15,1})_-\\
T_{(0,2)} & {\bf\overline{16}}_-+2\times {\bf 10}_+
& ({\bf 6,2})_-+2({\bf\overline{15},1})_+\\
\hline
\end{array}\,
\ee
where the subscripts $\pm$ represent \textit{intrinsic parities}, \be
p=\gamma\phi\,.\label{ip} \ee $p$ depends on the twist eigenvalue,
$\gamma$, and the oscillator phase, $\phi$; they are defined in
appendix~\ref{review}. Note that $p=+$ for gauge and
untwisted-sector states, and $p=+$ and $-$ have
multiplicities $2$ and $1$ respectively for non-oscillator
$T_{(01)}/T_{(02)}$ states.

Massless states in the untwisted and $T_{(0,1)},\,T_{(0,2)}$ twisted
sectors of model A1 are the intersections of those of the
$\lgp{SO}{10}$ and $\lgp{SU}{6}\times\lgp{SU}{2R}$ models. This can
be seen from the group branching rules. For example, the
$T_{(0,1)}$-sector matter has the following branchings,
\bea
\lgp{SO}{10} &\rightarrow& \ps  \nn\\
{\bf 16}_+&=&({\bf 4,2,1})_++({\bf\overline 4,1,2})_+\,,\nn\\
{\bf 10}_-&=&({\bf 6,1,1})_-+({\bf 1,2,2})_-\,,\\
\lgp{SU}{6}\times\lgp{SU}{2R}
&\rightarrow& \ps \nn\\
({\bf\overline 6,2})_+&=&({\bf\overline 4,1,2})_++({\bf 1,2,2})_+\,,\nn\\
({\bf 15,1})_-&=&({\bf 4,2,1})_-+({\bf 6,1,1})_-+({\bf
1,1,1})_-\,.
\eea
The states in common, $2({\bf\overline 4,1,2})_++({\bf 6,1,1})_-$, agree with that of the $T_2$-twisted sector in
eq.~\ref{modelA1}.

Massless fields in the other, i.e. $T_{(1,2)} (=T_1)$ and $T_{(1,0)}
(=T_3)$, twisted sectors are the unions of those of the
$\lgp{SO}{10}$ and $\lgp{SU}{6}\times\lgp{SU}{2R}$ models. Therefore
there are two sets of states, furnishing complete representations of
$\lgp{SO}{10}$ and $\lgp{SU}{6}\times\lgp{SU}{2R}$ respectively. For
example, the $T_1$ sector of model A1 contains $({\bf
4,2,1})+({\bf\overline 4,1,2})$ and $({\bf 4,1,1})+({\bf 1,2,1})$,
they are in the complete representations ${\bf 16}$ of
$\lgp{SO}{10}$ and $({\bf 6,1})$ of $\lgp{SU}{6}\times
\lgp{SU}{2R}$. In the notation of appendix \ref{review}, these two
sets of states have quantum numbers $n_2=0$ and $n_2=1$. (These
quantum numbers are the winding numbers along the direction where
the ${\bf W}_2$ Wilson line is imposed.) The $n_2=0$ and $n_2=1$
fixed points are thus the $\lgp{SO}{10}$ and
$\lgp{SU}{6}\times\lgp{SU}{2R}$ branes in the orbifold GUT language.

\subsection{Identifying orbifold parities in string theory \label{parities}}

To a certain degree, the above $\lgp{E}{6}$ heterotic model gives a string
theoretical realization of the orbifold GUT in sect.~\ref{sec:E6}.
Better yet, we also achieve an understanding of the orbifold
parities in terms of string theoretical quantities. Specifically,
the analogue of orbifold parities, eq.~\ref{eq:pp'}, in our
${\mathbb Z}_6$ string models can be defined as follows \cite{KRZ0}
\be P=p {\rm e}^{2\pi{\rm i}({\bf P}\cdot{\bf V}_2-{\bf r}\cdot{\bf
v}_2)}\,,\qquad P'=p{\rm e}^{2\pi{\rm i}({\bf P}\cdot{\bf V}'_2-{\bf
r}\cdot{\bf v}_2)}\,,\label{PP'}
\ee
where ${\bf V}_2$ and ${\bf V}_2'$ are the two inequivalent gauge
embeddings of the ${\mathbb Z}_2$ twist in sect.~\ref{ogut}, and $p$
is the intrinsic parity.

\begin{figure}[ht]
\scalebox{0.65}{\includegraphics{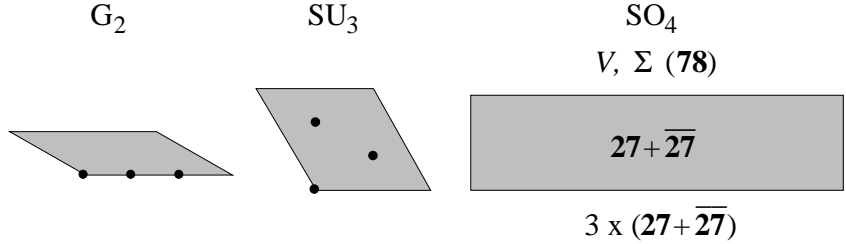} } \caption{$\lat$ lattice
with ${\mathbb Z}_3$ fixed points. The fields $V, \ \Sigma, \ $ and
${\bf 27} (\in U_1) + {\bf\overline{27}} (\in U_2)$ are bulk states
from the untwisted sectors. On the other hand, $3\times({\bf 27} +
{\bf\overline{27}})$ are ``bulk" states located on the
$T_{(0,1)}/T_{(0,2)}$ twisted sector ($\lgp{G}{2}$, $\lgp{SU}{3}$)
fixed points.}

\label{fig0}
\end{figure}
\begin{figure}[ht]
\scalebox{0.65}{\includegraphics{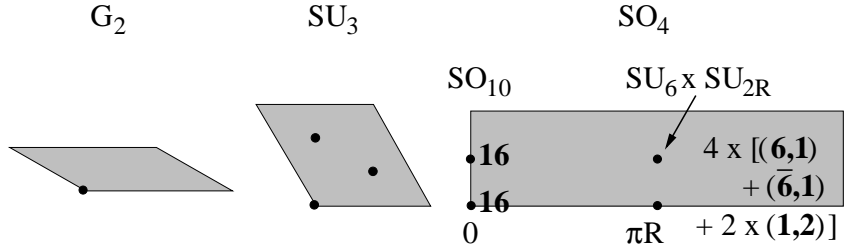} } \caption{$\lat$ lattice
with ${\mathbb Z}_6$ fixed points. The $T_{(1,1)}/ T_{(1,2)}$
twisted sector states sit at these fixed points.} \label{fig1}
\end{figure}
\begin{figure}[ht]
\scalebox{0.65}{\includegraphics{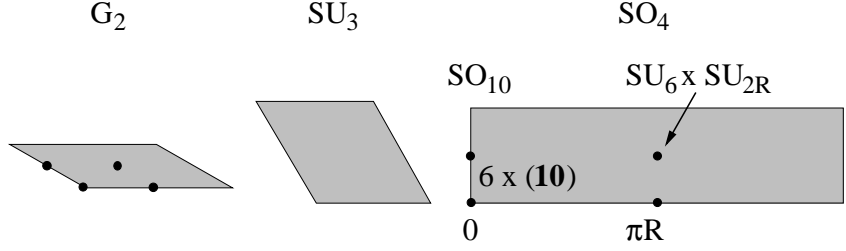} } \caption{$\lat$ lattice
with ${\mathbb Z}_2$ fixed points. The $T_{(1,0)}$ twisted sector
states sit at these fixed points.} \label{fig3}
\end{figure}

These parities can be deduced from the generalized Gliozzi-Scherk-Olive (GSO) projector
\cite{Gliozzi:1976qd,IMNQ2}, as in the paragraphs after eq.~\ref{GSO}. Since the terms in the exponents, ${\bf
P}\cdot{\bf V}_2-{\bf r}\cdot{\bf v}_2$ and ${\bf P}\cdot{\bf V}'_2-{\bf r}\cdot{\bf v}_2$, take integral or
half-integral values, $P$ and $P'$ are either $+$ or $-$. The orbifold translation corresponds to the difference
in $P$ and $P'$, i.e. $T={\rm e}^{2\pi{\rm i}{\bf P}\cdot{\bf W}_2}$.
The $P$, $P'$ and $T$ in string models have exactly
the same properties as that of the orbifold GUTs.

Evidently, in the $\lgp{E}{6}$ orbifold GUT model of sect.~\ref{sec:E6} states supported at the $\lgp{SO}{10}$
and $\lgp{SU}{6}\times\lgp{SU}{2R}$ branes are those with parities $P=+$ and $P'=+$, and states in the 4d
effective theory are those with parities $P=P'=+$; this agrees with the string theoretical interpretation, since
the parities in eq.~\ref{PP'} are nothing but the required GSO projections for the gauge, untwisted and
$T_{(01)}/T_{(02)}$ sector states (i.e. the bulk states) in string models. (The massless states, i.e.
$P=P^\prime=+$ modes from bulk and $T_{(0,1)}/T_{(0,2)}$ twisted sectors are shown in fig. \ref{fig0}.) From
information gathered in sect.~\ref{ogut} and appendix \ref{app:models}, we can also deduce the $P$ and $P'$
parities for the various bulk matter states. They are listed in table~\ref{table1}.
\begin{table}\caption{\label{table1}Parities for the bulk states in model A1,
computed from eq.~\ref{PP'}. The states have been decomposed to
the PS irreducible representations.}
\begin{ruledtabular}
\setlength{\tabcolsep}{0.2in}
\begin{tabular}{llcclcc}
Multiplicities & States & $P$ & $P'$ &States & $P$ & $P'$\\
\hline
1 & $V({\bf 15,1,1})$  & $+$ & $+$ &$\Sigma({\bf 15,1,1})$  & $-$ & $-$\\
1 & $V({\bf 1,3,1})$   & $+$ & $+$ &$\Sigma({\bf 1,3,1})$   & $-$ & $-$\\
1 & $V({\bf 1,1,3})$   & $+$ & $+$ &$\Sigma({\bf 1,1,3})$   & $-$ & $-$\\
1 & $V({\bf 6,2,2})$   & $+$ & $-$ &$\Sigma({\bf 6,2,2})$   & $-$ & $+$\\
1 & $V({\bf 4,2,1})$   & $-$ & $+$ &$\Sigma({\bf 4,2,1})$   & $+$ & $-$\\
1 & $V({\bf\overline 4,1,2})$& $-$ & $-$ &$\Sigma({\bf\overline 4,1,2})$& $+$ & $+$\\
1 & $V({\bf\overline 4,2,1})$& $-$& $+$ &$\Sigma({\bf\overline 4,2,1})$& $+$ & $-$\\
1 & $V({\bf 4,1,2})$   & $-$ & $-$ &$\Sigma({\bf 4,1,2})$   & $+$ & $+$\\
\hline
1 & $H({\bf 4,2,1})$   & $+$ & $+$ &$H^c({\bf\overline 4,2,1})$   & $-$ &$-$\\
1 & $H({\bf\overline 4,1,2})$& $+$ & $-$ &$H^c({\bf 4,1,2})$      & $-$& $+$\\
1 & $H({\bf 6,1,1})$   & $-$ & $+$ &$H^c({\bf 6,1,1})$      & $+$& $-$ \\
1 & $H({\bf 1,2,2})$   & $-$ & $-$ &$H^c({\bf 1,2,2})$      & $+$& $+$\\
\hline
2 & $H({\bf 4,2,1})_+$   & $+$ & $-$ &$H^c({\bf\overline 4,2,1})_+$   & $-$ &$+$\\
2 & $H({\bf\overline 4,1,2})_+$& $+$ & $+$ &$H^c({\bf 4,1,2})_+$      & $-$& $-$\\
2 & $H({\bf 6,1,1})_+$   & $-$ & $-$ &$H^c({\bf 6,1,1})_+$      & $+$& $+$ \\
2 & $H({\bf 1,2,2})_+$   & $-$ & $+$ &$H^c({\bf 1,2,2})_+$      & $+$& $-$\\
1 & $H({\bf 4,2,1})_-$   & $-$ & $+$ &$H^c({\bf\overline 4,2,1})_-$   & $+$ &$-$\\
1 & $H({\bf\overline 4,1,2})_-$& $-$ & $-$ &$H^c({\bf 4,1,2})_-$      & $+$& $+$\\
1 & $H({\bf 6,1,1})_-$   & $+$ & $+$ &$H^c({\bf 6,1,1})_-$      & $-$& $-$ \\
1 & $H({\bf 1,2,2})_-$   & $+$ & $-$ &$H^c({\bf 1,2,2})_-$      & $-$& $+$\\
\end{tabular}
\end{ruledtabular}
\end{table}

KK masses for these bulk states can also be derived in string models. The mode expansions of the coordinates
corresponding to the $\lgp{SO}{4}$ lattice are $X^i_{L,R}=x^i_{L,R}+p^i_{L,R}(\tau\pm\sigma)+{\rm oscillator\,
terms}$, with $p^i_L$, $p^i_R$ given by eq.~\ref{PLR}. The ${\mathbb Z}_2$ action maps $m$ to $-m$, $n$ to $-n$
and ${\bf W}_2$ to $-{\bf W}_2$, so physical states must contain linear combinations,
$|m,n\rangle\pm|-m,-n\rangle$; the eigenvalues $\pm 1$ correspond to the first ${\mathbb Z}_2$ parity of the
orbifold GUT models. The second embedding corresponds to a non-trivial Wilson line; it shifts the KK level by
$m\rightarrow m+{\bf P}\cdot{\bf W}_2$. Since $2{\bf W}_2$ is a vector of the integral
$\lgp{E}{8}\times\lgp{E}{8}$ lattice, the shift ${\bf P}\cdot {\bf W}_2$ must be an integer or half-integer. In
the orbifold GUT limit when the winding modes and the KK modes in the short direction of $\lgp{SO}{4}$ decouple,
eq.~\ref{PLR} reproduces the field theoretical mass formula in eq.~\ref{kkmass}.

As seen in sect.~\ref{ogut}, matter states in the $T_{(1,1)}/
T_{(1,2)}$ and $T_{(1,0)}$ twisted sectors, which may be identified
with the first two families, are localized on the two inequivalent
fixed points in the $\lgp{SO}{4}$ lattice. They are the
$\lgp{SO}{10}$ and $\lgp{SU}{6}\times \lgp{SU}{2R}$ brane states
(See figs. \ref{fig1} and \ref{fig3}). These twisted-sector states
are more tightly constrained than their orbifold GUT counterparts.
In orbifold GUT models the only consistency requirement is the
chiral anomaly cancellation, thus one can add arbitrary numbers of
vector-like representations to the branes. String models have to
satisfy more stringent modular invariance conditions
\cite{Dixon,vafa} (of course, one-loop modular invariance guarantees
the model is anomaly free, up to a possible anomalous abelian factor
\cite{LSW}), which also constrains any additional matter in
vector-like representations.

\subsection{Other models}

In this subsection, we discuss the other two models in appendix
\ref{app:models} in the orbifold GUT language, for completeness.
These models do not have matter-Higgs couplings at the renormalizable level,
and may have limited phenomenological interest.

Model A2 has already been analyzed in ref.~\cite{KRZ0}. In the 5d
bulk, it has an $\lgp{SO}{10}\times\lgp{SU}{2}$ gauge symmetry and
the following set of matter states,
\be
U\,{\rm sectors}: ({\bf 16,1})+({\bf 1,2}),\qquad T\,{\rm sectors}:
3({\bf 16, 1})+6({\bf 10,1})+15({\bf 1,2})\,.
\ee

The bulk gauge group is unbroken at the fixed point $x^5=0$ and
broken to the PS group at $x^5=\pi R$ respectively, the states
supported at these two points are
\be
\setlength{\arraycolsep}{0.25in}
\begin{array}{lll}
\hline {\rm Sectors} & \lgp{SO}{10}\times\lgp{SU}{2} & {\rm PS}\\\hline
U_1 & ({\bf 16,1}) & ({\bf 4,2,1})\\
U_2 & ({\bf 1,2}) & ({\bf 4,1,2})\\
U_3 & - & ({\bf 6,2,2})\\
T_{(0,1)} & 2({\bf 16,1})_++2({\bf 10,1})_-
& ({\bf 6,1,1})_-+2({\bf 1,2,2})_+\\
& +2({\bf 1,2})_++4({\bf 1,2})_- & +2({\bf 6,1,1})_++({\bf 1,2,2})_-\\
& & +({\bf 4,2,1})_-+2({\bf\overline 4,1,2})_+\\
T_{(0,2)} & ({\bf\overline{16},1})_-+4({\bf 10,1})_+
& 2({\bf 6,1,1})_++({\bf 1,2,2})_-\\
& +8({\bf 1,2})_++({\bf 1,2})_- & +({\bf 6,1,1})_-
+2({\bf 1,2,2})_+\\
& & +({\bf 4,1,2})_-+2({\bf\overline
4,2,1})_+\\
\hline
\end{array}\,.
\ee

Model B is similar to model A1, with an $\lgp{E}{6}$ bulk group and the
same set of bulk states as in eq.~\ref{bulkmatter}. The $\lgp{E}{6}$
group is broken to $\lgp{SO}{10}$ and $\lgp{SU}{6}\times\lgp{SU}{2L}$
respectively at the two fixed points, and the matter states are
\be
\setlength{\arraycolsep}{0.25in}
\begin{array}{lll}
\hline
{\rm Sectors} & \lgp{SO}{10} & \lgp{SU}{6}\times\lgp{SU}{2L}\\
\hline
U_1 & {\bf\overline{16}} & ({\bf 6,2})\\
U_2 & {\bf 10} & ({\bf 15,1})\\
U_3 & {\bf 16}+{\bf\overline{16}} & ({\bf 20,2})\\
T_{(0,1)} & 2\times {\bf\overline{16}}_++{\bf 10}_- &
({\bf 6,2})_-+2({\bf\overline{15},1})_+\\
T_{(0,2)} & {\bf 16}_-+2\times {\bf 10}_+
& 2({\bf\overline 6,2})_++({\bf 15,1})_-\\
\hline
\end{array}\,.
\ee

The 4d effective theories of models A2 and B have a PS symmetry, and
the complete matter content is listed in appendix \ref{app:models}.
Similar to model A1, matter fields in the untwisted and $T_2/T_4$
twisted sectors can be traced back to the states in the above two
tables, by using appropriate group branching rules. We also note
that both models contain two families of chiral matter from the
fixed point at $x^5=0$ and one family from the bulk, a common
feature to all our models. This feature predicts a non-abelian
dihedral $\lgp{D}{4}$ family symmetry -- {\em a novelty in string
model building} -- as we will see in the next section.

\subsection{The color-triplet problem}\label{sec:ct}

A major motivation for constructing orbifold GUT models is the well-known doublet-triplet splitting problem in
conventional 4d GUTs. Orbifold GUTs solve this problem by assigning appropriate orbifold parities to the Higgs
doublets and triplets, such that the triplets are automatically projected out of the effective theory
\cite{kawamura}.\footnote{Of course, more conventional field theoretical mechanisms \cite{md} have been widely
studied in the literature.} We have already seen this in the $\lgp{SO}{10}$ model in sect.~\ref{sec:so10}. This
mechanism, however, usually cannot be trivially implemented in heterotic models.

The difficulty is largely due to the intricate nature of string models. These models need to satisfy  delicate
modular invariance consistency conditions \cite{Dixon,vafa} and are physically more constrained than the orbifold
GUTs. Before imposing any Wilson line, the ${\mathbb Z}_6$ models of eqs.~\ref{gt1} and \ref{gt2} always contain
the ${\bf 10}$ representation of $\lgp{SO}{10}$ simultaneously in several sectors. We find it is impossible to
design modular-invariant Wilson lines to fulfill the following requirements: (a) break the gauge group to PS in
4d, (b) give rise to three chiral families, and (c) eliminate the color triplets altogether. Furthermore, the
${\bf 10}$ representation of the $T_3$ sector in model A1 does not suffer from additional projections even when
the ${\bf W}_2$ Wilson line is turned on. Indeed, it simply decomposes to the $({\bf 6},{\bf 1},{\bf 1}) +({\bf
1},{\bf 2},{\bf 2})$ representations under the PS group.

Although the presence of many color triplets is a nuisance, one
${\bf 3} + {\bf\overline 3}$ pair may be necessary to facilitate the
breaking of PS to the SM gauge group, as illustrated in the E$_6$
model in sect.~\ref{sec:E6}. Moreover, it is not entirely clear that
the color triplets in our models pose the same problems as in
conventional GUTs. Indeed, although there are color triplets (those
of the $T_1$/$T_3$ twisted sectors) with doublet companions having
exactly the same quantum numbers, in general we also have $({\bf
6,1,1})$ and $({\bf 1,2,2})$ states with different quantum numbers,
in all three models. The usual doublet-triplet problem does not
necessarily apply for the second situation. We need to check whether
it is possible to make all color triplets sufficiently heavy while
(hopefully) keeping one MSSM Higgs-doublet pair light; this requires
a better understanding of the effective actions of our models and
will be examined for model A1 in the next section.

\section{Phenomenology of model A1}\label{pheno}

We have seen in the previous sections that the E$_6$ orbifold GUT
and heterotic A1 model match
nicely in the low energy regime. In this section, we study some
phenomenological issues for the A1 model. We first study
gauge coupling unification, relying on a simplification due to the
correspondence between the field and string theoretical models. We
then study Yukawa couplings (including both renormalizable and
non-renormalizable couplings), concentrating on several immediate
phenomenological questions: (i) breaking of the PS symmetry to
that of the SM, (ii) mass generation for the color triplets,
(iii) proton stability, and (iv) matter-Higgs Yukawa
couplings. These couplings are introduced by hand in the orbifold GUTs.
In string models we no longer enjoy the same
kind of freedom. In fact, these couplings are determined by string
selection rules (reviewed in appendix \ref{sec:sr}). The low energy
phenomenology also depends crucially on the flat directions of the
effective ${\mathcal N}=1$ model.  However, we do not attempt to solve
this complicated problem here.

\subsection{Gauge coupling unification and proton decay \label{unif}}

As discussed earlier, since the first two families are located on
the $\lgp{SO}{10}$ brane, proton decay constraints require that the
5d compactification scale $M_c$ be greater than $O(10^{16})$ GeV.
However all these GUT scale thresholds must be consistent with low
energy gauge coupling unification.  Consider the solution to the 5d
renormalization group (RG) equations, i.e., the
GQW equations, in the orbifold GUT
limit,\footnote{In principle these equations can be derived from a
string theory calculation, following refs.~\cite{kaplunovsky,gauge,stieberger,dienes,MNS}.
However, it is difficult to obtain the GQW equations in analytic
form in string models with discrete Wilson lines \cite{MNS}, which
makes the calculation less practical for our purposes. Instead, in
deriving eq.~\ref{eq:5drg}, we have worked in the orbifold GUT
limit, and assumed the most important contributions to the gauge
threshold corrections come from the KK tower of the large
dimension of the $\lgp{SO}{4}$ lattice, with a physical cutoff at
the string scale, $\ms$. See appendix \ref{sec:RG} for more
details.} given by \bea
\frac{2\pi}{\alpha_i(\mu)}&\simeq&\frac{2\pi}{\alpha_{\rm string}}
+b^{\rm MSSM}_i\log\frac{M_{\rm PS}}{\mu} +(b^{\rm PS}_{++}+b_{\rm
brane})_i\log\frac{\ms}{M_{\rm PS}} \nn\\
&-&\frac{1}{2}(b^{\rm PS}_{++}+b^{\rm PS}_{--})_i\log\frac{\ms}{M_c}
+b^{{\rm E}_6}\left(\frac{\ms}{M_c} -1\right), \label{eq:5drg}
\eea where $M_{\rm PS}$ is the PS breaking scale and $\alpha_{\rm
string}$ is the gauge coupling at the string scale. In addition, in
the weakly coupled heterotic string we have the boundary condition
\be  \frac{2 \pi}{\alpha_{\rm string}} = \frac{\pi}{4}
\left(\frac{M_{\rm Pl}}{M_{\rm string}}\right)^2 + \frac{1}{2} \
\Delta^{\rm univ}, \label{eq:hetbc} \ee where the first term is the
tree level result and the second is a universal one loop stringy
correction. The latter correction depends on the value of the ${\cal
T}_3, \ {\cal U}_3$ moduli. Following ref.~\cite{KL}
we see that $\Delta^{\rm univ}$ is a finite
function of its argument (with a mild singularity when ${\cal T}_3 =
{\cal U}_3$, modulo PSL$(2, {\mathbb Z})$ transformations).  Since
the universal correction is not significant, we use the tree level
formula in the following.

Eq.~\ref{eq:5drg} can be compared mathematically to the 4d
equations given by \be
\frac{2\pi}{\alpha_i(\mu)} \simeq \frac{2\pi}{\alpha^{}_{\rm GUT}}+
b^{\rm MSSM}_i\log\frac{M_{\rm GUT}}{\mu} + 6 \; \delta_{i 3},
\label{eq:4drg}
\ee where $M_{\rm GUT} \simeq 3 \times 10^{16}$ GeV,  $\alpha_{\rm
GUT}^{-1} \simeq 24$ and we have included a threshold correction at
$M_{\rm GUT}$, required in order to fit the low energy data.

With the bulk field parities given in table~\ref{table1}, we find
the beta function coefficients, $\frac{1}{2}(b_{++}^{\rm
PS}+b_{--}^{\rm PS})=(1,1,1)$, $\;\; b_{++}^{\rm
PS}=(\frac{9}{5},-3,-3)$. The brane contributions include that of
the two PS families, $b_{\rm brane}=(4,4,4)$, and those from extra
matter fields, $n'_{{\bf 6}+{\bf\overline 6}}(\frac{2}{5},
1,1)+n_{\bf 10}(1,1,1)+n'_{{\bf 2}_R}(\frac{3}{10},0,0)$ (with
$n'_{{\bf 6}+{\bf\overline 6}}\leq 4$, $n_{\bf 10}\leq 6$ and
$n'_{{\bf 2}_R}\leq 12$). Equating the difference
$2\pi/\alpha_3(\mu) - 2\pi/\alpha_2(\mu)$, eqs. \ref{eq:5drg} and
\ref{eq:4drg} gives \be M_{\rm PS} \simeq {\rm e}^{-3/2} M_{\rm GUT}
\simeq 7 \times 10^{15} \ {\rm GeV} . \ee Then equating the
difference $2\pi/\alpha_2(\mu) - 2\pi/\alpha_1(\mu)$, we find \be
\log \frac{M_{\rm string}}{M_{\rm GUT}} \simeq \frac{8-3n'_{{\bf
2}_R}+6n'_{\bf 6+\overline{6}}}{32+2n'_{{\bf 2}_R}-4n'_{\bf
6+\overline{6}}} , \ee which results in a maximum value for $M_{\rm
string}$ for $n'_{{\bf 2}_R} = 0$, $n'_{\bf 6+\overline{6}} = 4$,
given by \be M_{\rm string}^{\rm max}\simeq {\rm e}^2 M_{\rm GUT}
\simeq 2 \times 10^{17} \ {\rm GeV} . \ee Finally we have \be
\frac{\alpha^{}_{\rm GUT}}{\alpha_{\rm string}} - 1 \simeq
\frac{\alpha^{}_{\rm GUT}}{2\pi} \left[ \log\frac{M_{\rm GUT}}{M_c}
- (n_{10} + n'_{\bf 6+\overline{6}}) \left( \log \frac{M_{\rm
string}}{M_{\rm GUT}} + \frac{3}{2} \right) \right].
\label{eq:alphas} \ee Using $M_{\rm string} = M_{\rm string}^{\rm
max}$ and eq.~\ref{eq:hetbc} for the tree level heterotic string
boundary condition,  we find there is \textit{no} solution,
consistent with $M_{\rm string} > M_c \simeq M_{\rm GUT} > M_{\rm
PS}$. The problem is that the value of $\alpha_{\rm string}$ given
by eq.~\ref{eq:hetbc} is much too small ($\alpha_{\rm string} \ll
\alpha_{\rm GUT}$) and it cannot be obtained by logarithmic running
above the compactification scale (note, $b^{\lgp{E}{6}} = 0$ and
thus there is no power-law running). This problem suggests that
non-trivial (perhaps non-perturbative) string boundary conditions
are required for consistency.

We have considered the 11d Ho\v{r}ava-Witten extension \cite{hw} of
the perturbative heterotic string boundary condition given by
\cite{witten} \be \frac{2 \pi}{\alpha_{\rm string}} = \frac{1}{2 (4
\pi)^{5/3}M_*\rho} \ \left(\frac{M_{\rm Pl}}{M_*}\right)^2,
\label{eq:witten} \ee where $M_*$ is given in terms of the 11d
Newton's constant by $\kappa^{2/3} = M_*^{-3}$ and $\rho$ is the
size of the eleventh dimension.  Now using eq.~\ref{eq:witten}, we
find solutions for $M_{\rm string} \simeq M_* = 2 M_{\rm GUT}, \;
M_c \simeq M_{\rm PS} \simeq {\rm e}^{-3/2} M_{\rm GUT}$ with $n'_{{\bf 2}_R} =
n'_{\bf 6+\overline{6}} = 4$ and $M_* \rho \simeq 2$.  Of course,
this solution provides an enhanced proton decay rate due to
dimension-6 operators with the dominant decay mode $p \rightarrow
e^+ \pi^0$. The decay rate for dimension-6 operators is given by
\cite{pdecay} \bea \tau (p \rightarrow e^+\pi^0) &\simeq& 1.25
\times 10^{36} \; \left(\frac{M_X}{3 \times 10^{16} \ {\rm
GeV}}\right)^4 \; \left(\frac{0.015 \ {\rm GeV}^3}{\beta_{\rm
lattice}}\right)^2 \;\; {\rm yrs}  \nn\\
&\simeq& 3 \times 10^{33} \; \left(\frac{0.015 \ {\rm GeV}^3}{\beta_{\rm
lattice}}\right)^2 \;\; {\rm yrs}, \eea where $\beta_{\rm lattice}$
is an input from lattice calculations of the three quark matrix
element.\footnote{To obtain this result we have taken the decay rate
for $\lgp{SU}{5}$ \cite{pdecay} and multiplied the amplitude by an
additional factor of two to account for the extra gauge exchange
present in $\lgp{SO}{10}$.} Recent results give a range of central
values $\beta_{\rm lattice} = 0.007 - 0.015$ \cite{lattice}. Note,
the present experimental bound for this decay mode from
Super-Kamiokande is $5.7 \times 10^{33}$ years at $90\%$ confidence
levels \cite{superk}. Thus this prediction is not yet excluded by
the data, but it should be observed soon.

\subsection{Yukawa couplings}\label{sec:yukawa}

\subsubsection{PS symmetry breaking, mass generation for
color-triplets and proton stability}

To successfully break the PS symmetry to that of the SM and generate mass for unwanted color triplet states, the
5d $\lgp{E}{6}$ heterotic model should contain non-trivial couplings of the form in eq.~\ref{eq:sp}. The model,
however, contains additional color triplets.  They could, in principle, develop mass through non-trivial Yukawa
couplings to, say, singlet fields.  In order to verify if this is a possibility, we need to know whether the
required couplings exist in the 4d effective theory of the string model. For this purpose we are particularly
interested in non-trivial couplings containing PS invariant operators, $({\bf 6,1,1})({\bf 6,1,1})$, $({\bf
6,1,1})({\bf 4,1,2})({\bf 4,1,2})$ and $({\bf 6,1,1})({\bf \overline{4},1,2})({\bf\overline{4},1,2)}$, that are
allowed by string selection rules. (In field theory, all gauge invariant operators would be allowed.) These rules
are reviewed in appendix \ref{sec:sr} and the relevant operators are given in eqs. \ref{eq:66} -- \ref{eq:64b4b}.

Cubic, renormalizable, couplings in model A1 are determined in eq.~\ref{eq:3pt}, they contain the following
operators of interest (we label the fields according to table \ref{tab1}),
\be
{\cal W}^{(3)}\supset S_1 (C_3)_{A\alpha}(C_3)_{B\beta} +
C_1\overline\chi_1^c\overline\chi_2^c+(C_3)_{A\alpha}f^c_B\chi^c_\beta+
(C_4)_\alpha \chi^c_\beta f_3^c+ (C_4)_\alpha f_A^cf_B^c,
\label{eq:3ptt}
\ee
where $\alpha,\beta=1,2$ labels the two $\gamma=1$ eigenstates in
the $T_{2,4}$ sectors, $A, B=1,2$ indicate degeneracies associated
with the $n'_2$ winding number (which corresponds to a hidden
$\lgp{S}{2}$ permutation symmetry). The string selection rules
require $\alpha+\beta=0\,{\rm mod}\,2$, $A+B=0\,{\rm mod}\,2$, and for
the last term $\alpha$ is unrestricted. Apparently these
couplings are not sufficient to break the PS symmetry and give mass
to all the color triplets contained in the $({\bf 6,1,1})$ and
$({\bf 4,1,2})+({\bf\overline 4,1,2})$ states. Thus, to achieve our
goal, we must also take into consideration higher-dimensional
operators. Some of them are listed in appendix \ref{sec:ac}. (Of
course, there are many more operators with even higher dimensions.
It is not obvious that ``stringy zeroes" exist. We omitted these
operators. They may or may not disrupt the following discussion.)

A few observations are appropriate.  The fields  $C_i$, $i = 1, \cdots, 4$ transform as a ${\bf 3}_{-2/3} +
{\bf\overline 3}_{2/3}$ under $\lgp{SU}{3C} \times \lgp{U}{1Y}$ or as $\overline D^c + \overline D$ where
$\overline D$ has the quantum number of an anti-down quark. These color triplets have multiplicities [in
brackets], $C_1 [1], \ C_2 [2], \ C_3 [4], \ C_4 [2]$. In addition $C_2, \ C_3$ appear in complete $\lgp{SO}{10}$
{\bf 10}-plets. We also have states contained in $\chi^c_\alpha, \ \overline\chi^c_{1,2}$ with quantum numbers of
$\overline D, \ \overline D^c, \ \overline U, \ \overline U^c, \ \overline E,\, \overline E^c$. Finally we have
the color triplet states in $q_1, \ q_2, \ \overline q_1, \ \overline q_2$ with multiplicity 2 each. These latter
are exotic states with fractional charge $\pm 1/2$ for the color singlets and  $\pm 1/6$ for the color triplets.

For the exotic states we find operators of the form $\overline q_1 (q_1 + q_2)$ multiplied by products of
singlets $S^n$ (up to sixth order), but no operators of the form $\overline q_2 (q_1 + q_2) S^n$ to order $n =
9$. Hence $\overline q_2$ and one linear combination of $q_1, \ q_2$ remain massless. This is a serious problem
for the model, since these states are absolutely stable and should have been observed. It remains to be seen
whether the operator $\overline q_2 (q_1 + q_2)  S^n$ is generated at order $n \geq 10$ or if it is forbidden by
the string selection rules to all orders.

Now consider the fields  $C_i$, $i = 1,\cdots, 4$ and
$\chi^c_\alpha, \ \overline\chi^c_{1,2}$.  In this sector we need to
both find a way of spontaneously breaking PS to the SM, as well as
giving all color triplets mass.   A related issue is the potential
problem of rapid proton decay mediated by these color triplets.   In
particular we must eliminate or greatly suppress the following
baryon/lepton-number violating effective operators \be
f f f^c \langle \chi^c S^n \rangle  \Longrightarrow  Q  L  \overline
D +  L  L  \overline E, \qquad
 f^c f^c f^c \langle \chi^c  S^n \rangle \Longrightarrow  \overline U \, \overline D \, \overline D  \label{bv}.
\ee Note we have checked that these operators are not generated prior to integrating out the color triplets, for
$n \leq 3$.\footnote{Of course, it would be better check to any order in $n$, or better yet, find a symmetry
which forbids them to all orders.} Nevertheless there is a danger that they will be generated in the effective
theory below the color triplet mass. In fact, consider the renormalizable couplings in eq.~\ref{eq:3ptt}. It is
evident that an effective mass term of the form $\langle S \rangle(C_4)^2$ combined with the coupling $C_ 4
(\chi^c_\alpha f_3^c+ f_A^cf_B^c)$ leads to the effective operator of the form \be \frac{1}{\langle S \rangle}
\chi^c_\alpha f_3^c f_A^c  f_B^c = \frac{M_{\rm PS}}{\langle S \rangle} \overline U_3 \overline D_A \overline D_B
. \ee Similarly, additional baryon/lepton-number violating operators are obtained from an effective $C_3C_4$ mass
term. These baryon-number violating operator may be phenomenologically acceptable {\em if} the coefficient is
sufficiently small. However it seems prudent to eliminate the offending mass terms by choosing a vacuum
configuration where the appropriate scalar vevs vanish. For example, given the superpotential terms in appendix
C, eq. \ref{eq:66}, we demand that the following vevs (i.e. the coefficients of $(C_4)^2$ and $C_3C_4$) vanish
$S_2S_{24}+{\cal S}_1^{(2)}{\cal S}_2^{(2)}=  {\cal S}^{(2)}_1+S_2S_{9}S_{24}+S_1{\cal S}^{(2)}_2 +
S_{19}S_{25}{\cal S}^{(2)}_2=0$.

In the following we consider the possibility of obtaining a baryon/lepton-number conserving low energy effective
theory.\footnote{The conventional wisdom of field theory is to use an R-parity (or family reflection symmetry) to
eliminate these baryon/lepton-number violating operators. (The R-parity has a bonus of predicting a generic
stable neutral fermionic superpartner, which makes it even more appealing phenomenologically.) Although from the
start our string models contain several discrete R symmetries at the level of 4d effective action, it is not
clear a priori whether any of them can survive symmetry breaking. Note that in our models an unbroken R parity
(which is capable of distinguishing the Higgs fields, $\chi^c$, from the matter, $f^c$) does not exist because
both $C_4\chi_\alpha^cf_3^c$ and $C_4f_A^cf_B^c$ couplings are allowed by string selection rules at the
renormalizable level.
} As a possible proof of existence, we suggest an ansatz where the
only singlets with non-vanishing vevs are \be \{S_1\},\quad \{S_{10}
{\cal S}_4^{(2)}\} , \quad [S_6S_7S_{14}S_{18}], \quad [S_7
S_{14}S_{16} ( S_6 S_{26} + S_2 S_{27})], \quad \{S_{26}\},
\label{eq:singlet} \ee where the curly braces represent classes of
singlets with the same transformation properties under all string
symmetries and we use square brackets when we explicitly present a
finite set of fields in the same class. The corresponding
superpotential from appendix \ref{sec:ac} is \bea {\cal W} &\supset
&\,\,\,\, \biggr(\{S_1\} (C_3)^2 + \{S_2 S_{22} {\cal S}_9^{(2)}\}
C_2 C_3
+ \{S_{10}  {\cal S}_4^{(2)}\}  C_1  C_4 \biggr)   \nn\\
& &+ \biggr(
C_1\overline\chi_1^c\overline\chi_2^c+C_3f^c_A\chi^c_\alpha+
C_4 (\chi^c_\alpha f_3^c+ f_A^cf_B^c) \biggr)  \nn \\
& &+ \biggr( \{S_2 S_{24} S_{26}\} C_4 (\overline\chi_1^c )^2 +
[S_{10}{\cal S}_1^{(2)} +S_6S_7S_{14}S_{18}+S_1S_{10}{\cal
S}^{(2)}_2+S_2{\cal S}^{(3)}_4]C_4(\overline\chi_2^c)^2  \nn \\
& &\qquad + [S_2S_3S_{12}{\cal S}_3^{(2)} + S_7 S_{14}S_{16} ( S_6
S_{26} + S_2 S_{27})] C_4 \chi_\alpha^c \chi_\beta^c \biggr) +
\{S_{26}\} \chi_\alpha^c \overline \chi_1^c . \label{eq:csp} \eea
Note
the coefficient of the $C_2 C_3$ term and the first term linear in $C_4$ vanishes due to the vevs we have chosen,
but there may be other higher-dimensional terms which replace them.  For example, the first element in the second
term linear in $C_4$ also vanishes, but the other terms may be non-zero.

The following vevs are assumed to vanish.
\bea
& [S_2S_{24}+{\cal S}_1^{(2)}{\cal S}_2^{(2)}], &
[{\cal S}^{(2)}_1+S_2S_{9}S_{24}+S_1{\cal S}^{(2)}_2 + S_{19}S_{25}{\cal S}^{(2)}_2], \nn  \\
& \{S_2 S_{9} S_{22}\}, &
[S_{10}{\cal S}^{(2)}_3+S_{13}{\cal S}^{(2)}_2+ S_1S_{10}S_{12}S_{32}+S_9 S_{10}{\cal S}^{(2)}_4],  \nn \\
& \{S_9\} \equiv
S_9 + S_1 S_{10} S_{21} S_{22} + {\cal S}^{(2)}_2 S_{11}^{2} + \cdots,
& [{\cal S}_9^{(2)}+S_{10}(S_{30}{\cal S}^{(2)}_5+S_{13}{\cal S}^{(2)}_4)],
\nn \\
& [{\cal S}^{(2)}_2+S_{26}{\cal S}^{(3)}_2],
&  [S_9{\cal S}_9^{(2)}+S_{13}(S_{10}{\cal S}^{(2)}_3+S_{13}{\cal S}^{(2)}_2)],  \nn \\
& \{S_{10}S_{13}S_{21}S_{22}\}, &  \{S_9S_{10}S_{12}S_{32}\},
\qquad  \{S_{10} {\cal S}_1^{(3)}\}
\eea With this choice we guarantee, at least to the order we have checked, that we do not generate
baryon/lepton-number violating operators in the low energy theory obtained by integrating out the color triplets.
A self-consistent solution to the necessary set of vevs is given by \be S_4 = S_9 = S_{11} = S_{13} = S_{17} =
S_{21} = S_{24} = S_{25} = S_{29} = S_{30} = S_{32} = 0, \label{ssol} \ee and all other vevs non-zero.   In
addition we require $[S_5 S_{33} + S_{10} S_{26}] = 0$, which may or may not require fine-tuning.  Unfortunately
we are not able to identify a symmetry which would extend this result to all orders in string perturbation.  This
is a serious problem for the model.

The first term in parentheses of eq.~\ref{eq:csp} gives mass to the color triplets $C_2$, $C_3$, $C_1$ and one
triplet in $C_4$. Since the doublets $h_2$ and $h_3$ have exactly the same quantum numbers as that of $C_2$ and
$C_3$, they acquire the same mass and also decouple from the low energy effective theory. In this way, we obtain
just one doublet field $h_1$ (from the $U_2$ sector).
It is a good candidate for the MSSM Higgs-doublet pair.

The last term of eq.~\ref{eq:csp} is in the form of eq.~\ref{eq:sp},
with two pairs of $\chi^c+\overline\chi^c$ fields. With $\langle
S_{26} \rangle \neq 0$ an F-flat solution is found with
$\langle\chi_1^c\rangle=\langle\overline\chi^c_1\rangle=0$, then
this part of the superpotential reduces to eq.~\ref{eq:sp}.
Employing F and D flatness conditions, the potential has a flat
direction along the right-handed-neutrino direction of
$\chi_2^c+\overline\chi_2^c$, as in eq.~\ref{eq:nuvev}. The
non-vanishing vevs break the PS to the SM gauge group and
subsequently gives mass to the remaining massless color triplets in
$C_4$ and the $D^c+\overline{D}^c$ components in
$\chi^c_2+\overline\chi^c_2$. The remaining charged states in
$\chi^c_2+\overline\chi^c_2$ obtain mass via the super-Higgs
mechanism. Lastly, the $\chi^c_1+\overline\chi^c_1$ also acquire
mass via the $\langle S_{26} \rangle$ vev in eq.~\ref{eq:csp}.

The remaining central problem is whether the required singlets, such as those in eq.~\ref{eq:singlet}, could
develop the appropriate vevs along F- and D-flat directions.  There are five abelian factors in the A1 model, one
of them is anomalous (which is cancelled by the generalized Green-Schwarz mechanism \cite{GS,au1}, as usual). In
general, the anomalous $\lgp{U}{1}$ factor destabilizes the original
vacua and contributes vevs to some of the singlet
fields. It requires further investigation to determine whether our assumptions in the previous paragraph are
substantiated, and moreover whether all the abelian symmetries may be broken with the vacuum solution. (The
${\mathcal N}=1$ supersymmetry, however, is generically preserved in the effective theory.) Refs.~\cite{FIQS,GCEEL}
have already obtained some necessary conditions for analyzing non-trivial singlet vevs. We leave a detailed
investigation for the future.

\subsubsection{$\lgp{D}{4}$ family symmetry \label{sec:d4}}

Before discussing the Yukawa matrices for quarks and leptons, we
consider the family symmetry of model A1.  The third family is a
bulk field, while the first two families are located on the two
$\mz_2$ fixed points in the $\lgp{SO}{4}$ torus with an
$\lgp{SO}{10}$ gauge symmetry.   One family sits at each fixed point
(see fig.~\ref{fig1}). Since the Wilson line in the $\lgp{SO}{4}$
torus lies in the orthogonal direction to these two fixed points,
the theory is invariant under the permutation of the first two
families, labelled by an index $n_2' = 0,1$ (or $A=1,2$).  In
addition, the string selection rule, eq. \ref{sg1}, requires that
every effective fermion mass operator include an even number of
fields with $n_2' = 1$. Hence these effective operators are
invariant under a $\mz_2$ parity $n'_2 \rightarrow -n'_2$. The two
operations are generated by the two Pauli matrices $\sigma_1 =
\left(
\begin{array}{cc} 0 & 1 \\ 1 & 0 \end{array} \right)$ and $\sigma_3 = \left( \begin{array}{cc} 1 & 0 \\ 0 & -1
\end{array} \right)$ acting on a real two dimensional vector.   The complete set of operations
closes on the discrete non-abelian family symmetry group $\lgp{D}{4}
= \{ \pm I, \ \pm \sigma_1, \ \pm \sigma_3, \  \mp {\rm i} \sigma_2
\}$. Note that the eight-element finite (dihedral) group
$\lgp{D}{4}$ is the symmetry group of a square. It has five
conjugacy classes and five faithful representations. The character
table is \be \setlength{\arraycolsep}{0.2in}
\begin{array}{lrrrrr}
\hline
 {\rm Classes} & I & -I &  \pm\sigma_1 &  \pm\sigma_3 &  \mp{\rm i}\sigma_2\\
\hline
{\rm Doublet} - D &  2 & -2 & 0 & 0 & 0\\
{\rm Singlet} - A_1 & 1 & 1 & 1 & 1 & 1\\
{\rm Singlet} - B_1 & 1 & 1 & 1 & -1 & -1 \\
{\rm Singlet} - B_2 & 1 & 1 & -1 & 1 & -1 \\
{\rm Singlet} - A_2 & 1 & 1 & -1 & -1 & 1 \\
\hline
\end{array}.
\ee
In our models, the first two families transform as the doublet, while the third
family transforms as the trivial singlet.

We have many $\lgp{SO}{10}$ singlets in our models, 
transforming as doublets under $\lgp{D}{4}$.  They appear in effective
higher dimension fermion mass operators.   
Consider, for example, two doublets under $\lgp{D}{4}$ given by 
\{$S_A,\tilde S_A$\}. 
Then in terms of these two doublets we can define bilinear combinations transforming as \{$A_1,A_2,B_1,B_2$\}. We have
\begin{eqnarray}  
&& S_1 \tilde S_1 +  S_2 \tilde S_2    \sim  A_1  \nn\\
&& S_1 \tilde S_2  - S_2 \tilde S_1    \sim  A_2  \nonumber \\
&& S_1 \tilde S_2 +  S_2 \tilde S_1    \sim  B_1   \nonumber \\
&& S_1 \tilde S_1 -  S_2 \tilde S_2    \sim  B_2 
\end{eqnarray}
The effective Yukawa couplings are then constructed in terms of 
$\lgp{D}{4}$ invariants. Define the $\lgp{D}{4}$ doublet
left-handed quarks and leptons $({\bf 4,2,1})$ [$= f_A$] 
and left-handed anti-quarks and anti-leptons $({\bf\bar 4,1,2})$ [$=
 f^c_A$] for the first two families and the Higgs multiplet 
$({\bf 1,2,2})$ [$= h$].  We then have the PS and $\lgp{D}{4}$
invariants:
\begin{eqnarray}  
&& h A_1  (f_1  f^c_1 + f_2 f^c_2) \equiv   h  A_1  (f_A  f^c_A)\nn \\
&& h  A_2  (f_1  f^c_2 - f_2  f^c_1)  \nn \\
&& h  B_1  (f_1  f^c_2 + f_2  f^c_1)  \nonumber \\
&& h  B_2  (f_1  f^c_1 - f_2  f^c_2)   \label{eq:other}
\end{eqnarray}
We can also have operators of the form
\be
h  (f_A   S_A) ( f^c_B   S_B)
=  h   [ f_1   f^c_1  S_1^2 + f_2  f^c_2  S_2^2 
+  (f_1  f^c_2 + f_2  f^c_1 )  S_1  S_2 ]
\label{eq:12mass}
\ee
Unfortunately there are, in principle, 
several possible ways of constructing $\lgp{D}{4}$ invariants.  We are not able
to determine, without further string calculations, how to contract the 
$\lgp{D}{4}$ indices.   In the following we
assume, for illustrative purposes, that only the simplest invariants,  
$A_1$ and $B_1$, appear in the effective
Yukawa couplings.

\subsubsection{Fermion masses}

The only Yukawa coupling in model A1 present at leading order is for
the third family, given by the first term in eq.~\ref{eq:3pt}. From
discussions in sect.~\ref{sec:E6}, we conclude that this coupling
unifies with the GUT gauge coupling at the 5d compactification
scale, as in eq.~\ref{eq:yg}.  Yukawa couplings for the first two
families come from higher-dimensional non-renormalizable operators.
In addition they are constrained by the $\lgp{D}{4}$ family
symmetry.   In principle, there are at least two possible types of
operators. The first type involves operators of the form $h_1
(f^{}_3 f_A^c, \ f^{}_A f_3^c, \ f^{}_Af_B^c)$, multiplied by
suitable singlets, and the second type also involves composite
singlets $\overline\chi^c\chi^c$. The second type of operators is
particularly important since it has the potential to discriminate
up-type  quarks and charged leptons from the down-type; this is
necessary to obtain a realistic CKM matrix and also resolve the
``bad" GUT relation $m_s/m_d = m_\mu/m_e$.

We define the following two composite operators \be {\cal O}_1
=\overline\chi_1^c\chi^c_\alpha,\qquad {\cal O}_2
=\overline\chi_2^c\chi^c_\alpha, \ee where the group indices are
arranged in all possible ways. From the string selection rules, it
is straightforward to show that the Yukawa matrix is (we only keep
representative terms, see eq.~\ref{yukmatapp} for more complete
expressions), \be (f_1\,f_2\,f_3)  h_1 \left( {\footnotesize
\begin{array}{ccc}
  {\cal O}_2 S^{(3,9,12)}_{\rm e} + S^{(10,22,22,23)}_{\rm e}  &
  {\cal O}_2 S^{(3,9,12)}_{\rm o} + S^{(10,22,22,23)}_{\rm o}  &
 {\cal O}_1 {\cal O}_2 S^{(3,12)}_{\rm e} +  S^{(9,10,22,22,23)}_{e} \\
  {\cal O}_2 S^{(3,9,12)}_{\rm o} + S^{(10,22,22,23)}_{\rm o}  &
 {\cal O}_2 S^{(3,9,12)}_{\rm e} + S^{(10,22,22,23)}_{\rm e}  &
 {\cal O}_1 {\cal O}_2 S^{(3,12)}_{\rm o} +  S^{(9,10,22,22,23)}_{\rm o} \\
 S^{(10,26)}_{e} &S^{(10,26)}_{\rm o} &   1
\end{array}} \right)\left(
\begin{array}{c}
f_1^c\\
f_2^c\\
f_3^c
\end{array}\right), \label{yukmat}
\ee where \be S^{(a,b,\cdots)}_{\rm e} =\sum_{\sum A={\rm even}}
S^a_{A}S^b_{B}\cdots,\qquad S^{(a,b,\cdots)}_{\rm o} =\sum_{\sum
A={\rm odd}} S^a_{A}S^b_{B}\cdots, \ee with $A$'s are the family
indices of the corresponding singlets (of the $T_{1,3}$ sectors).
Several comments on eq. \ref{yukmat} are now in order.
\begin{itemize}
\item  The structure of the Yukawa matrix is determined by a $\lgp{D}{4}$ family symmetry.  (See the caveat at
the end of Section \ref{sec:d4}.)

\item  Given the superpotential for color triplets, eq. \ref{eq:csp}, an F-flat direction requires $\langle
\overline \chi_1^c \rangle = 0$ which gives ${\cal O}_1 = 0$. If
however there is a higher-dimensional operator of the form $S_{26} S
\chi_\alpha^c \overline \chi_2^c$, then the combined terms  $S_{26}
(\chi_\alpha^c \overline \chi_1^c + S  \chi_\alpha^c \overline
\chi_2^c)$ has an F-flat solution with $\langle \overline \chi_1^c
\rangle, \langle \overline \chi_2^c \rangle \neq 0$ and thus ${\cal
O}_1, {\cal O}_2 \neq 0$. We will analyze the more general case.

\item
Given the superpotential for color triplet masses, our previous solution
eq.~\ref{ssol} requires $\langle S_9 \rangle =
0$.  Hence the composite operators $S^{(3,9,12)}$, $S^{(9,10,22,22,23)}$ vanish.   However, it is again possible
that these terms may still be present when higher order products of operators are considered.

\item It is crucial to understand how the PS group indices are contracted and what
the corresponding Clebsch-Gordon (CG) coefficients are.
In orbifold models, the massless matter fields correspond to the
(integral) highest weight
representations of the level-one Ka\v{c}-Moody algebra.
In principle one may extract the desired
information from the conformal blocks. We have not attempted to perform such a string theoretical analysis. Instead
we shall adopt a simpler field theoretical approach, following ref.~\cite{king}.
\end{itemize}

Our aim is to examine phenomenological implications of eq.~\ref{yukmat} in a simple setting. We shall consider
two simple cases in the following.

\vspace*{2mm}

\noindent{\textit{Case A --}}

First neglect the $(13)$ and $(23)$ entries of eq.~\ref{yukmat} (it
may be reasonable to do so, because they are higher order terms),
and consider the $2 \times 2$ sub-matrix corresponding to the second
and third families, i.e. \be \setlength{\arraycolsep}{0.1in} \left(
\begin{array}{cc}
{\cal O}_2 S^{(3,9,12)}_{\rm e} + S^{(10,22,22,23)}_{\rm e}  &
0 \\
 S^{(10,26)}_{\rm o} &   1
\end{array} \right).
\ee We may take ${\cal O}_2$ to be in the form of the ${\cal O}^W$
operator of ref.~\cite{king}. This operator has a vanishing
(non-vanishing) CG coefficient for the up (down) type fields. One
may require \be \langle {\cal O}_2 S^{(3,9,12)}_{\rm e}\rangle\sim
\frac{m_s}{m_b} \sim \lambda^2,\qquad \langle S_{\rm
e}^{(10,22,22,23)}\rangle\sim\frac{m_c}{m_t}\sim \lambda^3, \ee
where $\lambda\simeq 0.22$ is the Cabibbo angle. The mixing angle
$V_{cb}$ is then approximately $\lambda^2\langle S^{(10,26)}_{\rm
o}\rangle$, implying $S^{(10,26)}_{\rm o}\sim {\cal O}(1)$.

Next consider the $(2 \times 2)$
sub-matrix corresponding to the first and second families. Note that
this part always has the following form,
\begin{eqnarray}\setlength{\arraycolsep}{0.1in}
\left(
\begin{array}{cc}
a_{u,d} &
b_{u,d} \\
b_{u,d} &   a_{u,d}
\end{array} \right).
\end{eqnarray}
The democratic form may lead to realistic values for quark masses and
mixings. Taking $a_{u,d} = b_{u,d} (1+ \varepsilon_{u,d})$
with $\varepsilon_d \approx\lambda$
(which implies an approximate $\mz_2$
symmetry between vevs,
${\cal O}_2 S^{(3,9,12)}_{\rm e}
= {\cal O}_2 S^{(3,9,12)}_{\rm o}  (1 + \varepsilon_d)$,
$ S^{(10,22,22,23)}_{\rm e}  =
 S^{(10,22,22,23)}_{\rm o} (1 + \varepsilon_u)$ and
$ S^{(10,26)}_{\rm e}\simeq S^{(10,26)}_{\rm o}$),
we obtain the mass ratio $m_d/m_s$ and CKM angle
$V_{us}$ at correct orders.
More suppressed value for $\varepsilon_u$, e.g.
$\varepsilon_u \sim 10^{-3}$, is required for $m_u/m_c$.
Finally, it is also possible to obtain correct mass relations for
the charged leptons,
$m_\mu/m_\tau\sim m_s/m_b$ and $m_e / m_\mu\sim m_d/m_s$.

\vspace*{2mm}
\noindent{\textit{Case B --}}

Consider now the case that the (13) and (23) entries are not negligible.
Let us parameterize the (23) entry in the down sector by $y_{23}^d$
and assume the corresponding entry in the up sector is smaller (or comparable).
For simplicity, we also assume $S^{(10,26)}=0$.
The $(2 \times 2)$ sub-matrix of the second and third
families is
\begin{eqnarray}\setlength{\arraycolsep}{0.1in}
\left(
\begin{array}{cc}
{\cal O}_2 S^{(3,9,12)}_{\rm e} + S^{(10,22,22,23)}_{\rm e}  &
y^d_{23} \\
 0 &   1
\end{array} \right).
\end{eqnarray}
We can obtain correct mass ratios $m_c/m_t$ and
$m_s/m_b$ and mixing angle $V_{cb}$ if
$y^d_{23} \sim  {\cal O}_2 S^{(3,9,12)}_{\rm e}\sim\lambda^2$
and $S^{(10,22,22,23)}_{\rm e} \sim\lambda^3$. The discussion on
the first and second families follows essentially in the same way as
in case A. However, we now need to tune ${\cal O}_2 S^{(3,9,12)} +
S^{(10,22,22,23)}$ appropriately.

\vspace*{2mm}

Finally consider neutrino masses. The Dirac neutrino mass matrix has the same form as that of eq.
\ref{yukmatapp}.  Effective Majorana neutrino masses are obtained in eq.~\ref{eq:maj1}, where the non-trivial
effective operators have the form $f_a^c f_b^c \overline\chi_i^c \overline\chi^c_j$ ($a,b=1,2,3$, $i,j=1,2$) with
suitable powers of singlets. The non-vanishing vevs of $\overline \chi^c_i$ project out the right-handed
neutrinos in $f_3^c, \ f_A^c$.  One then obtains a Majorana mass of order $< M_{\rm PS}^2/M_{\rm string} \simeq
M_{\rm GUT}/(2{\rm e}^3) \simeq 7 \times 10^{14}$ GeV, which is just right for generating acceptable light
neutrino masses via the see-saw mechanism. Although at the present operator order the Majorana mass terms vanish
with the non-vanishing vevs discussed earlier, non-trivial operators may exist at higher order.

\section{Conclusion \label{con}}

In this paper we construct three-family PS models in the ${\mathbb
Z}_6$ abelian symmetric orbifold. Our models are mainly motivated by
recent discussions on orbifold GUTs. We are able to realize some
features of the orbifold GUTs in the string compactification limit
where the compactified space is effectively 5d. The
breaking of the ${\mathcal N}=2$ to
${\mathcal N}=1$ supersymmetry in 4d and the $\lgp{E}{6}$ (or
$\lgp{SO}{10}$) gauge symmetry to that of PS are realized similarly as in
orbifold GUTs. We find three family chiral matter fields, two of
them can be regarded as ``brane'' states and one as a ``bulk''
state, in the terminology of orbifold GUTs. These models extend
three-family orbifold string model building to non-prime-order
orbifolds, and we find some new features in the matter spectra when
compared to that of the prime-order orbifolds. Matter fields arise
not only from the untwisted but also from twisted sectors, and
typically there is a horizontal $2+1$ splitting in the family space.
This splitting may have the potential to better facilitate the description of
fermion masses and mixings.

We find one of our string models, with an $\lgp{E}{6}$ gauge group in 5d,
is particularly interesting. It has the following
properties:
\begin{enumerate}
\item Renormalizable Yukawa couplings exist only for the third family. Moreover the model predicts a unification
relation among the third family Yukawa couplings and GUT gauge coupling at the 5d compactification scale. \item
The renormalizable and non-renormalizable couplings can affect a spontaneous breaking of the PS symmetry to that
of the SM with fields in the $({\bf 6,1,1)}$ and $({\bf 4,1,2})+({\bf\overline{4},1,2})$ representations.
Moreover, after the symmetry breaking, many unwanted states can develop large mass. \item There is a non-abelian
$\lgp{D}{4}$ family symmetry for the first two families, which makes the model a good playground for studying
fermion mass hierarchy and the flavor problem in supersymmetry.
\end{enumerate}
We regard these phenomenological features as merits of our models.

On the other hand, there are two main problems for the models (of course, they are not unique to our models).
\begin{itemize}
\item Exotics: There is a pair of SM vector-like exotic particles at the low energy scale. To the order of
our analysis, we have not found any Yukawa coupling that can give them mass.

\item Proton stability: We have Higgs fields transforming in the same (or conjugate) PS representation as the SM
right-handed matter. In general, baryon/lepton-number violating effective operators are induced after PS symmetry
breaking. We have not yet found a symmetry that can distinguish the PS breaking Higgses from matter and
effectively eliminate the dangerous operators to all orders.
\end{itemize}

These problems have only been examined at a rather primitive and qualitative level. Specifically we have looked
for non-trivial Yukawa couplings allowed by string selection rules to certain orders. We have shown that the
baryon/lepton-number violating effective operators can be avoided if one prudently chooses appropriate values for
the singlet vevs. Unfortunately, we are not able to extend this argument to all orders in string perturbation.
The problem apparently results from the presence of {\em only the trivial fixed point} in the $T_1$ twisted
sector on the $\lgp{G}{2}$ torus.  This has the consequence of nullifying any space group selection rules for the $\lgp{G}{2}$
torus whenever any $T_1$ twisted sector operator is present.

In any case, these problems may point to the direction for refining our models.  For example, it would be
desirable to search for models with fewer color triplets and, if possible, no exotics.  The analysis presented
here is clearly just the beginning. It would certainly be useful to expand the search to other ${\mathbb
Z}_2\times {\mathbb Z}_N$ orbifolds with one (or two) Wilson lines in the $\lgp{SO}{4}$ direction in order to
find more effective 5 or 6d orbifold GUTs. (The ${\mathbb Z}_2\times {\mathbb Z}_2$ model might be particularly
interesting and it has been studied recently in ref.~\cite{Forste:2004ie}. Presumably the model is simpler than
ours because it has only three twisted sectors and the number of modular invariant gauge embeddings is more
limited. However, realistic three-family model have yet to be constructed.) In brief, we believe our analysis has
opened up a promising new direction for string model building.

\begin{acknowledgments}

This work has been presented at the Planck '04 and String
Phenomenology '04 conferences and the Aspen ``String and Real World"
workshop. We would like to thank the participants of these meetings,
especially A. Faraggi, H. P. Nilles and S.~Stieberger, for
stimulating conversations. T.~K.\/ was supported in part by the
Grant-in-Aid for Scientific Research (\#16028211) and the
Grant-in-Aid for the 21st Century COE ``The Center for Diversity and
Universality in Physics'' from Ministry of Education, Science,
Sports and Culture of Japan. S.~R. and R.-J.~Z. were supported in
part by DOE grants DOE/ER/01545-858 and DE-FG02-95ER40893
respectively. They also thank the Aspen Center for Physics for
hospitality during the final stage of this work.

\end{acknowledgments}

\appendix

\section{Recipes for constructing non-prime-order orbifold models
in the presence of discrete Wilson lines \label{review}}

In this appendix we review the construction of non-prime-order
orbifold models with discrete Wilson lines.

Our starting point is the 10d heterotic string theory, which
consists of a 26d left-moving bosonic string and a 10d
right-moving superstring. Modular invariance requires the momenta of
the internal left-moving bosonic degrees of freedom (16 of
them) lie in a 16d Euclidean even self-dual lattice, we choose to
be the $\lgp{E}{8}\times\lgp{E}{8}$ root lattice.\footnote{For an
orthonormal basis, the $\lgp{E}{8}$ root lattice consists of
following vectors, $(n_1,n_2,\cdots,n_8)$ and
$(n_1+\frac{1}{2},n_2+\frac{1}{2},\cdots,n_8+\frac{1}{2})$, where
$n_1, n_2,\cdots n_8$ are integers and $\sum_{i=1}^8 n_i=0\,{\rm
mod}\,2$. }

To make a connection to the 4d world, we must compactify 6 spatial
dimensions of the 10 remaining space-time dimensions. There are
many ways (see, e.g., ref.~\cite{Schellekens} for a collection of
early works) to achieve a 4d ${\mathcal N}=1$ supersymmetric spectrum,
among them the most studied in the literature is the orbifold
construction \cite{Dixon,BL,IMNQ,IMNQ2,FIQS,katsuki}. We will use
the simplest abelian symmetric orbifold construction, where we
differentiate the space-time and internal degrees of freedom, and
realize the orbifold twists and Wilson lines by shifts in the
$\lgp{E}{8}\times\lgp{E}{8}$ lattice. This type of construction
admits a clear space-time interpretation.

The full definition of an orbifold model requires the specification of a six-torus T$^6$, corresponding to the
compactified spatial dimensions, a point group, (such as the cyclic groups ${\mathbb Z}_N$ or ${\mathbb
Z}_N\times{\mathbb Z}_M$ with $N,\ M =3,4,6,7,8,12$ \cite{Dixon}), corresponding to the automorphism of the
T$^6$-lattice, and an embedding of the space group, (which consists of the point group and the lattice
translations), in the $\lgp{E}{8}\times\lgp{E}{8}$ lattice. Normally one denotes the generator of the discrete
group ${\mathbb Z}_N$ by a \textit{twist vector} ${\bf v}=(v_1,v_2,v_3)$, acting on the three complex planes by
$\theta: Z_i\rightarrow {\rm e}^{2\pi {\rm i} v_i}Z_i$ $(i=1,2,3)$. To ensure that one space-time supersymmetry
survives in 4d, the twist vector needs to satisfy $\pm v_1\pm v_2\pm v_3=0\,{\rm mod}\,2$,\footnote{The signs are
arbitrary. We will use the convention that all signs are positive.} and none of the $v_i$'s vanishes. In the
abelian orbifold construction, embeddings of the space group in the $\lgp{E}{8}\times\lgp{E}{8}$ are realized by
shifts of the corresponding lattice, ${\bf P}\rightarrow {\bf P} + k{\bf V}+l{\bf W}$, where $k$, $l$ are
integers, ${\bf P}$ are vectors of the ${\rm E_8\times E_8}$ root lattice, ${\bf V}$ the \textit{gauge twists},
realizing the point group, and ${\bf W}$ the \textit{discrete Wilson lines}, realizing the lattice translations.
The cyclic group multiplication rules require $N{\bf V}$ and $N_W{\bf W}$ to be in the
$\lgp{E}{8}\times\lgp{E}{8}$ lattice. (In general the degrees of the Wilson lines, $N_W$, divide the degree of
the orbifold twist, $N$.)

String states closed on T$^6$, i.e. those satisfying the condition
$Z_i(\tau,\sigma+\pi)=Z_i(\tau,\sigma)$ modulo lattice translations,
give rise to the untwisted-sector states. Besides the ${\mathcal N}=1$
supergravity multiplet and modulus fields that parameterize
deformations of the background fields, the untwisted-sector states
also give rise to gauge and matter fields. Embeddings of the point
group in $\lgp{E}{8}\times\lgp{E}{8}$ break the gauge symmetry down
to its commutator subgroups, i.e., the surviving non-zero roots
satisfy
\be {\bf P}^2=2\,,\qquad {\bf P}\cdot {\bf V}\in{\mathbb Z}\,.
\label{gauge}\ee
Note that we cannot lower the rank of the surviving groups in
abelian orbifold models, since by construction the gauge twists and
Wilson lines commute with the $\lgp{E}{8}\times\lgp{E}{8}$ Cartan
subalgebra.

The conditions for the untwisted-sector matter states are similar.
It is convenient to bosonize the right-moving fermionic degrees of
freedom and denote their $\lgp{SO}{8}$ momenta in the light-cone gauge by
${\bf r}$.\footnote{That is, ${\bf r}$'s are in the $\lgp{SO}{8}$ weight
lattice, the integral and half-integral weights correspond to the
Neveu-Schwarz (NS) and Ramond (R) sector fermions. For R-sector
weights, the first components, $r_0$, indicate the helicities of
space-time fermions. In this notation, the first component of the
twist vector, $v_0$, is zero. ${\bf r}$ are commonly referred to as
the H momenta.} The right-moving NS sectors ($b^i_{-1/2}$ for
$i=1,2,3$) of the untwisted matter states have $\lgp{SO}{8}$ weights ${\bf
r}_1=(0,1,0,0)$, ${\bf r}_2=(0,0,1,0)$ and ${\bf r}_3=(0,0,0,1)$ and
they pick up phases ${\rm e}^{-2\pi{\rm i}{\bf r}_i\cdot{\bf v}}$ under the
orbifold twists, so the corresponding roots must satisfy
\be {\bf P}^2=2\,,\qquad {\bf P}\cdot {\bf V} - {\bf r}_i\cdot {\bf
v} \in {\mathbb Z} \qquad (i=1,2,3) \,. \label{matter}\ee
This gives three untwisted-matter sectors $U_i$, one for each
complex plane. (Eq.~\ref{gauge} can also be written in the same
form with ${\bf r}_0=(1,0,0,0)$, where the non-zero entry lies in
the uncompactified direction).

The gauge groups and untwisted-sector matter spectra are modified when discrete Wilson lines are turned on. In
the presence of general background fields $G_{ij}$, $B_{ij}$ and ${\bf W}_i$, the canonical momenta conjugate to
the compactified coordinates and the gauge coordinates are $\Pi^i=p_L^i+p_R^i+\frac{1}{2}{\bf
W}^i\cdot({\bm\Pi}+{\bf p}^{}_L)+B^{ij}(p^{}_{Lj}-p^{}_{Rj})$ and ${\bm\Pi}={\bf p}^{}_L-\frac{1}{2}{\bf
W}_i(p_L^i-p_R^i)$. Since $p_L^i-p_R^i=2n^i$, $\Pi^i=m^i$ and ${\bm\Pi}={\bf P}$, where the integers $m^i, n^i$
are the momentum (or KK modes) and winding quantum numbers, we have \cite{narain,wl} (the string unit is
$\alpha'=1/2$)
\bea
p_L^i&=&\frac{m^i}{2}+\left(G^{ij}-B^{ij}-\frac{1}{4}{\bf W}^i\cdot
{\bf W}^j\right)n_j-\frac{1}{2}{\bf P}\cdot{\bf W}^i \,,\nn\\
p_R^i&=&p_L^i-2n^i\,,\qquad {\bf p}^{}_L\,=\,{\bf P}+n^i {\bf
W}_i\,.\label{PLR}
\eea
In the models studied in this paper, we take $B_{ij} = 0$ and
$G_{ij} = \frac{1}{2}R_iR_j{\bf e}_i \cdot {\bf e}_j$ is defined by
the geometry of the internal space T$^6$ with basis vectors given by
${\bf e}_i$ and dimensions by $R_i$. In this case, these equations,
along with the masslessness conditions
\bea
& &\frac{1}{4}m_{R}^2=N_R + \frac{1}{2} G_{ij}p_R^ip_R^j + \frac{1}{2} {\bf r}^2 - \frac{1}{2} =0 \label{eq:hmom} \\
& &\frac{1}{4}m_L^2=N_L + \frac{1}{2} G_{ij}p_L^ip_L^j +
\frac{1}{2}{\bf p}_L^2 -1 =0
\eea
where $N_L, \ N_R$ are integral oscillator mode numbers and the last
two terms in eq. \ref{eq:hmom} are the contribution of the
bosonized NSR fermions, require the winding number for massless
states to be zero, and the last term of $p_L^i$ an integer (for both
gauge and matter fields), i.e.,
\be {\bf P}\cdot {\bf W} \in{\mathbb Z}\,. \label{PA}\ee

String states closed on themselves under the identification of a non-trivial element of the point group (modulo
translations by lattice vectors) give rise to the twisted-sector states. For the $k^{\rm th}$ twisted-sector
$T_k$ (for which the complex compactified coordinates satisfy $Z_i(\tau,\sigma+\pi)={\rm e}^{2\pi{\rm i}
kv_i}Z_i(\tau,\sigma)$), the $\lgp{E}{8}\times\lgp{E}{8}$ and $\lgp{SO}{8}$ momenta are shifted according to
${\bf P}\rightarrow {\bf P}+ k {\bf X}_{n_{\bf f}}$ and ${\bf r}\rightarrow {\bf r}+k{\bf v}$, where ${\bf
X}_{n_{\bf f}}={\bf V}+n^i_{\bf f}{\bf W}_i$ with $n^i_{\bf f}$ being fixed-point dependent winding
numbers.\footnote{Discrete Wilson lines break the degeneracies of the fixed points. The integers $n^i_{\bf f}$
should be chosen appropriately, depending on the direction and degree of the Wilson line
\cite{wl,kobayashi,kobayashi2}.} The massless states satisfy the following equations
\cite{Dixon,IMNQ,IMNQ2,FIQS,katsuki},
\bea
\frac{1}{4}m_{R}^2=\sum_{i=1}^6 N^R_i\omega_i^{(k)}
+\frac{1}{2}\left({\bf r}+k{\bf v}\right)^2+a^{(k)}_R=0\,,&&\label{masscond1}\\
\frac{1}{4}m_L^2=\sum_{i=1}^6
N_i^L\omega_i^{(k)}+\frac{1}{2}\left({\bf P}+k{\bf X}_{n_{\bf
f}}\right)^2+a_L^{(k)}=0\,, &&\label{masscond2}
\eea
where $N^R_i$ and $N^L_i$ are intergral numbers of the right- and
left-moving (bosonic) oscillators, $a^{(k)}_R$, $a_L^{(k)}$ are the
normal ordering constants,
\bea
a^{(k)}_R
&=&-\frac{1}{2}+\frac{1}{2}\sum_{i=1}^3|{\widehat{kv_i}}|\left(1-|{\widehat{kv_i}}|\right)\,,\nn\\
a_L^{(k)}
&=&-1+\frac{1}{2}\sum_{i=1}^3|{\widehat{kv_i}}|\left(1-|{\widehat{kv_i}}|\right)\,,
\eea
with $\widehat{kv_i}=kv_i\,{\rm mod}\,1$, and
$\omega_i^{(k)}=\widehat{kv_i}$ if $\widehat{kv_i}> 0$ and
$1-\widehat{kv_i}$ if $\widehat{kv_i}\leq 0$ are oscillator
energies. Thus, a twisted-sector matter state can be labelled by the
twisted sector, $k$, the fixed point it is localized on, ${\bf f}$,
the number of windings associated with the fixed point, $n_{\bf
f}^i$, and the number of right- and left-moving oscillators, $N_i^R$
and $N^L_i$.

Note, however, that not all gauge twists and discrete Wilson lines
are physically allowed. To ensure modular invariance of the one-loop
partition function (or the level-matching condition) for the right-
and left-movers, one needs to require
\cite{Dixon,vafa,IMNQ,IMNQ2,FIQS,katsuki}
\be
N({\bf X}_{n_{\bf f}}^2-{\bf v}^2)=0\,{\rm mod}\,2\,.\label{mod1}
\ee
In addition, in non-prime-order orbifold models, the degrees of
Wilson lines are in general the divisors of that of the orbifold
twist; they need to satisfy more stringent modular-invariance
requirements. For example, in the ${\mathbb Z}_6$ model where ${\bf
v}_6=\frac{1}{6}(1,2,-3)$ there are at most three admissible Wilson
lines, two of degree-2 (${\bf W}^{(i)}_{2},\, i=1,2$) and one of
degree-3 (${\bf W}_{3}$)
\cite{kobayashi,kobayashi2}.\footnote{Actually, the number of
admissible Wilson lines also depends on the compactified lattice. If
we restrict it to a Lie algebra root lattice, then there are four
possibilities,
(A) $\lgp{SU}{6}\oplus\lgp{SU}{2}$, (B) $\lgp{SU}{3}\oplus\lgp{SO}{8}$,
(C) $\lgp{SU}{3}\oplus\lgp{SO}{7}\oplus\lgp{SU}{2}$, and (D) $\lat$, whose
Coxeter elements realize the $\mz_6$ orbifolding. (Note that the
lattices C and D are the sublattices of B, where the
quotients are isomorphic to $\mz_2$ and $\mz_2\times\mz_2$. The
$\mz_2\times\mz_2$ is the center of $\lgp{SO}{8}$. There are two
additional lattices,
$\lgp{SU}{4}^{[2]}\oplus\lgp{SU}{3}\oplus\lgp{SU}{2}$ and $\lgp{SU}{3}^{[2]}\oplus\lgp{SU}{3}\oplus \lgp{SO}{4}$,
which also involve outer automorphism of the root lattice and give identical results to C and D.) The A (B/C)
lattice has at most one degree-2 (one degree-2 and one degree-3) Wilson line(s). The D lattice corresponds to the
case in the main text and is the most intuitive one. For our three-family models, we choose to turn on two Wilson
lines, one degree 2 and one degree 3. The low energy phenomenology for the last three lattices are different,
since both the quantum numbers (which label the states) and the selection rules (which determine allowed
couplings) depend on the lattice choice. In particular, the $\lgp{D}{4}$ family symmetry does not subsist for the
B and C lattices.} Modular-invariance conditions involving these Wilson lines are the following more restrictive
set (where $i=1,2$),
\be
\{2({\bf W}^{(i)}_{2})^2\,,\,\,3({\bf W}_{3})^2\,,\,\,4{\bf
W}_2^{(1)}\cdot {\bf W}^{(2)}_2\,,\,\,12 {\bf W}^{(i)}_{2}\cdot {\bf
W}_{3}\}=0\,{\rm mod}\,2.\label{mod2}
\ee

Further complications arise for non-prime-order orbifold models.
Unlike the prime-order orbifolds such as the ${\mathbb Z}_3$ and
${\mathbb Z}_7$ models, fixed points of the higher-twisted sectors
are not always invariant under the defining orbifold twist,
$\theta$. However, one can find linear combinations of the states
corresponding to these fixed points such that they have definite
eigenvalues under the $\theta$ rotation. As it has been shown in
ref.~\cite{kobayashi3,kobayashi2}, for any fixed point represented
by a space-group element $(\theta^k, {\bf l})$ (i.e., a fixed point
${\bf f}$ satisfying ${\bf f}=\theta^k{\bf f}+{\bf l}$, where ${\bf
l}$ is a vector of the T$^6$ lattice and called the {\it equivalent
shift vector}), if it can be written as a power of a prime element
$(\theta^m, {\bf l}')$ (a prime element is the one that cannot be
written as a power of any other element), then the linear
combinations,
\be
|k,\gamma\rangle=\sum_{\ell=0}^{m-1}
\gamma^{-\ell}|\theta^k,\theta^\ell {\bf l}\rangle\,,\label{eigen}
\ee
have eigenvalues $\gamma={\rm e}^{2\pi{\rm i} n/m}$
and $n=0,1,\cdots,m-1$ under
the $\theta$ rotation. We can therefore use this notation to
appropriately label twisted-sector states arising from different
fixed points according to their $\theta$-eigenvalues.

Finally, in the ${\mathbb Z}_6$ orbifold that we are most interested in (and other orbifolds containing
sub-orbifolds with fixed tori), special care must also be taken in the presence of discrete Wilson lines. This
orbifold contains ${\mathbb Z}_2$ and ${\mathbb Z}_3$ sub-orbifolds with twists $\{1,\theta^3\}$ and
$\{1,\theta^2,\theta^4\}$ respectively. The second and third complex planes are invariant tori of the $\theta^3$
and $\theta^{2}/\theta^{4}$ twists, i.e., they are unrotated under the respective orbifold twists. Consequently
these directions have mode expansions of a toroidal coordinate in the $T_3$ and $T_{2,4}$ sectors, as in
eq.~\ref{PLR}, with gauge momenta replaced by the shifted momenta, ${\bf P}+3{\bf X}_{n_{\bf f}}$ in the $T_3$
and ${\bf P}+2{\bf X}_{n_{\bf f}}$, ${\bf P}+4{\bf X}_{n_{\bf f}}$ in the $T_2$, $T_4$ sectors. There are also
Wilson-line dependent terms. The mode expansions of the second direction in the $T_3$ sector and the third
direction in the $T_{2,4}$ sectors contain the $2{\bf W}_3$ and $3{\bf W}_2^{(i)}$ ($i=1,2$) Wilson lines
respectively. (The numeric factors are due to the fact that the only non-trivial action for the ${\bf W}_3$ and
${\bf W}_2^{(i)}$ Wilson lines are the ${\mathbb Z}_2$ and ${\mathbb Z}_3$ sub-orbifold twists.) Just like the
untwisted-sector states, the winding numbers for massless states along these unrotated directions in respective
twisted sectors must be set to zero, and the shifted momenta must satisfy
the (necessary) projection conditions $2({\bf
P}+3{\bf X}_{n_{\bf f}})\cdot {\bf W}_3\in{\mathbb Z}$ for the $T_3$ and $3({\bf P}+2{\bf X}_{n_{\bf f}})\cdot
{\bf W}_2^{(i)}\in{\mathbb Z}$, $3({\bf P}+4{\bf X}_{n_{\bf f}})\cdot {\bf W}_2^{(i)}\in{\mathbb Z}$ for the
$T_2$, $T_4$ sector states. Using the modular invariance conditions, eqs.~\ref{mod1} and \ref{mod2}, these
conditions reduce to
\be
{\bf P}\cdot{\bf W}_3\in{\mathbb Z}\,,\quad{\rm for}~T_3~{\rm
sector}\,;\qquad {\bf P}\cdot{\bf W}_2^{(i)}\in{\mathbb
Z}\,,\quad{\rm for}~T_{2,4}~{\rm sector}\,.\label{proj3}
\ee
These projections are crucial for rendering an anomaly-free mass
spectrum and have not been properly addressed in the literature.

Multiplicities of the twisted-sector states are computed from the
generalized  Gliozzi-Scherk-Olive (GSO) projector
\cite{Gliozzi:1976qd,IMNQ2}, which can be derived from the one-loop
partition function by power-series expansions. In our notation, for
a twisted-sector state labelled by $\{k,\gamma,n_{\bf f}\}$ and
appropriate oscillator numbers, the projector is \cite{kobayashi}
\be
P(k,\gamma,n_{\bf f})
=\frac{1}{N}\sum_{\ell=0}^{N-1}\left[\Delta(k,\gamma,n_{\bf
f})\right]^\ell\,,\label{GSO}
\ee
where
\be
\Delta(k,\gamma,n_{\bf
f})=\phi\gamma\exp\left\{{\rm i}\pi\biggl[(2{\bf P}+k{\bf X}_{n_{\bf
f}})\cdot{\bf X}_{n_{\bf f}}-(2{\bf r}+k{\bf v})\cdot {\bf
v}\biggr]\right\}\,.
\label{Delta}
\ee
In this expression, $\phi=\exp[2\pi{\rm i}\sum_{i=1}^6 (N^L_i-N^R_i){\hat
v}_i]$ is the oscillator phase, with $\hat{v}_i={\rm
sgn}({\widehat{kv_i}})v_i$, $\hat{v}_{i+3}=-{\rm
sgn}({\widehat{kv_i}})v_i$ if ${\widehat{kv_i}}\neq 0$, and
$\hat{v}_{i},\hat{v}_{i+3}=0$ if ${\widehat{kv_i}}=0$, for
$i=1,2,3$. Only states with $\Delta(k,\gamma, n_{\bf f})=1$ will
survive the projection.  (Equivalently, the generalized GSO
projector can be defined as in ref.~\cite{IMNQ2}, in a slightly
different fashion. We find it is more convenient to implement
eq.~\ref{GSO} in the non-prime-order orbifolds.)

The GSO projector in eqs.~\ref{GSO} and \ref{Delta} needs to be
modified slightly for states in the untwisted sectors and those
twisted sectors with fixed tori. In the untwisted case where $k=0$,
$\gamma=\phi=1$, $n_{\bf f}=0$, the required modification is simply
an additional projection with respect to the Wilson line,
\be
\Delta^\ell\rightarrow
 \frac{1}{N_W}\sum_{m_{\bf
f}=0}^{N_W-1} \exp\biggl[{\rm i}\pi\ell(2{\bf P}\cdot{\bf X}_{m_{\bf
f}}-2{\bf r}\cdot {\bf v})\biggr]\,.
\ee
Then $\Delta=1$ gives ${\bf P}\cdot{\bf V}-{\bf r}\cdot {\bf
v}\in{\mathbb Z}$, ${\bf P}\cdot {\bf W}\in\mz$, the same projection
conditions as in eqs.~\ref{gauge}, \ref{matter} and {\ref{PA}.

The second case is more complicated. Assume that the $T_k$ sector
has a fixed torus, and the Wilson line of degree $N_W$ also lies in
this torus. Then we must have $N_W$ divides $k$, and the required
modification to the GSO projector is
\be
\Delta^\ell\rightarrow \frac{(\phi\gamma)^\ell}{N_W}\sum_{m_{\bf
f}=0}^{N_W-1} \exp\left\{{\rm i}\pi\ell\biggl[(2{\bf P}+k{\bf X}_{m_{\bf
f}})\cdot{\bf
X}_{m_{\bf f}}\right.
\left.-(2{\bf r}+k{\bf v})\cdot {\bf
v})\biggr]\right\}\,.\label{modification2}
\ee
In this expression, it must be understood that each shifted
momentum, ${\bf P}+k{\bf X}_{m_{\bf f}}$, satisfies its own
masslessness condition, eq.~\ref{masscond2}. Therefore the momenta
${\bf P}$ in different terms differ by a multiple of $k{\bf W}$,
which is a vector in the ${\rm E_8\times E_8}$ root lattice; these
states are isomorphic to each other. Requiring that $\Delta=1$ in
eq.~\ref{modification2} then implies an additional projection
${\bf P}\cdot{\bf W}\in\mz$ for the $T_k$ twisted sector states. It
can also be seen that with this condition eq.~\ref{modification2}
reduces to eqs.~\ref{GSO} and \ref{Delta} with ${\bf X}_{n_{\bf
f}}={\bf V}$. These discussions can be easily generalized to the
cases with several Wilson lines. In the ${\mathbb Z}_6$ model, they
give the same projection conditions as in eq.~\ref{proj3}.

While it is quite tedious to carry out all these procedures in
practice, they can be easily computerized, as follows. First we
check the modular invariance of the gauge twists and Wilson lines,
and find all the ${\rm E_8\times E_8}$ roots that satisfy eqs.
\ref{gauge} and \ref{PA}. From these we find the set of linearly
independent positive roots (i.e., the simple roots) and determine
the unbroken gauge groups. Next we find the roots corresponding to
the untwisted- and twisted-sector matter states, from eqs.
\ref{matter}, \ref{PA}, and eqs. \ref{masscond1},
\ref{masscond2} respectively. For the twisted-sector matter, we
only keep states of negative helicity (i.e. left-handed), and
compute their multiplicities using the generalized GSO projector in
eq.~\ref{GSO}. We also need to take special care of the
twisted-sector states in the presence of Wilson lines, as noted in
the paragraph after eq.~\ref{eigen}. We then decompose the roots
into highest weight representations, and determine their Dynkin
indices and dimensions using the standard group theoretical method.
Finally, the remaining abelian factors are found by searching for
orthogonal roots, and the charges for matter states are determined
accordingly by projecting the matter states onto these abelian
roots.

\section{Three-family Pati-Salam models \label{app:models}}

\subsection{The models \label{model}}

In this appendix, we apply the procedure outlined in
appendix~\ref{review} to construct three-family PS models in the
${\mathbb Z}_6$ orbifold with ${\bf v}_6=\frac{1}{6}(1,2,-3)$. The
${\mathbb Z}_6$ is equivalent to the ${\mathbb Z}_2\times {\mathbb
Z}_3$ orbifold, where the two twist vectors are ${\bf v}_2=3{\bf
v}_6=\frac{1}{2}(1,0,-1)$ and ${\bf v}_3=2{\bf
v}_6=\frac{1}{3}(1,-1,0)$.

There are in total $61$ inequivalent modular invariant choices for
the gauge twists in the ${\mathbb Z}_6$ orbifold model
\cite{katsuki}.  To narrow down the possibilities, we demand the
models we start with (before imposing any Wilson line) contain an
$\lgp{SO}{10}$ gauge group and some matter fields in ${\bf 16}/\overline{\bf
16}$ representations in the first or third twisted sectors. Although
this step makes our results less generic, it greatly reduces the
large number of possible models to a manageable subset. We choose
the following two gauge twists,

\noindent{$\bullet$~model A}:
\be
{\bf
V}_6=\frac{1}{6}\left(22200000\right)\left(11000000\right)\,,\label{gt1}
\ee \noindent{$\bullet$~model B}: \be {\bf
V}_6=\frac{1}{6}\left(41100000\right)\left(22000000\right)\,,\label{gt2}
\ee
which break the $\lgp{E}{8}\times\lgp{E}{8}$ gauge symmetry down to
$\lgp{SO}{10}\times\lgp{SU}{3}\times\lgp{E}{7}^\prime$
and $\lgp{SO}{10}\times\lgp{SU}{2}\times
\lgp{E}{7}'$ respectively. The model A (B) contains four (two) ${\bf 16}$
and one (three) $\overline{\bf 16}$ in the untwisted sectors, and
eighteen (fourteen) ${\bf 16}$ and three (seven) $\overline{\bf 16}$
in the twisted sectors; in total there are eighteen (six) $\lgp{SO}{10}$
families \cite{katsuki}.

To further break the gauge symmetries and reduce the number of
families, we impose discrete Wilson lines. As previously mentioned,
there are at most one degree-3 Wilson line in the second complex
plane, and two degree-2 Wilson lines in the third. We choose to
add two of them, one of degree-2 and one of degree-3, as
follows,

\noindent{$\bullet$~model A1}: \bea & & {\bf
W}_2=\frac{1}{2}\left(100001  1  1\right) \left( 0000 0000\right)\,,
\qquad {\bf W}_3=\frac{1}{3}\left( 1  -1  00 0000\right)\left( 0020
0000 \right)\,,\label{wla1} \eea \noindent{$\bullet$~model A2}: \bea
& & {\bf W}_2=\frac{1}{2}\left( 1  0000 1  1  1\right) \left( 0000
0000\right)\,, \qquad {\bf W}_3=\frac{1}{3}\left(2  1 -1
00000\right)\left(
 0  2  1  1  0000\right)\,,\label{wla2}
\eea \noindent{$\bullet$~model B}: \bea & & {\bf
W}_2=\frac{1}{2}\left(0001 1  000\right) \left( 1  0  1
00000\right)\,, \qquad {\bf W}_3=\frac{1}{3}\left( 0  1 -1
00000\right)\left( 00 2 00000\right)\,. \label{wlb}\eea
One can easily verify our choices satisfy the modular-invariance
requirements, eqs. \ref{mod1} and \ref{mod2}. The ${\bf W}_3$
Wilson line of models A1/B and A2 breaks the gauge group in the
observable sector to $\lgp{SO}{10}$ and $\lgp{SO}{10}\times\lgp{SU}{2}$,
respectively, and the ${\bf W}_2$ breaks them further down to the PS
gauge group, in all three models.

The remaining unbroken gauge groups are
$\ps\times\lgp{SO}{10}^\prime\times\lgp{SU}{2}^\prime\times(\lgp{U}{1})^5$
for models A1/A2 and $\ps\times\lgp{SO}{10}'\times(\lgp{U}{1})^6$
for model B. The untwisted- and twisted-sector matter provide the
following irreducible representations of the PS gauge group (modulo
some singlets),

\noindent{$\bullet$~model A1}: \bea U_1: && ({\bf 4},{\bf 2}, {\bf
1}) \,,\,\, U_2: ({\bf 1},{\bf 2},{\bf 2})\,,\,\,
U_3: ({\bf 4},{\bf 1},{\bf 2})+(\overline{\bf 4},{\bf 1},{\bf 2})\,,\nn\\
T_1: && 2({\bf 4}, {\bf 2}, {\bf 1})+2(\overline{\bf 4},{\bf 1},{\bf
2})+4({\bf 4},{\bf 1},{\bf 1})+4(\overline{\bf 4},{\bf 1},{\bf 1})
+8({\bf 1,2,1})
+8({\bf 1,1,2})+2({\bf 1,1,2;1,2})\,,\nn\\
T_2:&&2(\overline{\bf 4},{\bf 1},{\bf 2})+({\bf 6},{\bf 1},{\bf
1})\,,\qquad T_3:\,6({\bf 6},{\bf 1}, {\bf 1})+6({\bf 1},{\bf
2},{\bf 2})\,,\qquad T_4:\,({\bf 4}, {\bf 1}, {\bf 2}) +2({\bf
6},{\bf 1},{\bf 1})\,, \label{modelA1} \eea
\noindent{$\bullet$~model A2}: \bea U_1: && ({\bf 4},{\bf 2}, {\bf
1})
\,,\nonumber\\
T_1: && 2({\bf 4}, {\bf 2}, {\bf 1})+2(\overline{\bf 4},{\bf 1},{\bf
2}) +4({\bf 4},{\bf 1},{\bf 1})+4(\overline{\bf 4},{\bf 1},{\bf
1})+8({\bf
1,2,1}) +6({\bf 1,1,2})\nn\\
&+&2({\bf 1,2,1;1,2})+2({\bf 1,1,2;1,2})\,,\nn\\
T_2: && 2(\overline{\bf 4},{\bf 1},{\bf 2})+({\bf 6},{\bf 1},{\bf
1})+({\bf 1},{\bf 2},{\bf 2})\,,\qquad T_3:\, 2({\bf 4},{\bf 1},
{\bf 1})+2(\overline{\bf 4},{\bf 1},{\bf
1})+6({\bf 1,1,2})\,, \nn\\
T_4: && ({\bf 4}, {\bf 1}, {\bf 2}) +2({\bf 6},{\bf 1},{\bf
1})+2({\bf 1},{\bf 2},{\bf 2}) \,,
\eea
\noindent{$\bullet$\,model
B}:
\bea U_1: && ({\bf 4,2,1})\,,\,\, U_2: ({\bf 6}, {\bf 1},{\bf
1})\,,\,\,
U_3: ({\bf \overline{4},2,1})+({\bf 4},{\bf 2},{\bf 1})\,,\nonumber\\
T_1: && 2({\bf 6}, {\bf 1}, {\bf 1})+2({\bf 1},{\bf 2},{\bf
2})+4({\bf 4},{\bf 1},{\bf 1}) +4(\overline{\bf 4},{\bf 1},{\bf
1})+10({\bf 1,2,1})
+8({\bf 1,1,2})\,,\nn\\
T_2: && 2({\bf \overline{4},1,2})+({\bf 1},{\bf 2},{\bf 2})\,,\qquad
T_3:\,4({\bf 4},{\bf 2},{\bf 1})+4(\overline{\bf 4},{\bf 1},{\bf
2})+2(\overline{\bf 4},{\bf 2}, {\bf 1})+2({\bf 4},{\bf 1},{\bf
2})+6({\bf 1,2,1})\,,\nn\\
T_4:&&({\bf 4},{\bf 1},{\bf 2})+2({\bf 1},{\bf 2},{\bf 2}) \,.
\eea
In general, the $T_{6-k}$-sector is the CPT conjugate of the
$T_k$-sector, therefore $T_5$ does not give rise to additional
states. However we need to keep those states from the $T_4$-sector
whose CPT conjugates in the $T_2$-sector have positive helicities.

We see there are indeed three chiral families ($({\bf
4,2,1})+({\bf{\overline 4},1,2})$ under the PS group), two from the
$T_1$ ($T_3$) twisted sector and one from the untwisted and the
$T_{2,4}$ twisted sectors in model A1/A2 (B), modulo some $({\bf
4,1,2})+({\bf {\overline 4},1,2})$ (and $({\bf 4,2,1})+({\bf
{\overline 4},2,1})$ in model B) vector-like pairs. Each $({\bf
4,2,1})+({\bf{\overline 4},1,2})$ family encompasses one complete
family of the SM quarks and leptons (plus the right-handed
neutrino), since they decompose under the SM gauge group,
$\lgp{SU}{3C}\times\lgp{SU}{2L}\times\lgp{U}{1Y}$, as follows:
$({\bf 4,2,1})=({\bf 3,2})_{1/6}+({\bf 1,2})_{-1/2}$,
$({\bf{\overline
4},1,2})=({\bf\overline{3},1})_{1/3}+({\bf\overline{3},1})_{-2/3}+({\bf
1,1})_{1}+({\bf 1,1})_0$. The complete matter spectra for these
models is listed in appendix \ref{notation}, where the notation for
the twisted-sector states is also explained. Three-family PS models
have been previously constructed in heterotic string models using
the free-fermionic method \cite{ALR}, but to our knowledge, has not
been realized with abelian orbifolds.

Like all the three-family orbifold models in the literature
\cite{IMNQ,IMNQ2,FIQS}, our models suffer from the
embarrassment-of-riches syndrome, i.e., they contain many candidates
for SM exotic particles, $({\bf 4,1,1})$, $({\bf{\overline 4},1,1})$
and $({\bf 1,2,1})$, $({\bf 1,1,2})$. (These states decompose under
the SM gauge group as follows: $({\bf 4,1,1})=({\bf
3,1})_{1/6}+({\bf 1,1})_{-1/2}$,
$({\bf\overline{4},1,1})=({\bf\overline{3},1})_{-1/6}+({\bf
1,1})_{1/2}$, $({\bf 1,2,1})=({\bf 1,2})_0$ and $({\bf 1,1,2})=({\bf
1,1})_{1/2}+({\bf 1,1})_{-1/2}$, and have exotic charge
assignments.)  Nevertheless, these exotic states are necessary for
the quantum consistency of the models. Models A1/A2 (B) contain five
(six) additional abelian symmetries (their charges for the matter
fields are listed in appendix \ref{notation}), one of them is
anomalous. This $\lgp{U}{1A}$ anomaly satisfies the following
universal condition,
\be
{\rm Tr} T(R){\widetilde Q}_A={\rm Tr} {\widetilde Q}_i^2
{\widetilde Q}_A=\frac{1}{3}{\rm Tr}{\widetilde Q}_A^3
=\frac{1}{24}{\rm Tr}{\widetilde Q}_A\nn\\
=\left\{\begin{array}{ll}
-\frac{1}{2\sqrt{3}}\,,&\quad {\rm Model~A1}\,,\\[1mm]
1\,,&\quad {\rm Model~A2}\,,\\[1mm]
\frac{4}{3}\,,&\quad{\rm Model~B}\,,
\end{array}\right.
\ee
where ${\widetilde Q}_{i, A}=Q_{i, A}/k^{1/2}_{i, A}$ with $k_{i}, k_A$ are the normalization factors (``levels")
of the corresponding abelian groups, and $2T(R)$ is the index of the representation $R$ under a specific
non-abelian factor. This condition guarantees the anomalies are cancelled by the generalized Green-Schwarz
mechanism \cite{GS} in 4d \cite{au1}. In fact, considerations of these anomalies provide useful consistency
checks for our computer program. One can also check that all the 4d chiral anomalies for the non-abelian gauge
groups and global anomalies for the $\lgp{SU}{2}$ groups vanish.

We note that our models differ from the three-family SM-like models constructed from ${\mathbb Z}_3$ orbifolds in
several respects. For example, in the simplest model of the first paper in ref.~\cite{IMNQ}, the left-handed
quarks arise solely from the untwisted sectors, one family each from the three untwisted sectors. The
right-handed quarks and left-handed leptons arise from the twisted sectors, and the multiplicities of three come
from the three equivalent fixed points in one of the complex planes. As such, the three families are completely
degenerate in the horizontal family space. Our models exhibit quite different spectral patterns. Two complete
families come from the $T_1/T_3$ twisted sectors, and one family comes from the combination of the untwisted and
$T_2/T_4$ twisted sectors. This type of pattern breaks the degeneracies among the three families, and may have
better prospects for explaining the observed fermion mass hierarchy and CP phases.  It will also be interesting
to generalize our models to other non-prime-order and ${\mathbb Z}_N\times {\mathbb Z}_M$ orbifolds to see
whether this feature persists.

The above three-family PS models have some advantages over string-based SM-like and GUT models which have been
the foci of model buildings in the past. The SM-like models usually contain many extra abelian gauge symmetries,
and in general it is difficult to reproduce the standard hypercharge normalization as in GUT theories. (For an
extensive search of $\lgp{U}{1Y}$ in the ${\mathbb Z}_3$ orbifold, see ref.~\cite{giedt}.) This may upset the
successful prediction of the weak-mixing angle in the context of supersymmetric unification theories (see,
however, ref.~\cite{kim} for alternative tri-unification models). In the PS model, on the other hand, the
hypercharge normalization is standard, by construction. The hypercharge $\lgp{U}{1Y}$ is a diagonal subgroup of
the $\lgp{U}{1T_3}\subset \lgp{SU}{2R}$ and the $\lgp{U}{1(B-L)}\subset\lgp{SU}{4C}$, i.e.
$Q_Y=T_{3R}+\frac{1}{2}(B-L)$, hence $k_Y=1+\frac{1}{4}\times\frac{8}{3} =\frac{5}{3}$.  Moreover, all our models
contain $\lgp{SO}{10}$ spinor representations, which may be a welcoming ingredient for better low energy
phenomenology \cite{Nilles}. (In contrast, the spinor representations are absent in SM-like string models and
models based on the Spin$(32)/\mz_2$ heterotic string and superstrings.) Finally, when compared with string GUT
models based on $\lgp{SU}{5}$ or other larger gauge symmetries \cite{sgut}, the PS models generically do not need
large Higgs representations (such as those in the adjoint representation) to break the GUT gauge symmetries.
These large Higgs representations require one to realize the current algebra of the gauge symmetries at higher
Ka\v{c}-Moody levels \cite{hl} and render the models more complicated.

\subsection{Complete matter spectra}\label{notation}

\begingroup
\squeezetable
\begin{table*}[th]\caption{\label{tab1}Matter spectrum of model A1.
The levels of the $\lgp{U}{1}$ groups are $24$, $24$, $72$, $36$ and $48$.}
\begin{ruledtabular}
\begin{tabular}{|l|l|rrrrr|r|l|l|rrrrr|r|}
Sectors& {\tiny ${\rm
PS\times SO_{10}'\times SU_2'}$} & $Q_1$ &
$Q_2$ & $Q_3$ & $Q_4$ & $Q_A$ & Labels & Sectors & {\tiny ${\rm
PS\times SO_{10}'\times SU_2'}$} & $Q_1$ & $Q_2$ &
$Q_3$ & $Q_4$ & $Q_A$ & Labels \\
\hline\hline
$U_1$ & (${\bf{4}}, {\bf 2},{\bf 1},{\bf 1},{\bf 1}$)
& $1$ & $1$ & $0$ & $3$ & $-2$ & $f_3$ & $T_{2(-1)(1)(00)(000)}$ &(${\bf
1}, {\bf 1}, {\bf 1}, {\bf 1},{\bf 1}$) &
$0$ & $-2$ & $-4$ & $-2$ & $0$ & $S_{14}$ \\
&(${\bf 1}, {\bf 1},{\bf 1},{\bf 10},{\bf 2}$) & $0$ & $0$ & $0$ &
$1$ & $0$ & $B'$ & $T_{2(-1)(2)(00)(000)}$ &(${\bf 1}, {\bf 1}, {\bf 1},
{\bf 1},{\bf 1}$) &
$0$ & $0$ & $4$ & $-2$ & $4$ & $S_{15}$ \\
\cline{1-8}
$U_2$ &(${\bf 1}, {\bf 2},{\bf 2},{\bf 1},{\bf 1}$) &
$2$ & $-1$ & $0$ & $-2$ & $2$ & $h_1$ &  &(${\bf 1}, {\bf 1}, {\bf 1}, {\bf
1},{\bf 1}$) &
$0$ & $0$ & $0$ & $-2$ & $-2$ & $S_{16}$ \\
&(${\bf 1}, {\bf 1}, {\bf 1}, {\bf 1},{\bf 1}$) & $0$ & $0$ & $0$
& $2$ & $0$ & $S_1$ & &(${\bf 1}, {\bf 1}, {\bf 1}, {\bf 10},{\bf 1}$) &
$0$ & $0$ & $-2$ & $2$ & $-2$ &  $A'_1$ \\
\cline{1-8} $U_3$ & (${\bf 4}, {\bf 1}, {\bf 2}, {\bf 1},{\bf 1}$) &
$3$ & $0$ & $0$ & $1$ & $0$ & $\overline\chi^c_1$ &
$T_{2(1)(0)(00)(000)}$ &($\overline{\bf 4}, {\bf 1}, {\bf 2}, {\bf
1}, {\bf 1}$) &
$1$ & $0$ & $0$ & $1$ & $0$ & $\chi^c_{1,2}$ \\
& ($\overline{\bf 4},{\bf 1}, {\bf 2}, {\bf 1},{\bf 1}$)  & $-3$ &
$0$ & $0$ & $-1$ & $0$ & $f_3^c$ & $T_{2(1)(1)(00)(000)}$ &(${\bf
1}, {\bf 1}, {\bf 1}, {\bf 1},{\bf 2}$) &
$0$ & $-2$ & $-2$ & $-5$ & $0$ & $D'_5$\\
\cline{1-8}
$T_{1(1)(0)(0n')(000)}$ & (${\bf 4},{\bf
2},{\bf 1},{\bf 1},{\bf 1}$) & $-1$ & $0$ & $0$ & $0$ & $0$ & $f_{1,2}$ &
$T_{2(1)(2)(00)(000)}$ &(${\bf 1}, {\bf 1}, {\bf 1}, {\bf 1}, {\bf
2}$) & $0$ & $0$ & $2$ & $-1$ & $4$ & $D'_6$ \\
& ($\overline{\bf 4}, {\bf 1}, {\bf 2},{\bf 1},{\bf 1}$) & $-1$ &
$0$ & $0$ & $0$ & $0$ & $f_{1,2}^c$ & &(${\bf 1}, {\bf 1}, {\bf 1},
{\bf 16}, {\bf 1}$) & $0$ & $0$ & $-1$ & $0$ & $-2$ & $F'$
\\
$T_{1(1)(1)(0n')(000)}$ & (${\bf 1}, {\bf 1}, {\bf 1}, {\bf 1},{\bf
2}$) & $-2$ & $0$ & $-2$ & $0$ & $-2$ & $D'_1$ &
$T_{2(-1)(0)(00)(0{\overline 1}0)}$ &(${\bf 1}, {\bf 1}, {\bf 1},
{\bf 1},{\bf 1}$) & $0$ & $2$ & $4$ & $4$ & $2$ & $S_{17}$
\\
& (${\bf 1}, {\bf 1}, {\bf 1}, {\bf 1},{\bf 2}$) & $2$ & $1$ & $2$ &
$2$ & $2$ & $D'_2$ & $T_{2(-1)(2)(00)(0{\overline 1}0)}$ &(${\bf 1},
{\bf 1}, {\bf 1}, {\bf 1},{\bf 1}$) & $0$ & $0$ & $0$ & $0$ & $-2$ &
$S_{18}$
\\
$T_{1(1)(2)(0n')(000)}$ &(${\bf 1}, {\bf 1}, {\bf 1}, {\bf 1},{\bf
2}$) & $-2$ & $1$ & $2$ & $0$ & $0$ & $D'_3$ & $T_{2(-1)(2)(00)(100)}$
&(${\bf 1}, {\bf 1}, {\bf 1}, {\bf 1},{\bf 2}$) & $0$ & $0$ & $-2$
& $1$ & $-2$ & $D'_7$
\\
&(${\bf 1}, {\bf 1}, {\bf 1}, {\bf 1},{\bf 2}$) & $2$ & $-1$ &
$-2$ & $-2$ & $-2$ & $D'_4$ & $T_{2(1)(0)(00)(100)}$ & (${\bf 1}, {\bf 1},
{\bf 1}, {\bf 1}, {\bf 1}$) & $0$ & $2$ & $4$ & $4$ & $2$ & $S_{19}$
\\
$T_{1(1)(0)(1n')(000)}$ &(${\bf 4}, {\bf 1}, {\bf 1},{\bf 1}, {\bf
1}$) & $-1$ & $0$ & $2$ & $-1$ & $3$ & $q_1$ &
$T_{2(1)(2)(00)(100)}$&(${\bf 1}, {\bf 1}, {\bf 1}, {\bf 1},{\bf
1}$) & $0$ & $0$ & $0$ & $0$ & $-2$ & $S_{20}$
\\
&(${\bf  4}, {\bf 1}, {\bf 1}, {\bf 1},{\bf 1}$) & $-1$ & $0$ & $-2$
& $1$ & $-3$ & $q_2$ & $T_{2(1)(2)(00)(0{\overline 1}0)}$ &(${\bf
1}, {\bf 1}, {\bf 1}, {\bf
1},{\bf 2}$) & $0$ & $0$ & $-2$ & $1$ & $-2$ & $D'_8$ \\
\cline{9-16} &(${\bf 1}, {\bf 2}, {\bf 1}, {\bf
1},{\bf 1}$) & $2$ & $0$ & $2$ & $0$ & $3$ & $D_1^\ell$ &
$T_{3(\omega)(0)(0n')(000)}$ &(${\bf 6}, {\bf 1}, {\bf 1}, {\bf
1}, {\bf 1}$) & $0$ & $0$ & $0$ & $1$ & $0$ & $C_2$
\\
&(${\bf 1}, {\bf 2}, {\bf 1}, {\bf 1},{\bf 1}$) & $2$ & $0$ & $-2$
& $2$ & $-3$ & $D_2^\ell$ & & (${\bf 1}, {\bf 2}, {\bf 2}, {\bf 1}, {\bf 1}$) &
$0$ & $0$ & $0$ & $1$ & $0$ & $h_2$
\\
$T_{1(1)(1)(1n')(000)}$ &(${\bf 1}, {\bf 2}, {\bf 1}, {\bf 1},{\bf
1}$) & $-2$ & $0$ & $2$ & $-2$ & $1$ & $D_3^\ell$ & & (${\bf 1}, {\bf 1}, {\bf
1}, {\bf 1}, {\bf 1}$)  & $2$ & $-1$ & $0$ & $-3$ & $2$ & $S_{21}$
\\
&(${\bf 1}, {\bf 1}, {\bf 2}, {\bf 1},{\bf 1}$) & $0$ & $-1$ &
$-2$ & $-2$ & $-3$ & $D_1^r$ & $T_{3(\omega^2)(0)(0n')(000)}$ &(${\bf 1},
{\bf 1}, {\bf 1}, {\bf 1},{\bf 1}$) & $-2$ & $1$ & $0$ & $1$ &
$-2$ & $S_{22}$
\\
&($\overline{\bf 4}, {\bf 1}, {\bf 1}, {\bf 1},{\bf 1}$) & $1$ & $0$
& $2$ & $-1$ & $1$ & $\overline{q}_1$ & &(${\bf 1}, {\bf 1}, {\bf
1}, {\bf 1},{\bf 1}$) & $2$ & $-1$ & $0$ & $-1$ & $2$ & $S_{23}$
\\
$T_{1(1)(2)(1n')(000)}$ &(${\bf 1}, {\bf 1}, {\bf 2}, {\bf 1},
{\bf 2}$) & $0$ & $0$ & $0$ & $-1$ & $-1$  & $\Delta$ &
$T_{3(1)(0)(0n')(000)}$ &(${\bf 1}, {\bf 1}, {\bf 1}, {\bf 1},{\bf
1}$) & $-2$ & $1$ & $0$ & $3$ & $-2$ & $S_{24}$
\\
$T_{1(1)(0)(0n')(100)}$ &(${\bf 1}, {\bf 1}, {\bf 1}, {\bf 1},{\bf
1}$) & $-2$ & $-1$ & $0$ & $-3$ & $2$ & $S_2$ & & (${\bf 6}, {\bf 1}, {\bf
1}, {\bf 1}, {\bf 1}$) & $0$ & $0$ & $0$ & $-1$ & $0$ & $C_3$
\\
&(${\bf 1}, {\bf 1}, {\bf 1}, {\bf 1},{\bf 1}$) & $-2$ & $-1$ &
$-4$ & $-1$ & $-4$ & $S_3$ & & (${\bf 1}, {\bf 2}, {\bf 2}, {\bf 1}, {\bf
1}$) & $0$ & $0$& $0$ & $-1$ & $0$ & $h_3$
\\
\cline{9-16} &(${\bf 1}, {\bf 1}, {\bf 1}, {\bf 1},{\bf 1}$) & $-2$
& $2$ & $4$ & $3$ & $2$ & $S_4$ &  $T_{4(-1)(0)(00)(000)}$ &(${\bf
4}, {\bf 1}, {\bf 2}, {\bf 1},{\bf 1}$) & $-1$ & $0$ & $0$ & $-1$ &
$0$ & $\overline\chi_{2}^c$
\\
$T_{1(1)(1)(0n')(100)}$ &(${\bf 1}, {\bf 1}, {\bf 1}, {\bf 1},{\bf
1}$) & $-2$ & $0$ & $0$ & $-1$ & $-2$ & $S_5$  & $T_{4(-1)(1)(00)(000)}$
&(${\bf 1}, {\bf 1}, {\bf 1}, {\bf 1},{\bf 2}$) & $0$ & $2$ & $2$
& $5$ & $0$ & $D'_9$
\\
&(${\bf 1}, {\bf 1}, {\bf 1}, {\bf 1},{\bf 1}$) & $2$ & $1$ & $4$
& $1$ & $2$ & $S_6$ & $T_{4(-1)(2)(00)(000)}$ &(${\bf 1}, {\bf 1}, {\bf
1}, {\bf 1},{\bf 2}$) & $0$ & $0$ & $-2$ & $1$ & $-4$ & $D'_{10}$
\\
$T_{1(1)(2)(0n')(100)}$ &(${\bf 1}, {\bf 1}, {\bf 1}, {\bf 1},{\bf
1}$) & $-2$ & $1$ & $0$ & $3$ & $0$ & $S_7$ &  &(${\bf 1}, {\bf 1}, {\bf
1}, \overline{\bf 16},{\bf 1}$) & $0$ & $0$ & $1$ & $0$ & $2$ & $\overline{F}'$\\
&(${\bf 1}, {\bf 1}, {\bf 1}, {\bf 1},{\bf 1}$) & $2$ & $-1$ &
$-4$ & $1$ & $-2$ & $S_8$ & $T_{4(1)(0)(00)(000)}$ &(${\bf 6}, {\bf 1},
{\bf 1}, {\bf 1}, {\bf 1}$) & $2$ & $0$ & $0$ & $0$ & $0$ & $C_4$
\\
$T_{1(1)(0)(1n')(100)}$ &(${\bf 1}, {\bf 2}, {\bf 1}, {\bf 1},{\bf
1}$) & $-2$ & $-1$ & $-2$ & $-2$ & $-1$ & $D_4^\ell$ & &(${\bf 1}, {\bf 1},
{\bf 1}, {\bf 1},{\bf 1}$) & $0$ & $-2$ & $-4$ & $-2$ & $-2$ & $S_{25}$
\\
&($\overline{\bf 4}, {\bf 1}, {\bf 1}, {\bf 1},{\bf 1}$) & $1$ &
$-1$ & $-2$ & $-1$ & $-1$ & $\overline{q}_2$ & &(${\bf 1}, {\bf 1},
{\bf 1}, {\bf 1},{\bf 1}$) & $-4$ & $0$ & $0$ & $-2$ & $0$ &
$S_{26}$
\\
$T_{1(1)(2)(1n')(100)}$ &(${\bf 1}, {\bf 1}, {\bf 2}, {\bf 1},
{\bf 1}$) & $0$ & $0$ & $-2$ & $2$ & $-1$ & $D_2^r$ & $T_{4(1)(1)(00)(000)}$
&(${\bf 1}, {\bf 1}, {\bf 1}, {\bf 1},{\bf 1}$)& $0$ & $2$ & $4$ &
$2$ & $0$ & $S_{27}$
\\
$T_{1(1)(0)(1n')(010)}$ &(${\bf 1}, {\bf 1}, {\bf 2}, {\bf 1},
{\bf 1}$) & $0$ & $1$ & $2$ & $2$ & $1$ & $D_3^r$ & $T_{4(1)(2)(00)(000)}$
&(${\bf 1}, {\bf 1}, {\bf 1}, {\bf 1},{\bf 1}$)& $0$ & $0$ & $-4$
& $2$ & $-4$ & $S_{28}$
\\
$T_{1(1)(0)(0n')(001)}$ &(${\bf 1}, {\bf 1}, {\bf 1}, {\bf 1},{\bf
1}$) & $2$ & $0$ & $0$ & $1$ & $0$ & $S_9$ & &(${\bf 1}, {\bf 1}, {\bf 1},
{\bf 1},{\bf 1}$) & $0$ & $0$ & $0$ & $2$ & $2$ & $S_{29}$
\\
$T_{1(1)(0)(0n')(00{\overline 1})}$ &(${\bf 1}, {\bf 1}, {\bf 1},
{\bf 1},{\bf 1}$) & $2$ & $0$ & $0$ & $1$ & $0$  & $S_9'$ &  &(${\bf
1}, {\bf 1}, {\bf 1}, {\bf 10}, {\bf 1}$) & $0$ & $0$ & $2$ & $-2$ &
$2$ & $A'_2$
\\
$T_{1(1)(0)(1n')(200)}$ &(${\bf 1}, {\bf 1}, {\bf 2}, {\bf 1},{\bf
1}$) & $0$ & $1$ & $2$ & $2$ & $1$ & $D_4^r$ &
$T_{4(-1)(0)(00)(\overline{1}00)}$ &(${\bf 1}, {\bf 1}, {\bf 1},
{\bf 1},{\bf 1}$) & $0$ & $-2$ & $-4$ & $-4$ & $-2$ & $S_{30}$
\\
$T_{1(1)(0)(0n')(110)}$ &(${\bf 1}, {\bf 1}, {\bf 1}, {\bf 1},{\bf
1}$) & $2$ & $0$ & $0$ & $1$ & $0$ & $S_{10}$ &
$T_{4(-1)(2)(00)(\overline{1}00)}$ &(${\bf 1}, {\bf 1}, {\bf 1},
{\bf 1},{\bf 1}$) & $0$ & $0$ & $0$ & $0$ & $2$ & $S_{31}$
\\
$T_{1(1)(0)(0n')(300)}$ &(${\bf 1}, {\bf 1}, {\bf 1}, {\bf 1},{\bf
1}$) & $2$ & $0$ & $0$ & $1$ & $0$ & $S_{11}$ & $T_{4(-1)(2)(00)(010)}$
&(${\bf 1}, {\bf 1}, {\bf 1}, {\bf 1},{\bf 2}$) & $0$ & $0$ & $2$
& $-1$ & $2$ & $D'_{11}$
\\
\cline{1-8}
$T_{2(-1)(0)(00)(000)}$ &(${\bf 6}, {\bf 1}, {\bf 1}, {\bf 1},{\bf
1}$) & $-2$ & $0$ & $0$ & $0$ & $0$ & $C_1$ & $T_{4(1)(0)(00)(010)}$
&(${\bf 1}, {\bf 1}, {\bf 1}, {\bf 1},{\bf 1}$) & $0$ & $-2$ &
$-4$ & $-4$ & $-2$ & $S_{32}$
\\
&(${\bf 1}, {\bf 1}, {\bf 1}, {\bf 1},{\bf 1}$) & $0$ & $2$ & $4$
& $2$ & $2$  & $S_{12}$ & $T_{4(1)(2)(00)(010)}$ &(${\bf 1}, {\bf 1}, {\bf
1}, {\bf 1},{\bf 1}$) & $0$ & $0$ & $0$ & $0$ & $2$ & $S_{33}$
\\
&(${\bf 1}, {\bf 1}, {\bf 1}, {\bf 1},{\bf 1}$) & $4$ & $0$ & $0$ &
$2$ & $0$ &$S_{13}$ & $T_{4(1)(2)(00)(\overline{1}00)}$ &(${\bf 1},
{\bf 1}, {\bf 1}, {\bf 1},{\bf 2}$) &
$0$ & $0$ & $2$ & $-1$ & $2$  & $D'_{12}$ \\
\end{tabular}
\end{ruledtabular}
\end{table*}
\endgroup

\begingroup
\squeezetable
\begin{table*}[th]\caption{\label{tab2}
Matter spectrum of model A2. The levels of the $\lgp{U}{1}$ groups are $16$,
$72$, $32$, $80$ and $36$.}
\begin{ruledtabular}
\begin{tabular}{|l|l|rrrrr|l|l|rrrrr|}
Sectors& ${\rm PS\times\lgp{SO}{10}'\times\lgp{SU}{2}'}$ & $Q_1$ & $Q_2$ &
$Q_3$ & $Q_4$ & $Q_A$ & Sectors& ${\rm PS\times\lgp{SO}{10}'\times
\lgp{SU}{2}'}$ & $Q_1$ & $Q_2$ &
$Q_3$ & $Q_4$ & $Q_A$ \\
\hline\hline $U_1$ & (${\bf{4}}, {\bf 2},{\bf 1},{\bf 1},{\bf 1}$)
& $0$ & $3$ & $1$ & $-5$ & $0$ & $T_{2(-1)(1)(00)(000)}$&(${\bf
1}, {\bf 1}, {\bf 1}, {\bf 1},{\bf 1}$) &
$2$ & $-6$ & $0$ & $4$ & $0$ \\
& (${\bf 1}, {\bf 1},{\bf 1},{\bf 16},{\bf 1}$) & $0$ & $0$ & $1$
& $1$ & $3$  & &(${\bf 1}, {\bf 2}, {\bf 2}, {\bf 1},{\bf 1}$) &
$0$ & $0$ & $2$ & $0$ & $0$
\\
 &(${\bf 1}, {\bf 1},{\bf 1},{\bf 1},{\bf 2}$) &
$0$ & $0$ & $-2$ & $4$ & $3$  & &(${\bf 1}, {\bf 1}, {\bf 1}, {\bf
1},{\bf 1}$) & $-2$ & $6$ & $4$ & $-4$ & $0$
\\
\cline{1-7} $U_2$ &(${\bf 1}, {\bf 1}, {\bf 1}, {\bf 1},{\bf 1}$)
& $2$ & $-6$ & $-4$ & $0$ & $0$   & $T_{2(-1)(2)(00)(000)}$
&(${\bf 1}, {\bf 1}, {\bf 1}, {\bf 1},{\bf 1}$) & $0$ & $-4$ &
$-4$ & $4$ & $-2$
\\
\cline{1-7}
$T_{1(1)(0)(0n')(000)}$ & (${\bf 4},{\bf 2},{\bf 1},{\bf 1},{\bf
1}$) & $0$ & $-1$ & $-1$ & $1$ & $1$ & $T_{2(1)(0)(00)(000)}$
&($\overline{\bf 4}, {\bf 1}, {\bf 2}, {\bf 1}, {\bf 1}$) & $0$ &
$1$ & $1$ & $1$ & $2$
\\
& ($\overline{\bf 4}, {\bf 1}, {\bf 2},{\bf 1},{\bf 1}$) & $0$ &
$-1$ & $-1$ & $1$ & $1$ & $T_{2(1)(1)(00)(000)}$ &(${\bf 1}, {\bf
1}, {\bf 1}, {\bf 1},{\bf 2}$) & $0$ & $0$ & $-2$ & $-4$ & $-3$
\\
$T_{1(1)(1)(0n')(000)}$ & (${\bf 1}, {\bf 1}, {\bf 1}, {\bf
16},{\bf 1}$) & $0$ & $0$ & $-1$ & $-1$ & $0$  & &(${\bf 1}, {\bf
1}, {\bf 1}, {\bf 1},{\bf 2}$) & $0$ & $0$ & $-2$ & $0$ & $3$
\\
$T_{1(1)(2)(0n')(000)}$ &(${\bf 1}, {\bf 1}, {\bf 1}, {\bf 1},{\bf
2}$) & $-2$ & $4$ & $0$ & $0$ & $2$  & $T_{2(1)(2)(00)(000)}$
&(${\bf 1}, {\bf 1}, {\bf 1}, {\bf 1}, {\bf 2}$) & $0$ & $-4$ &
$-2$ & $4$ & $1$
\\
&(${\bf 1}, {\bf 1}, {\bf 1}, {\bf 1},{\bf 2}$) & $2$ & $-2$ & $0$ &
$2$ & $2$ & $T_{2(-1)(0)(00)(0{\overline 1}0)}$ &(${\bf 1}, {\bf 1},
{\bf 1}, {\bf 1},{\bf 1}$) & $2$ & $-2$ & $0$ & $0$ & $2$
\\
$T_{1(1)(0)(1n')(000)}$ &(${\bf 4}, {\bf 1}, {\bf 1},{\bf 1}, {\bf
1}$) & $-1$ & $2$ & $-1$ & $-3$ & $1$ & $T_{2(-1)(1)(00)(0{\overline
1}0)}$ &(${\bf 1}, {\bf 1}, {\bf 1}, {\bf 1},{\bf 1}$) & $0$ & $0$ &
$0$ & $-4$ & $0$
 \\
&(${\bf  4}, {\bf 1}, {\bf 1}, {\bf 1},{\bf 1}$) & $1$ & $-4$ &
$-1$ & $5$ & $1$ & $T_{2(1)(0)(00)(100)}$ & (${\bf 1}, {\bf 1},
{\bf 1}, {\bf 1}, {\bf 1}$) & $2$ & $-2$ & $0$ & $0$ & $2$
 \\
&(${\bf 1}, {\bf 2}, {\bf 1}, {\bf 1},{\bf 1}$) & $-1$ & $5$ & $2$
& $-4$ & $1$  & $T_{2(1)(1)(00)(100)}$ &(${\bf 1}, {\bf 1}, {\bf
1}, {\bf 1},{\bf 1}$) & $0$ & $0$ & $0$ & $-4$ & $0$
\\\cline{8-14}
&(${\bf 1}, {\bf 2}, {\bf 1}, {\bf 1},{\bf 1}$) & $1$ & $-1$ & $2$
& $4$ & $1$  & $T_{3(\omega)(0)(0n')(000)}$ &(${\bf 1}, {\bf 1},
{\bf 1}, {\bf 1}, {\bf 1}$) & $0$ & $0$ & $2$ & $2$ & $-3$
\\
$T_{1(1)(1)(1n')(000)}$ &(${\bf 1}, {\bf 2}, {\bf 1}, {\bf 1},{\bf
2}$) & $1$ & $-3$ & $0$ & $2$ & $0$  &
$T_{3(\omega)(0)(1n')(000)}$ &(${\bf 1}, {\bf 1}, {\bf 2}, {\bf
1}, {\bf 1}$) & $-1$ & $3$ & $0$ & $-2$ & $3$
\\
&(${\bf 1}, {\bf 1}, {\bf 2}, {\bf 1},{\bf 2}$) & $-1$ & $3$ & $2$
& $-2$ & $0$  & $T_{3(\omega^2)(0)(0n')(000)}$ &(${\bf 1}, {\bf
1}, {\bf 1}, {\bf 1},{\bf 1}$) & $-2$ & $0$ & $-2$ & $0$ & $-3$
\\
$T_{1(1)(2)(1n')(000)}$ &(${\bf 1}, {\bf 1}, {\bf 2}, {\bf 1},
{\bf 1}$) & $-1$ & $1$ & $0$ & $-4$ & $-1$ & &(${\bf 1}, {\bf 1},
{\bf 1}, {\bf 1},{\bf 1}$) & $2$ & $-6$ & $-2$ & $2$ & $-3$
 \\
&(${\bf 1}, {\bf 2}, {\bf 1}, {\bf 1}, {\bf 1}$) & $-1$ & $1$ &
$2$ & $0$ & $-1$  & &(${\bf 1}, {\bf 1}, {\bf 1}, {\bf 1},{\bf
1}$) & $-2$ & $6$ & $2$ & $-2$ & $3$
\\
&($\overline{\bf 4}, {\bf 1}, {\bf 1}, {\bf 1}, {\bf 1}$) & $1$ &
$-2$ & $1$ & $-1$ & $-1$  & &(${\bf 1}, {\bf 1}, {\bf 1}, {\bf
1},{\bf 1}$) & $2$ & $0$ & $2$ & $0$ & $3$
\\
$T_{1(1)(0)(0n')(100)}$ &(${\bf 1}, {\bf 1}, {\bf 1}, {\bf 1},{\bf
1}$) & $-2$ & $2$ & $-2$ & $-2$ & $1$ &
$T_{3(\omega^2)(0)(1n')(000)}$ &(${\bf 4}, {\bf 1}, {\bf 1}, {\bf
1},{\bf 1}$) & $-1$ & $0$ & $-1$ & $3$ & $3$
\\
&(${\bf 1}, {\bf 1}, {\bf 1}, {\bf 1},{\bf 1}$) & $0$ & $-4$ & $-2$
& $6$ & $1$ & &($\overline{\bf 4}, {\bf 1}, {\bf 1}, {\bf 1},{\bf
1}$) & $1$ & $0$ & $1$ & $-3$ & $-3$
\\
&(${\bf 1}, {\bf 1}, {\bf 1}, {\bf 1},{\bf 1}$) & $2$ & $-4$ &
$-2$ & $0$ & $1$ & $T_{3(1)(0)(0n')(000)}$ &(${\bf 1}, {\bf 1},
{\bf 1}, {\bf 1},{\bf 1}$) & $0$ & $0$ & $-2$ & $-2$ & $3$
\\
$T_{1(1)(1)(0n')(100)}$ & (${\bf 1}, {\bf 1}, {\bf 1}, {\bf
1},{\bf 1}$) & $0$ & $0$ & $-2$ & $-2$ & $-3$ &
$T_{3(1)(0)(1n')(000)}$ &(${\bf 1}, {\bf 1}, {\bf 2}, {\bf 1},{\bf
1}$) & $1$ & $-3$ & $0$ & $2$ & $-3$
\\\cline{8-14}
&(${\bf 1}, {\bf 1}, {\bf 1}, {\bf 1},{\bf 1}$) & $0$ & $0$ & $-2$
& $2$ & $3$   & $T_{4(-1)(0)(00)(000)}$ &(${\bf 4}, {\bf 1}, {\bf
2}, {\bf 1},{\bf 1}$) & $0$ & $-1$ & $-1$ & $-1$ & $-2$
\\
$T_{1(1)(2)(0n')(100)}$ & (${\bf 1}, {\bf 1}, {\bf 1}, {\bf
1},{\bf 1}$) & $-2$ & $4$ & $2$ & $-4$ & $-1$ &
$T_{4(-1)(1)(00)(000)}$ &(${\bf 1}, {\bf 1}, {\bf 1}, {\bf 1},{\bf
2}$) & $0$ & $0$ & $2$ & $4$ & $3$
\\
&(${\bf 1}, {\bf 1}, {\bf 1}, {\bf 1},{\bf 1}$) & $2$ & $-2$ & $2$
& $-2$ & $-1$  & &(${\bf 1}, {\bf 1}, {\bf 1}, {\bf 1},{\bf 2}$) &
$0$ & $0$ & $2$ & $0$ & $-3$
\\
$T_{1(1)(0)(1n')(100)}$ & (${\bf 1}, {\bf 2}, {\bf 1}, {\bf
1},{\bf 1}$) & $-1$ & $-1$ & $-2$ & $2$ & $1$ &
$T_{4(-1)(2)(00)(000)}$ &(${\bf 1}, {\bf 1}, {\bf 1}, {\bf 1},{\bf
2}$) & $0$ & $4$ & $2$ & $-4$ & $-1$
\\
&($\overline{\bf 4}, {\bf 1}, {\bf 1}, {\bf 1},{\bf 1}$) & $-1$ &
$2$ & $1$ & $1$ & $1$ & $T_{4(1)(0)(00)(000)}$ &(${\bf 6}, {\bf 1},
{\bf 1}, {\bf 1}, {\bf 1}$) & $0$ & $2$ & $2$ & $-2$ & $-2$
\\
$T_{1(1)(1)(0n')(010)}$ &(${\bf 1}, {\bf 1}, {\bf 1}, {\bf 1},
{\bf 2}$) & $0$ & $0$ & $0$ & $-2$ & $0$ & &(${\bf 1}, {\bf 1},
{\bf 1}, {\bf 1},{\bf 1}$) & $-2$ & $2$ & $0$ & $4$ & $4$
\\
$T_{1(1)(0)(1n')(010)}$ &(${\bf 1}, {\bf 1}, {\bf 2}, {\bf 1},
{\bf 1}$) & $1$ & $-1$ & $0$ & $0$ & $1$ & &(${\bf 1}, {\bf 1},
{\bf 1}, {\bf 1},{\bf 1}$) & $0$ & $-4$ & $-4$ & $0$ & $-2$
\\
$T_{1(1)(0)(0n')(001)}$ &(${\bf 1}, {\bf 1}, {\bf 1}, {\bf 1},
{\bf 1}$) & $0$ & $2$ & $2$ & $0$ & $1$ & $T_{4(1)(1)(00)(000)}$
&(${\bf 1}, {\bf 1}, {\bf 1}, {\bf 10},{\bf 1}$)& $0$ & $0$ & $2$
& $2$ & $0$
\\
$T_{1(1)(0)(0n')(00{\overline 1})}$ &(${\bf 1}, {\bf 1}, {\bf 1},
{\bf 1}, {\bf 1}$) & $0$ & $2$ & $2$ & $0$ & $1$ & &(${\bf 1}, {\bf
1}, {\bf 1}, {\bf 1},{\bf 1}$)& $-2$ & $6$ & $0$ & $-4$ & $0$
\\
$T_{1(1)(1)(0n')(200)}$ &(${\bf 1}, {\bf 1}, {\bf 1}, {\bf 1},{\bf
2}$) & $0$ & $0$ & $0$ & $-2$ & $0$ & &(${\bf 1}, {\bf 2}, {\bf
2}, {\bf 1},{\bf 1}$)& $0$ & $0$ & $-2$ & $0$ & $0$
\\
$T_{1(1)(0)(1n')(200)}$ &(${\bf 1}, {\bf 1}, {\bf 2}, {\bf 1},{\bf
1}$) & $1$ & $-1$ & $0$ & $0$ & $1$ & &(${\bf 1}, {\bf 1}, {\bf
1}, {\bf 1},{\bf 1}$)& $2$ & $-6$ & $-4$ & $4$ & $0$
\\
$T_{1(1)(0)(0n')(110)}$ &(${\bf 1}, {\bf 1}, {\bf 1}, {\bf 1},{\bf
1}$) & $0$ & $2$ & $2$ & $0$ & $1$ & $T_{4(1)(2)(00)(000)}$
&(${\bf 1}, {\bf 1}, {\bf 1}, {\bf 1},{\bf 1}$)& $0$ & $4$ & $4$ &
$-4$ & $2$
\\
$T_{1(1)(0)(0n')(300)}$ &(${\bf 1}, {\bf 1}, {\bf 1}, {\bf 1},{\bf
1}$) & $0$ & $2$ & $2$ & $0$ & $1$ &
$T_{4(-1)(0)(00)(\overline{1}00)}$ &(${\bf 1}, {\bf 1}, {\bf 1},
{\bf 1},{\bf 1}$)& $-2$ & $2$ & $0$ & $0$ & $-2$
\\
\cline{1-7}
$T_{2(-1)(0)(00)(000)}$ &(${\bf 6}, {\bf 1}, {\bf 1}, {\bf 1},{\bf
1}$) & $0$ & $-2$ & $-2$ & $2$ & $2$ &
$T_{4(-1)(1)(00)(\overline{1}00)}$ &(${\bf 1}, {\bf 1}, {\bf 1},
{\bf 1},{\bf 1}$) & $0$ & $0$ & $0$ & $4$ & $0$
\\
&(${\bf 1}, {\bf 1}, {\bf 1}, {\bf 1},{\bf 1}$) & $2$ & $-2$ & $0$
& $-4$ & $-4$ & $T_{4(1)(0)(00)(010)}$ &(${\bf 1}, {\bf 1}, {\bf
1}, {\bf 1},{\bf 1}$) & $-2$ & $2$ & $0$ & $0$ & $-2$
\\
&(${\bf 1}, {\bf 1}, {\bf 1}, {\bf 1},{\bf 1}$) & $0$ & $4$ & $4$
& $0$ & $2$  & $T_{4(1)(1)(00)(010)}$ &(${\bf 1}, {\bf 1}, {\bf
1}, {\bf 1},{\bf 1}$) & $0$ & $0$ & $0$ & $4$ & $0$
\\
$T_{2(-1)(1)(00)(000)}$ &(${\bf 1}, {\bf 1}, {\bf 1}, {\bf
10},{\bf 1}$) &
$0$ & $0$ & $-2$ & $-2$ & $0$ & & & & & & & \\
\end{tabular}
\end{ruledtabular}
\end{table*}
\endgroup

\begingroup
\squeezetable
\begin{table*}[th]\caption{Matter spectrum of model B. The levels of the
$\lgp{U}{1}$ groups are $24$, $88$, $40$, $216$, $88$
and $576$.\label{tab3}}
\begin{ruledtabular}
\begin{tabular}{|l|l|rrrrrr|l|l|rrrrrr|}
Sectors& ${\rm PS\times\lgp{SO}{10}'} $ &  $Q_1$ & $Q_2$ & $Q_3$ & $Q_4$
& $Q_5$ & $Q_A$ & Sectors& ${\rm PS\times\lgp{SO}{10}'} $ & $Q_1$ & $Q_2$
& $Q_3$
& $Q_4$ & $Q_5$ & $Q_A$ \\
\hline\hline $U_1$ & (${\bf 4}, {\bf 2},{\bf 1},  {\bf 1}$) & $-1$ &
$0$ & $1$ & $0$ & $0$  & $12$ & $T_{2(1)(0)(00)(000)}$
&(${\bf\overline 4}, {\bf 1}, {\bf 2}, {\bf 1}$) &
$-1$ & $0$ & $-1$ & $-4$ & $0$ & $-1$  \\
\cline{1-8} $U_2$ & (${\bf 1}, {\bf 1},{\bf 1},  {\bf 1}$) & $4$ &
$0$ & $0$ & $0$ & $0$  & $-12$ & &(${\bf 1}, {\bf 1}, {\bf 1},
{\bf 1}$) & $0$ & $-4$ & $4$ & $8$ & $0$ & $5$
\\
&(${\bf 6}, {\bf 1},{\bf 1},  {\bf 1}$) & $-2$ & $0$ & $-2$ & $0$
& $0$ & $-12$ & &(${\bf 1}, {\bf 1}, {\bf 1}, {\bf 1}$) & $0$ &
$4$ & $0$ & $8$ & $0$ & $11$
\\
&(${\bf 1}, {\bf 1}, {\bf 1}, {\bf 10}$) & $0$ & $-2$ & $0$ & $6$
& $2$ & $6$  & $T_{2(1)(1)(00)(000)}$ &(${\bf 1}, {\bf 1}, {\bf
1}, {\bf 10}$) & $0$ & $2$ & $0$ & $-2$ & $2$ & $5$
\\
\cline{1-8} $U_3$ & (${\bf 4}, {\bf 2}, {\bf 1}, {\bf 1}$) & $3$ &
$0$ & $1$ & $0$ & $0$ & $0$  & &(${\bf 1}, {\bf 1}, {\bf 1},
\overline{\bf 16}$) & $0$ & $1$ & $0$ & $1$ & $-1$ & $8$
\\
& ($\overline{\bf 4}, {\bf 2}, {\bf 1}, {\bf 1}$) & $-3$ & $0$ &
$-1$ & $0$ & $0$ & $0$  & &(${\bf 1}, {\bf 1}, {\bf 1}, {\bf 1}$) &
$0$ & $0$ & $0$ & $4$ & $-4$ & $11$
\\
\cline{1-8}
$T_{1(1)(0)(0n')(000)}$ & (${\bf 6}, {\bf 1}, {\bf 1}, {\bf 1}$) &
$0$ & $0$ & $0$ & $4$ & $0$ & $-5$ & $T_{2(1)(2)(00)(000)}$
&(${\bf 1}, {\bf 1}, {\bf 1}, {\bf 10}$) & $0$ & $-2$ & $0$ & $-6$
& $-2$ & $5$
\\
& (${\bf 1}, {\bf 2}, {\bf 2}, {\bf 1}$) & $0$ & $0$ & $0$& $4$ &
$0$  & $-5$ & &(${\bf 1}, {\bf 1}, {\bf 1}, {\bf 1}$) & $0$ & $0$
& $4$ & $12$ & $4$  & $5$
\\
& (${\bf 1}, {\bf 1}, {\bf 1}, {\bf 1}$) & $2$ & $0$ & $0$ & $-8$
& $0$ & $4$ & $T_{2(-1)(1)(00)(100)}$ &(${\bf 1}, {\bf 1}, {\bf
1}, {\bf 1}$) & $0$ & $0$ & $0$ & $4$ & $4$ & $11$
\\
$T_{1(1)(1)(0n')(000)}$ & (${\bf 1}, {\bf 1}, {\bf 1}, {\bf 1}$) &
$2$ & $0$ & $-2$ & $-4$ & $0$ & $13$  & $T_{2(-1)(2)(00)(100)}$
&(${\bf 1}, {\bf 1}, {\bf 1}, {\bf 1}$) & $0$ & $-4$ & $0$ & $0$ &
$0$ & $11$
\\
& (${\bf 1}, {\bf 1}, {\bf 1}, {\bf 1}$) & $2$ & $4$ & $0$ & $-4$ &
$-4$ & $4$ & $T_{2(1)(1)(00)(0{\overline 1}0)}$ & (${\bf 1}, {\bf
1}, {\bf 1}, {\bf 1}$) & $0$ & $0$ & $0$ & $4$ & $4$ & $11$
\\
& (${\bf 1}, {\bf 1}, {\bf 1}, {\bf 1}$) & $2$ & $4$ & $2$ & $8$ &
$4$ & $7$ & $T_{2(1)(2)(00)(0{\overline 1}0)}$ &(${\bf 1}, {\bf 1},
{\bf 1}, {\bf 1}$) & $0$ & $-4$ & $0$ & $0$ & $0$ & $11$
\\
\cline{9-16}
$T_{1(1)(2)(0n')(000)}$ &(${\bf 1}, {\bf 1}, {\bf 1}, {\bf 1}$) &
$2$ & $-4$ & $2$ & $0$ & $4$ & $7$ & $T_{3(\omega)(0)(0n')(000)}$
&(${\bf\overline 4}, {\bf 2}, {\bf 1}, {\bf 1}$) & $-1$ & $0$ & $-1$
& $0$ & $0$ & $-6$
\\
&(${\bf 1}, {\bf 1}, {\bf 1}, {\bf 1}$) & $2$ & $0$ & $4$ & $0$ &
$0$ & $-2$  & &(${\bf 4}, {\bf 1}, {\bf 2}, {\bf 1}$) & $-1$ & $0$
& $-1$ & $0$ & $0$ & $-6$
\\
&(${\bf 1}, {\bf 1}, {\bf 1}, {\bf 1}$) & $2$ & $-4$ & $2$ & $0$ &
$-4$ & $7$ & $T_{3(\omega)(0)(1n')(000)}$ &(${\bf 1}, {\bf 2},
{\bf 1}, {\bf 1}$) & $0$ & $-2$ & $4$ & $6$ & $2$ & $-6$
\\
$T_{1(1)(0)(1n')(000)}$ &(${\bf 1}, {\bf 2}, {\bf 1}, {\bf 1}$) &
$0$ & $2$ & $0$ & $-2$ & $2$ & $-11$ & $T_{3(1)(0)(0n')(000)}$
&(${\bf 4}, {\bf 2}, {\bf 1}, {\bf 1}$) & $1$ & $0$ & $1$ & $0$ &
$0$ & $6$\\
&(${\bf 4}, {\bf 1}, {\bf 1}, {\bf 1}$) & $-1$ & $-2$ & $-1$ & $-2$
& $-2$ & $10$ & &(${\bf\overline 4}, {\bf 1}, {\bf 2}, {\bf 1}$) &
$1$ & $0$ & $1$ & $0$ & $0$ & $6$
\\
&(${\bf 1}, {\bf 1}, {\bf 2}, {\bf 1}$) & $2$ & $-2$ & $0$ & $-2$
& $-2$  & $10$ & $T_{3(1)(0)(0n')(000)}$ &(${\bf 1}, {\bf 2}, {\bf
1}, {\bf 1}$) & $0$ & $2$ & $-4$ & $-6$ & $-2$ & $6$
\\
$T_{1(1)(1)(1n')(000)}$ &(${\bf 4}, {\bf 1}, {\bf 1}, {\bf 1}$) &
$-1$ & $2$ & $-1$ & $2$ & $2$ & $10$ &
$T_{3(\omega)(0)(0n')(001)}$ &(${\bf 1}, {\bf 1}, {\bf 1}, {\bf
1}$) & $-2$ & $0$ & $0$ & $0$ & $0$ & $6$
\\
&(${\bf 1}, {\bf 2}, {\bf 1}, {\bf 1}$) & $0$ & $-2$ & $2$ & $2$ &
$2$ & $13$ & $T_{3(\omega)(0)(0n')({\overline 1}00)}$ &(${\bf 1},
{\bf 1}, {\bf 1}, {\bf 1}$) & $2$ & $0$ & $0$ & $0$ & $0$ & $-6$
\\
&(${\bf 1}, {\bf 1}, {\bf 2}, {\bf 1}$) & $2$ & $2$ & $0$ & $2$ &
$2$ & $10$ & $T_{3(\omega)(0)(0n')(00{\overline 1})}$ &(${\bf 1},
{\bf 1}, {\bf 1}, {\bf 1}$) &$2$ & $0$ & $0$ & $0$ & $0$ & $-6$
\\
$T_{1(1)(2)(1n')(000)}$ &(${\bf 1}, {\bf 1}, {\bf 2}, {\bf 1}$) &
$-2$ & $2$ & $-2$ & $-6$ & $-2$ & $-2$ &
$T_{3(\omega^2)(0)(0n')(100)}$ &(${\bf 1}, {\bf 1}, {\bf 1}, {\bf
1}$) & $2$ & $0$ & $0$ & $0$ & $0$ & $-6$
\\
&(${\bf 1}, {\bf 2}, {\bf 1}, {\bf 1}$) & $0$ & $-2$ & $0$ & $-6$ &
$-2$ & $-11$ & $T_{3(\omega^2)(0)(0n')({\overline 1}00)}$ &(${\bf
1}, {\bf 1}, {\bf 1}, {\bf 1}$) & $-2$ & $0$ & $0$ & $0$ & $0$ & $6$
\\
&(${\bf \overline 4}, {\bf 1}, {\bf 1}, {\bf 1}$) & $1$ & $2$ & $-1$
& $-6$ & $-2$ & $-2$ & $T_{3(1)(0)(0n')(100)}$ &(${\bf 1}, {\bf 1},
{\bf 1}, {\bf 1}$) & $2$ & $0$ & $0$ & $0$ & $0$ & $-6$
\\
$T_{1(1)(0)(0n')(100)}$ &(${\bf 1}, {\bf 1}, {\bf 1}, {\bf 1}$) &
$-2$ & $-4$ & $2$ & $4$ & $0$ & $-2$ & $T_{3(1)(0)(0n')(001)}$
&(${\bf 1}, {\bf 1}, {\bf 1}, {\bf 1}$) & $-2$ & $0$ & $0$ & $0$ &
$0$ & $6$
\\
&(${\bf 1}, {\bf 1}, {\bf 1}, {\bf 1}$) & $-2$ & $4$ & $-2$ & $4$ &
$0$ & $4$ & $T_{3(1)(0)(0n')(00{\overline 1})}$ &(${\bf 1}, {\bf 1},
{\bf 1}, {\bf 1}$) & $-2$ & $0$ & $0$ & $0$ & $0$ & $6$
\\\cline{9-16}
$T_{1(1)(1)(0n')(100)}$ &(${\bf 1}, {\bf 1}, {\bf 1}, {\bf 1}$) &
$-2$ & $-4$ & $-2$ & $-4$ & $0$ & $4$ & $T_{4(-1)(0)(00)(000)}$
&(${\bf 4}, {\bf 1}, {\bf 2}, {\bf 1}$) & $1$ & $0$ & $1$ & $4$ &
$0$ & $1$
\\
&(${\bf 1}, {\bf 1}, {\bf 1}, {\bf 1}$) & $-2$ & $0$ & $0$ & $-4$
& $-4$  & $-5$ & &(${\bf 1}, {\bf 1}, {\bf 1}, {\bf 1}$) & $0$ &
$4$ & $-4$ & $-8$ & $0$ & $-5$
\\
&(${\bf 1}, {\bf 1}, {\bf 1}, {\bf 1}$) & $-2$ & $0$ & $2$ & $8$ &
$4$ & $-2$ & &(${\bf 1}, {\bf 1}, {\bf 1}, {\bf 1}$) & $0$ & $-4$
& $0$ & $-8$ & $0$ & $-11$
\\
$T_{1(1)(2)(0n')(100)}$ &(${\bf 1}, {\bf 1}, {\bf 1}, {\bf 1}$)
&$-2$ & $0$ & $-2$ & $0$ & $4$ & $4$ & $T_{4(-1)(1)(00)(000)}$
&(${\bf 1}, {\bf 1}, {\bf 1}, {\bf 10}$) & $0$ & $-2$ & $0$ & $2$
& $-2$ & $-5$
\\
&(${\bf 1}, {\bf 1}, {\bf 1}, {\bf 1}$) & $-2$ & $4$ & $0$ & $0$ &
$0$ & $-5$ & &(${\bf 1}, {\bf 1}, {\bf 1}, {\bf 16}$) & $0$ & $-1$
& $0$ & $-1$ & $1$ & $-8$
\\
&(${\bf 1}, {\bf 1}, {\bf 1}, {\bf 1}$) & $-2$ & $0$ & $-2$ & $0$
& $-4$ & $4$  & &(${\bf 1}, {\bf 1}, {\bf 1}, {\bf 1}$) & $0$ &
$0$ & $0$ & $-4$ & $4$ & $-11$
\\
$T_{1(1)(0)(1n')(100)}$ &(${\bf 1}, {\bf 2}, {\bf 1}, {\bf 1}$) &
$0$ & $-2$ & $-2$ & $-2$ & $-2$ & $-2$ & $T_{4(-1)(2)(00)(000)}$
&(${\bf 1}, {\bf 1}, {\bf 1}, {\bf 10}$) & $0$ & $2$ & $0$ & $6$ &
$2$ & $-5$
\\
$T_{1(1)(1)(1n')(100)}$ &(${\bf 1}, {\bf 1}, {\bf 2}, {\bf 1}$)
&$-2$ & $-2$ & $0$ & $2$ & $2$ & $1$ & &(${\bf 1}, {\bf 1}, {\bf
1}, {\bf 1}$) & $0$ & $0$ & $-4$ & $-12$ & $-4$ & $-5$
\\
&(${\bf 1}, {\bf 2}, {\bf 1}, {\bf 1}$) & $0$ & $2$ & $-2$ & $2$ &
$2$ & $-2$ & $T_{4(1)(0)(00)(000)}$ &(${\bf 1}, {\bf 1}, {\bf 1},
{\bf 1}$) & $0$ & $0$ & $0$ & $-8$ & $0$ & $10$
\\
&(${\bf\overline 4}, {\bf 1}, {\bf 1}, {\bf 1}$) & $1$ & $-2$ & $1$
& $2$ & $2$ & $1$ & &(${\bf 1}, {\bf 2}, {\bf 2}, {\bf 1}$) & $-2$ &
$0$ & $0$ & $4$ & $0$ & $1$
\\
$T_{1(1)(0)(0n')(200)}$ &(${\bf 1}, {\bf 1}, {\bf 1}, {\bf 1}$) &
$2$ & $0$ & $2$ & $4$ & $0$ & $7$ & $T_{4(1)(1)(00)(000)}$ &(${\bf
1}, {\bf 1}, {\bf 1}, {\bf 1}$) & $0$ & $4$ & $0$ & $-4$ & $-4$ &
$10$
\\
$T_{1(1)(0)(0n')(010)}$&(${\bf 1}, {\bf 1}, {\bf 1}, {\bf 1}$) &
$2$ & $0$ & $2$ & $4$ & $0$ & $7$ & $T_{4(1)(2)(00)(000)}$ &(${\bf
1}, {\bf 1}, {\bf 1}, {\bf 1}$) & $0$ & $0$ & $4$ & $0$ & $0$ &
$4$
 \\\cline{1-8}
$T_{2(-1)(0)(00)(000)}$ &(${\bf 1}, {\bf 1}, {\bf 1}, {\bf 1}$) &
$0$ & $0$ & $0$ & $8$ & $0$ & $-10$ & $T_{4(-1)(1)(00)(010)}$
&(${\bf 1}, {\bf 1}, {\bf 1}, {\bf 1}$) & $0$ & $0$ & $0$ & $-4$ &
$-4$ & $-11$
\\
&(${\bf 1}, {\bf 2}, {\bf 2}, {\bf 1}$) & $2$ & $0$ & $0$ & $-4$ &
$0$ & $-1$ & $T_{4(-1)(2)(00)(010)}$ &(${\bf 1}, {\bf 1}, {\bf 1},
{\bf 1}$) &
$0$ & $4$ & $0$ & $0$ & $0$ & $-11$  \\
$T_{2(-1)(1)(00)(000)}$ &(${\bf 1}, {\bf 1}, {\bf 1}, {\bf 1}$) &
$0$ & $-4$ & $0$ & $4$ & $4$ & $-10$ &
$T_{4(1)(1)(00)(\overline{1}00)}$ &(${\bf 1}, {\bf 1}, {\bf 1}, {\bf
1}$)
& $0$ & $0$ & $0$ & $-4$ & $-4$ & $-11$ \\
$T_{2(-1)(2)(00)(000)}$ &(${\bf 1}, {\bf 1}, {\bf 1}, {\bf 1}$) &
$0$ & $0$ & $-4$ & $0$ & $0$ & $-4$  &
$T_{4(1)(2)(00)(\overline{1}00)}$ &(${\bf 1}, {\bf 1}, {\bf 1}, {\bf
1}$) & $0$ & $4$ & $0$ & $0$ & $0$ & $-11$
\\
\end{tabular}
\end{ruledtabular}
\end{table*}
\endgroup

We list the complete matter content of the three-family PS models, including all the singlet fields, in tables
\ref{tab1}, \ref{tab2} and \ref{tab3}.

Let us explain our notation for the twisted-sector states somewhat. For the sake of presentation, we take the six
compactified dimensions to be a factorizable Lie algebra root lattice, $\lat$ (see fig.~\ref{fig:lattice}), as in
sect.~\ref{ogut}.\footnote{In fact, the matter spectra also depend on our choice of the compactifed lattice. The
Lie algebra root lattice sits at special points of the six-torus modulus space with enhanced gauge symmetries.
The spectra (if we restrict ourselves to the root lattice) should be regarded as truncated spectra of the
complete model, where there are additional states neutral under the observable gauge groups. The use of
non-simply-laced algebra $\lgp{G}{2}$ may also be worrisome. But it may be constructed from a simply-laced
algebra such as the $\lgp{SO}{8}$ by appropriate outer-automorphism, at the expense of reducing the rank of the hidden-sector groups. The $\lgp{G}{2}$ lattice
can also be replaced by the $\lgp{SU}{3}^{[2]}$ lattice without changing any
of our discussions on low energy phenomenology. The $\lgp{SU}{3}^{[2]}$
Coxeter element is generated by the Weyl reflections with respect
to ${\bf e}_1$, $s_1$, and with respect to ${\bf e}_2-{\bf e}_1$,
$s_{2\leftrightarrow 1}$,
$C_{\rm SU_3^{[2]}}
=s_1s_{2\leftrightarrow 1}=\left(\begin{array}{cc}
0&1\\
-1& 1\end{array}\right)$, where ${\bf e}_1$ and ${\bf e}_2$ are the
simple roots of the $\lgp{SU}{3}$.}
The notation, although pertaining to the particular case,
can be generalized to any other lattice.

The twisted sectors contain various matter states, arising from
different fixed points. It is useful to introduce the following
notation for the twisted-sector states (see appendix \ref{review},
also ref.~\cite{kobayashi2}),
\begin{equation}
  T_{k(\gamma)^{}_\alpha(n_3)(n_2n_2')N},
\end{equation}
where the first sub-index represents the $k^{\rm th}$ twisted
sector, the next three quantum numbers, $(\gamma)_\alpha$, $(n_3)$
and $(n_2,n_2')$, specify fixed points on the first, second and
third complex planes, respectively. The last sub-index
$N=(N_1N_2N_3)$ denotes an array of three integral oscillator
numbers for the left movers. For example, $(100)$ and $({\overline
1}00)$ represent creating one oscillator in the $Z_1$ and
${\overline Z}_1$ directions. We only need to keep the left-moving
oscillator numbers since all the massless states in our models have
zero numbers of right-moving oscillators.

The quantum numbers $(\gamma)_\alpha$, $(n_3)$ and $(n_2,n_2')$ label conjugacy classes of the fixed points. The
second complex plane, i.e., the $\lgp{SU}{3}$ lattice, is a ${\mathbb Z}_3$ orbifold, and has three fixed points,
${\bf f}={\bf 0}$, $\frac{1}{3}(2{\bf e}_3+{\bf e}_4)$ and $\frac{1}{3}({\bf e}_3+2{\bf e}_4)$, where ${\bf
e}_3$, ${\bf e}_4$ are the basis of the $\lgp{SU}{3}$ lattice. The conjugacy classes are labelled by equivalent
shift vectors $a{\bf e}_3+b{\bf e}_4$, where
$a+b\equiv n_3=0,1,2\,{\rm mod}\,3$ is the ``winding number." We can
introduce at most one degree-3 Wilson line, and specify these fixed points by $n_3=0,1,2$ for $k \neq 3$, and
$n_3=0$ for $k=3$ (since $Z_2$ is the fixed plane of the $\theta^3$ twist). Similarly, the third complex plane,
i.e., the $\lgp{SO}{4}$ lattice, is a ${\mathbb Z}_2$ orbifold, and has four fixed points, ${\bf
f}=\frac{1}{2}(n_2{\bf e}_5+n_2'{\bf e}_6)$, where $n_2, n_2'=0,1$ and ${\bf e}_5$, ${\bf e}_6$ are the basis
vectors of the $\lgp{SO}{4}$ lattice. The equivalent shift vectors for these fixed points are $-n_2{\bf
e}_5-n'_2{\bf e}_6$. We can introduce at most two independent degree-2 Wilson lines, and specify these fixed
points by a pair of winding numbers, $n_2,n_2'=0,1$ for $k=1,3,5$, and $n_2=n_2'=0$ for $k=2,4$ (since $Z_3$ is
the fixed plane of the $\theta^{2,4}$ twists). In the present models, we introduce only one non-vanishing
${\mathbb Z}_2$ Wilson line, so the degree-2 degeneracy associated with $n'_2$ is not resolved.

For the first complex plane (i.e., the $\lgp{G}{2}$ lattice), there
is only one fixed point in the $T_1$ twisted sector, ${\bf f}={\bf
0}$. The corresponding state will be denoted by $(\gamma)=(1)$. In
the $T_2$ and $T_4$ twisted sectors, there are three classes of
fixed points, ${\bf f}_a=\frac{1}{3}a{\bf e}_1$ with $a=0,1,2$. The
last two, ${\bf f}_{1,2}$, transform into each other
under the orbifold twist $\theta$, while the first, ${\bf f}_0$,
is invariant. For states associated with the non-invariant fixed
points, we can take linear combinations such that they have definite
$\theta$-eigenvalues, $\gamma = \pm 1$. We label the two
eigen-states with $\gamma = 1$ by $|{\bf f}_0 \rangle \equiv (1)_1$,
$\frac{1}{\sqrt{2}}(|{\bf f}_1\rangle + | {\bf f}_2 \rangle) \equiv
(1)_2$ and the one with $\gamma=-1$ by $\frac{1}{\sqrt{2}}( |{\bf
f}_1 \rangle - | {\bf f}_2 \rangle) \equiv (-1)$. Similarly, there
are four classes of fixed points in the $T_3$ twisted sector, ${\bf
f}_{ab}=\frac{1}{2}(a{\bf e}_1+b{\bf e}_2)$, with $a,b=0,1$. Three
of them, ${\bf f}_{01}$, ${\bf f}_{10}$ and ${\bf f}_{11}$,
transform into each other under the orbifold twist $\theta$, while
the other, ${\bf f}_{00}$, is invariant. After taking linear
combinations, we have two states with $\gamma =1$, $| {\bf f}_{00}
\rangle $ and $\frac{1}{\sqrt{3}}(| {\bf f}_{10} \rangle + | {\bf
f}_{01} \rangle + |{\bf f}_{11} \rangle ) $, denoted by $(1)_{1,2}$,
and two states $\frac{1}{\sqrt{3}}(| {\bf f}_{10} \rangle +\gamma |
{\bf f}_{01} \rangle + \gamma^2 |{\bf f}_{11} \rangle )$ with
eigenvalues $\gamma = \omega,\,\omega^2$, where $\omega = {\rm e}^{{\rm i}2 \pi
/3}$, denoted by $(\omega)$ and $(\omega^2)$.

Multiplicities of the twisted-sector states are computed from the generalized GSO projector of eq.~\ref{GSO}.
They equal the number of degenerate fixed points with quantum numbers $k$, $\gamma$, $\alpha$, $n_3$, $n_2$,
$n_2'$ and $N$. With two Wilson lines, one degree-2 and one degree-3, the $T_1$-sector states have multiplicities
two, due to the unresolved degree-2 degeneracies in $n'_2$. The $T_2$ and $T_4$-sector states have multiplicities
two and one, for states with eigenvalues $\gamma=1$ and $\gamma=-1$, since they have two and one associated fixed
points, $(1)_{1,2}$ and $(-1)$. Finally, the $T_3$-sector states have multiplicities four and two, for states
with eigenvalues $\gamma=1$ and $\gamma = \omega, \omega^2$, since the associated fixed points are $(1)_{1,2}$
and $(\omega), (\omega^2)$ and there are additional degeneracies in $n'_2=0,1$.

\section{Yukawa couplings in the ${\mathbb Z}_6$ models \label{app:yukawa}}

\subsection{Selection rules}\label{sec:sr}

In this subsection, we give the complete set of string selection
rules in the ${\mathbb Z}_6$ orbifold model. Although these
rules are well known, they are not always stated correctly
in the previous literature.

In heterotic string models, physical states corresponding to
space-time bosons can be written in terms of vertex operators in the
$(-1)$-ghost picture, they are
\begin{equation}
V_{-1}^{(\ell)} = {\rm e}^{-\phi} \prod_{i=1}^3 (\partial
Z_i)^{N_i^{(\ell)}} (\partial \overline Z_i)^{\overline
N_i^{(\ell)}} {\rm e}^{-2{\rm i} {\bf r}^{(\ell)}\cdot {\bf H}} {\rm
e}^{2{\rm i} {\bf P}^{(\ell)}\cdot{\bf X}} \sigma_{f}^{(\ell)},
\end{equation}
where $\phi$ and $(\partial Z_i)^{N_i^{(\ell)}}$, $(\partial
{\overline Z}_i)^{\overline{N}_i^{(\ell)}}$ denote the
superconformal ghost and $N_i^{(\ell)}$ and
$\overline{N}_i^{(\ell)}$ left-moving oscillators in the $Z_i$ and
$\overline{Z}_i$ directions. ${\bf P}^{(\ell)}$ and ${\bf
r}^{(\ell)}$ are the $\lgp{E}{8}\times\lgp{E}{8}$ and $\lgp{SO}{8}$
shifted momenta (${\bf X}$ and ${\bf H}$ are bosonic coordinates
parameterizing the respective maximal tori); the latter is commonly
known as the H-momentum. More explicitly, in the ${\mathbb Z}_6$
model the internal $\lgp{SO}{6}$ parts of the H-momenta for states
in different sectors are:
\begin{eqnarray}
&&U_1:~(1,0,0),    \qquad\quad U_2:~(0,1,0), \qquad\quad U_3:~(0,0,1),\nn\\
&&T_1:~(1,2,3)/6,  \qquad  T_2:~(2,4,0)/6, \qquad
T_3:~(3,0,3)/6,  \qquad  T_4:~(4,2,0)/6.
\end{eqnarray}
Finally, $\sigma_{f}^{(\ell)}$ denotes the twisted field, creating
twisted-sector vacua out of the untwisted ones; it is thus trivial
for the untwisted sector states.

Equivalently, due to the world-sheet superconformal symmetry for the
right-moving superstring, bosonic state vertex operators can be
written in other ghost pictures by acting with a picture changing
operator \cite{CFT}, $\supset {\rm e}^{\phi}
\sum_{i=1}^3(\psi_i\overline
\partial\overline{Z}_i+\overline{\psi}_i\overline \partial Z_i) = {\rm e}^{\phi}
\sum_{i=1}^3 ({\rm e}^{2{\rm i}{\bf r}^i_v\cdot{\bf H}}\overline
\partial\overline{Z}_i +{\rm e}^{-2{\rm i}{\bf r}^i_v\cdot{\bf H}}\overline
\partial Z_i)$, where ${\bf r}^1_v = (1,0,0)$, ${\bf r}_v^2=(0,1,0)$,
${\bf r}_v^3 =(0,0,1)$. Therefore in the $0$-ghost picture the H
momenta are reduced by ${\bf r}_v$'s and additional factors of
${\overline\partial}Z_i$ and ${\overline\partial}\overline{Z}_i$ are
introduced. However the unique combination (defined up to a modding
with respect to the degree of orbifolding) \be
R_i^{(\ell)}=r_i^{(\ell)}-N_i^{(\ell)}+\overline{N}_i^{(\ell)}\label{Rcharge}
\ee remains invariant under picture-changing operations.

For Yukawa couplings in string theory, one can use the standard
conformal field theory technique \cite{CFT,CFT2} to compute the
corresponding $n$-point correlation function,
\be
\langle V_{-1}^{(1)}
V_{-1/2}^{(2)} V_{-1/2}^{(3)} V_0^{(4)} V_0^{(5)} \cdots
V_0^{(n)}\rangle,
\ee
where $V_{-1/2}$'s are the vertex operators for
space-time fermions in the $(-1/2)$-ghost picture and all but one of
the bosonic vertex operators have been brought into the $0$-ghost
picture. Note that the total ghost charge must cancel with a
background ghost charge of $2$.

For our purposes, however, we do not need the exact form of the above correlation function;\footnote{The
functional dependence of Yukawa couplings on moduli can be determined by a field theoretical method \cite{kp}.
The low energy effective action of 4d strings enjoys a target space duality symmetry PSL$(2,{\mathbb Z})$
\cite{GPR} (or its congruence subgroups in cases with discrete Wilson lines \cite{td}), under which the matter
fields transform according to their modular weights.  For the $U_k$ untwisted-sector states, these weights are
$t_i=-1$ (if $i=k$) and $0$ (if $i\neq k$) for the three K\"ahler moduli. For twisted-sector states,
$t_i=-1+r_i-N_i+\overline{N}_i$, where $r_i, N_i, \overline{N}_i$ are the H-momenta and oscillator numbers. The
superpotential has a weight $(-1,-1,-1)$ (since the combination of K\"ahler potential and superpotential ${\cal
K}+\log|{\cal W}|^2$ has weight $(0,0,0)$), so the modulus-dependent Yukawa couplings must have certain definite
weights under the duality group and they are the modular forms. Physically, the modulus dependence accounts for
the world-sheet instanton effects \cite{DG} and may be important for suppressing some Yukawa couplings.} string
selection rules suffice to tell us whether certain types of Yukawa couplings are non-trivial, i.e., not
identically zero. These rules are provided by different parts of the vertex operator.

The gauge part requires conservation of the
$\lgp{E}{8}\times\lgp{E}{8}$ momenta, $\sum_\ell {\bf P}^{(\ell)} =
0$, which is nothing but the gauge invariance condition for matter
couplings. A similar rule also holds for the H-momenta, but we have
to be more careful because these momenta are not invariant under the
picture-changing operations. Instead of the H-momenta, the rules
must be imposed on the well-defined R-charges of eq. \ref{Rcharge},
they are \be \sum_\ell R_1^{(\ell)}=1\,{\rm mod}\,6,\qquad \sum_\ell
R_2^{(\ell)}=1\,{\rm mod}\,3,\qquad \sum_\ell R_3^{(\ell)}=1\,{\rm
mod}\,2.\label{Hmom} \ee We note that they can be understood as
conservations of discrete global R-charges in the field theory
language \cite{Rcharge}. The nature of R-symmetries can be seen from
the spacetime supersymmetry charge $Q_{\rm SUSY} =\int \frac{{\rm
d}z}{2\pi{\rm i}}{\rm e}^{-\phi/2}S{\rm e}^{2{\rm i}{\bf r}^{}_{\rm
SUSY}\cdot{\bf H}}$ where $S$ is the spin field for uncompactified
dimensions, ${\bf r}^{}_{\rm SUSY}=\frac{1}{2}(1,1,1)$, and the
integration is over the world-sheet coordinates. Under the orbifold
twist $Q_{\rm SUSY}\rightarrow {\rm e}^{{\rm i}\pi v_i}Q_{\rm SUSY}$
and the bosonic vertex operator $V^{(\ell)}\rightarrow {\rm
e}^{-2{\rm i}\pi R_i^{(\ell)}v_i}V^{(\ell)}$. Eq.~\ref{Hmom} is
equivalent to the statement that the superpotential is invariant
under this twist.

The part associated with the twisted fields leads to the so-called space group selection rule. Global monodromy
requires that the product of space group elements associated with the fixed points contain the unit $(1,{\bf
0})$, that is,
\bea
& & \sum_\ell k_\ell=0\,{\rm mod}\,6\,, \label{sg0}\\
& & \sum_\ell (1-\theta^{k_\ell}_{(\ell)}) {\bf f}_\ell = \sum_\ell
(1-\theta^{k_\ell}_{(\ell)}){\bm\Lambda}_{(\ell)}\,,\label{pg}
\eea
where the indices $\ell$ sum over the twisted-sector states (in the $k_\ell$-th sector), ${\bf f}_\ell$ denote
the fixed points under these twists, and ${\bm\Lambda}_\ell$ are arbitrary lattice vectors of the compactified
space. Eq. \ref{sg0} is the point group selection rule for the $\mathbb Z_6$ models.

Eq. \ref{pg} implicitly depends on the compactified lattice. For our choice of the root lattice $\lat$, using the
notation of appendix \ref{notation}, we find it is equivalent to \cite{kobayashi2} \be
\sum_\ell n_3^{(\ell)} =0\,{\rm mod}\,3\,,\qquad \sum_\ell
(n_2^{(\ell)},n_2'^{(\ell)})=(0,0)\,{\rm mod}\,2\,, \label{sg1}
\ee
for quantum numbers in the second and third complex planes.
We note the above rules can be
understood as discrete global symmetries (${\mathbb Z}_3$ and ${\mathbb
Z}_2$ respectively) in the field theory
language.

The conditions in the first plane are more complex since the fixed
points need to be reorganized in terms of the $\theta$-eigenstates.
Denoting the states by $(\gamma^{(\ell)})_{\alpha^{(\ell)}}$, in
addition to the apparent requirement that the product of
$\theta$-eigenvalues
\be
\prod_\ell\gamma^{(\ell)}=1\,,\label{sg2}
\ee
there are further conditions arising from eq.~\ref{pg}. We list
them as follows:
\begin{enumerate}
\item There is no additional selection rules if
the couplings involve states in the $T_1$ sector -- all
gauge-invariant couplings consistent with eqs.~\ref{Hmom},
\ref{sg0} and \ref{sg1} are allowed. This follows from the fact
that in the $T_1$ sector all $\lgp{G}{2}$ root vectors are in the
same conjugacy class.

\item For couplings not involving any $T_1$ twisted state, the selection rule is determined by a $\mz_3$
($\mz_2\times \mz_2$) discrete symmetry for the $T_{2,4}$ ($T_3$) states. The $T_{2,4}$-sector states are related
to the $\mz_3$ classes by $(\gamma=1)_1=[0]$, $(\gamma=1)_2=[1]+[2]$ and $(\gamma=-1)=[1]-[2]$, and the
$T_3$-sector states are related to the $\mz_2\times\mz_2$ classes by $(\gamma=1)_1=[00]$,
$(\gamma=1)_2=[10]+[01]+[11]$ and $(\gamma=\omega,\omega^2)=[10]+\gamma[01]+\gamma^2[11]$. The allowed couplings
can then be worked out by applying the appropriate multiplication tables of the conjugacy classes, $[a][b]=[(a+b)\,{\rm mod}\,3]$
and $[a_1b_1][a_2b_2]=[(a_1+a_2)\,{\rm mod}\,2,(b_1+b_2)\,{\rm mod}\,2]$.

\end{enumerate}

The above selection rules can be used to determine the non-trivial Yukawa couplings. For example, consider
three-point couplings (which correspond to renormalizable terms in the superpotential). From the H-momentum
conservation eq.~\ref{Hmom} and the point group rule eq.~\ref{sg0} we find the following set of allowed couplings
in the ${\mathbb Z}_6$ model (all of these states have zero oscillator numbers),
\be
U_1 U_2 U_3, \,\,\, T_1 T_2 T_3, \,\,\, T_1 T_1 T_4, \,\,\, T_3 T_3
U_2, \,\,\, T_2 T_4 U_3\,.\label{3-point}
\ee

Taking into account the space group rules, we find the complete list of three-point couplings in model A1
involving fields with non-trivial representations under the $\lgp{SU}{4C}$ factor of the PS group (we have used
the field notation of table \ref{tab1}),
\bea
&&h_1\,f_3\,f_3^c\,,\quad
\sum_{\alpha,\beta=0,1}(h_3)_{A\alpha}\,f^{}_B\,\chi^c_\beta\,,\quad
C_1\,\overline\chi_1^c\,\overline\chi^c_2\,,\quad
\sum_{\alpha,\beta=0,1}(C_3)_{A\alpha}\,f_B^c\,\chi_\beta^c\,,\nn\\
&&\sum_{\alpha+\beta=0\,{\rm mod}\,2}(C_4)_\alpha
f_3^c\,\chi_\beta^c\,,\quad
\sum_{\alpha=0,1}(C_4)_\alpha\,f_A\,f_B\, \,,\quad
\sum_{\alpha=0,1}(C_4)_\alpha\,f_A^c\,f_B^c\,,
\nn\\
&& \sum_{\alpha+\beta=0\,{\rm mod}\,2}
(S_{26})_{\alpha}\,\chi^c_\beta\,\overline\chi^c_1\,,\quad
S_{13}\,\overline\chi_2^c\,f_3^c\,,\quad
\sum_{\alpha=0,1}(C_4)_{\alpha}\,(q_1)_A\, (q_2)_B\, ,\nn\\
&&\sum_{\alpha=0,1}(S_{28})_{\alpha}\,(q_1)_A\,(\overline{q}_1)_B\,,\quad
\sum_{\alpha+\beta=0\,{\rm mod}\,2}
S_1\,(C_3)_{A\alpha}(C_3)_{B\beta}\,,\label{eq:3pt}
\eea
where the ``family" indices $A, B=1,2$ satisfy a $\mz_2$ condition,
$A+B=0\,{\rm mod}\,2$.

\subsection{Allowed Yukawa couplings in the A1 model}\label{sec:ac}

In this subsection, we list allowed Yukawa couplings in the A1 model, using the string selection rules of
appendix \ref{sec:sr}. These operators all have the form ${\cal O}S^n$, where ${\cal O}$ are composite singlet
operators that involve non-trivial PS fields, and $S$ are the singlet fields. (Of course, one can also consider
operators involving other observable- and hidden-sector fields.) We work out some of the lowest order
allowed operators. We shall also follow the field naming scheme
in table~\ref{tab1}.

First notice certain structures exist for these Yukawa couplings. A composite operator of singlet fields is
equivalent to the unit operator $\bf 1$ if it is inert under all the global and local symmetries of
appendix~\ref{sec:sr}. Thus the superpotential factorizes ${\cal W}=\{{\bf 1}\}\hat{\cal W}$, with the first few
terms of $\{{\bf 1}\}$ given by \bea \{{\bf 1}\}&=&{\bf 1} + S_{19}S_{32}+S_9(S_4S_{32}+S_5S_{33}+S_{10}S_{26})
+S_{19}^2S_{32}^2+\cdots\,\nn\\
&=&({\bf 1}-S_{19}S_{32})^{-1} [{\bf
1}-S_9(S_4S_{32}+S_5S_{33}+S_{10}S_{26})]^{-1}\times(\cdots), \eea
where we have performed a ``resummation" in the second line. Note
that $\{{\bf 1}\}$ can be absorbed into the K\"ahler potential by a
transformation, ${\cal K}\rightarrow {\cal K}+\log|\{{\bf 1}\}|^2$.
As a result, the sufficient conditions for F-flatness and vanishing
cosmological constant are $\partial\hat{\cal W} =\hat{\cal W}=0$. In
the following expressions, we shall omit the caret.

We list the relevant effective operators as follows, suppressing the
$A$ and $\alpha$ indices (defined in the last section) except for
those of the matter fields and the $\chi^c$, $\overline\chi^c$ Higgs
fields.

\vspace*{2mm} \noindent{\textit{A. Color triplet masses}}

\noindent{$\bullet$ The $({\bf 6,1,1})({\bf 6,1,1})$ operators}:
\bea
{\cal W}&=&(S_1+S_{19}S_{25}+S_9{\cal S}_1^{(2)} +S_2S_{24}{\cal
S}_8^{(2)})(C_3)^2+
(S_2S_{24}+{\cal S}_1^{(2)}{\cal S}_2^{(2)})(C_4)^2\nn\\
&+& ({\cal S}^{(2)}_1+S_2S_{9}S_{24}+S_1{\cal S}^{(2)}_2
+S_{19}S_{25}{\cal S}^{(2)}_2)\,C_3C_4\nn\\
&+&S_2 S_{9} S_{22}C_2 C_4+S_2 S_{22} {\cal S}_8^{(2)} C_2C_3
+S_{10}{\cal S}^{(3)}_1(C_1)^2+S_{10}{\cal S}^{(2)}_4\,C_1C_4\nn\\
&+&[S_{10}{\cal S}^{(2)}_3+S_{13}{\cal S}^{(2)}_2+ S_1S_{10}S_{12}S_{32}+S_9S_{10}{\cal
S}^{(2)}_4]\,C_1C_3+\cdots  \label{eq:66}
\eea

\noindent{$\bullet$ The $({\bf 6,1,1})({\bf 4,1,2})({\bf 4,1,2})$
operators:}
\bea
{\cal W}&=&C_1\overline\chi_1^c\overline\chi_2^c
+(S_2S_{12}S_{24}S_{32}+{\cal S}^{(2)}_2{\cal S}^{(2)}_4)
C_1(\overline\chi_1^c)^2\nn\\
&+&S_{10}(S_{10}{\cal S}^{(2)}_3 +S_{13}{\cal S}^{(2)}_2)C_1
(\overline\chi_2^c)^2
+S_2 S_9 S_{22} S_{26} C_2(\overline\chi_1^c)^2\nn\\
&+&S_{26}({\cal S}^{(2)}_1+S_2S_9S_{24}+S_1{\cal S}^{(2)}_2) C_3
(\overline\chi_1^c)^2 +(S_1 S_{10}+S_{10}S_{19}S_{25}+S_9S_{10}{\cal
S}^{(2)}_1 +S_1{\cal S}^{(3)}_2)C_3 (\overline\chi_2^c)^2
\nn\\
&+& S_2 S_{24} S_{26}C_4 (\overline\chi_1^c )^2 +(S_{10}{\cal
S}_1^{(2)} +S_6S_7S_{14}S_{18}+S_1S_{10}{\cal S}^{(2)}_2+S_2{\cal
S}^{(3)}_4)C_4(\overline\chi_2^c)^2+\cdots, \label{eq:644}
\eea

\noindent{$\bullet$ The $({\bf 6,1,1})({\bf\overline 4,1,2})({\bf
\overline 4,1,2})$ operators:}
\bea
{\cal W}&=&C_3f_A^c\chi_\alpha^c+C_4(f_3^c\chi_\alpha^c+f_A^cf_B^c)\nn\\
&+&(S_9+S_1S_{10}S_{21}S_{22}+{\cal S}^{(2)}_2 S_{11}^{2})
[C_4f_3^cf_A^c+C_3(f_3^c\chi_\alpha^c+f_A^cf_B^c)]\nn\\
&+&({\cal S}_8^{(2)}+S_{10}(S_{30}{\cal S}^{(2)}_5+S_{13}{\cal
S}^{(2)}_4))[C_3f_3^cf_A^c+C_4(f_3^c)^2] +({\cal
S}^{(2)}_2+S_{26}{\cal S}^{(3)}_2)
[C_3\chi_\alpha^c\chi_\beta^c+C_4f_A^c\chi_\alpha^c],\nn\\
&+&S_{10}S_{12}S_{32}C_1f_A^c\chi_\alpha^c +[S_9{\cal
S}_8^{(2)}+S_{13}(S_{10}{\cal S}^{(2)}_3+S_{13}{\cal S}^{(2)}_2)]
C_3(f_3^c)^2\nn\\
&+&S_{10}S_{13}S_{21}S_{22}C_1f_3^cf_A^c
+S_9S_{10}S_{12}S_{32}
C_1(f_3^c\chi_\alpha^c+f_A^cf_B^c)
+({\cal S}^{(2)}_2)^2C_4\chi_\alpha^c\chi_\beta^c,\\
{\cal W}^{(8)}&\supset&S_{12}(S_{11}S_{30}{\cal S}^{(2)}_4+ S_{10}S_{32}{\cal
S}^{(2)}_2)C_1\chi_\alpha^c\chi_\beta^c +S_2S_{10}S_{21}S_{22}^2C_2\chi_\alpha^c\chi_\beta^c. \label{eq:64b4b}
\eea
\noindent{$\bullet$ The $({\bf 6,1,1})({\bf 4,2,1})({\bf 4,2,1})$ operators:}
\bea
{\cal W}&=&S_2^2S_9S_{10}S_{12}S_{32}C_1f_3^2+S_2S_{21}^2{\cal
S}^{(2)}_7C_2f_3^2
+S_2^2S_9C_3f_3^2+S_2^2C_4f_3^2\nn\\
&+&S_2S_9S_{10}S_{12}S_{32}C_1f_3f_A+S_{21}^2{\cal S}^{(2)}_7C_2f_3f_A
+(S_2S_9+S_{21}S_{23}{\cal S}^{(2)}_7)C_3f_3f_A+S_2C_4f_3f_A\nn\\
&+&S_9S_{10}S_{12}S_{32}C_1f_Af_B+S_{10}(S_2S_{12}S_{13}S_{22}S_{32}
+S_3S_{12}^2S_{23}S_{32}+S_6S_{14}S_{16}S_{22}S_{33})C_2f_Af_B\nn\\
&+&(S_9+S_{11}^2{\cal
S}^{(2)}_2+S_1S_{10}S_{21}S_{22})C_3f_Af_B+C_4f_Af_B+\cdots, \eea

In the above equations we have used the following composite singlet
operators of dimensions two and three,
\bea
&&{\cal S}^{(2)}_1=S_4S_{25}+S_5S_{29},\qquad
{\cal S}^{(2)}_2=S_4S_{32}+S_5S_{33}+S_{10}S_{26},\nn\\
&&{\cal S}^{(2)}_3=S_{12}S_{25}+ S_{13}S_{26}+ S_{17}S_{32}+ S_{19}S_{30},
\qquad
{\cal S}^{(2)}_4=S_4S_{30}+ S_5S_{31},\nn\\
&&{\cal S}^{(2)}_5=S_4S_{13}+S_{10}S_{17},\qquad {\cal
S}^{(2)}_6=S_{21}S_{24}+ S_{22}S_{23},\nn\\
&&{\cal S}^{(2)}_7=S_3S_4+S_5S_7,\qquad {\cal S}^{(2)}_8=S_9^2+
S_{11}^2,\\
&&{\cal S}^{(3)}_1=S_9 S_{21}S_{22}+ S_{10} S_{12}S_{30},\qquad
{\cal S}^{(3)}_2=S_{10}S_{19}S_{32}+ S_{11}S_{17}S_{30},\nn\\
&&{\cal S}^{(3)}_3=S_{10}S_{20}S_{33}+ S_{11}S_{18}S_{31},\qquad
{\cal S}^{(3)}_4=S_3S_{13}S_{17}+S_9S_{10}S_{24}.
\eea

\vspace*{2mm} \noindent{\textit{B. Quark and lepton Yukawa
couplings}}

Quark and lepton Yukawa couplings are operators of the {$\bf
(4,2,1)(\overline 4,1,2)(1,2,2)$} type. They are given by
\bea & & {\cal O}_{(a,b)} = f_a  h_1  f_b^c,  \qquad a,\,b=1,2,3. \eea
In addition, the PS symmetry breaking fields may enter the effective
fermion mass operators at higher order. We thus consider the
operators of the {({\bf 4,1,2})(${\bf \overline 4}$,{\bf 1,2}) type}
given by
\be {\cal O}_1 = \overline \chi_1^c \chi_\alpha^c, \qquad {\cal O}_2
= \overline \chi_2^c  \chi_\alpha^c \ee
Using these operators, we search for all the allowed higher
dimensional operators of the form, ${\cal O}_{(a,b)} ({\cal
O}_i)^{n} S^{n'}$, i.e., $({\bf 4, 2,1})(\overline {\bf 4}, {\bf
1,2}) ({\bf 1,2,2}) [({\bf 4,1,2}) (\overline {\bf 4},{\bf 2,
1})]^nS^{n'}$, with the smallest possible $n$, $n'$ (for $n+n'\leq
4$ or $n'\leq 5$). The result is ($A, B=1,2$)
\bea
{\cal W}&=&{\cal S}_2^{(2)}{\cal
O}_{(3,A)}+(S_{10}^2S_{12}S_{24}S_{30}+S_9S_{10}S_{22}{\cal
S}_6^{(2)}+S_{21}S_{22}{\cal S}^{(3)}_4+S_3S_{12}{\cal O}_1{\cal
O}_2){\cal O}_{(A,3)}\nn\\
&+&(S_{10}S_{22}{\cal S}_6^{(2)}+S_3S_9S_{12}{\cal O}_2){\cal
O}_{(A,B)}.\label{yukmatapp}
\eea

\vspace*{2mm} \noindent{\textit{C. Neutrino masses}}

Consider the three PS breaking operators
\be {\cal O}_{i j} = \overline \chi_i^c \overline \chi_j^c, \qquad
i, j = 1,2. \ee  We then find higher dimension effective Majorana
neutrino mass operators of the form (with minimal number of
singlets)
\bea {\cal W}&=&
 ({\cal S}_8^{(2)}S_{26} + S_1S_{21}S_{22}) f_3^c  f_3^c  {\cal O}_{1 1}
+ S_{13}f_3^c f_3^c{\cal O}_{1 2}
+ S_9S_{10}{\cal S}^{(2)}_8f_3^c  f_3^c {\cal O}_{2 2 }
+ S_9S_{26}f_3^c  f_A^c  {\cal O}_{1 1}\nn\\
&+& (S_{10} {\cal S}_3^{(2)} + S_{13} {\cal S}_2^{(2)})f_3^c  f_A^c  {\cal O}_{1 2}
+S_{26}f_A^c  f_B^c  {\cal O}_{1 1}
 + S_{10} {\cal S}_4^{(2)}f_A^c  f_B^c  {\cal O}_{1 2}
+S_9S_{10}f_A^c  f_B^c  {\cal O}_{2 2}  .\label{eq:maj1}
 \eea

\section{Gauge coupling evolution\label{sec:RG}}

In this appendix, we derive the GQW
equations for gauge coupling unification in the ${\mathbb Z}_6$
string models. We shall work in the orbifold GUT limit and also make
some simplification assumptions of the matter spectra.

It is well known that gauge coupling unification in heterotic string
models has a serious problem in the perturbative regime
\cite{dienes, witten}. From a simple reduction of the 10d string
effective action, one finds $G_N={\rm e}^{2\phi}\alpha'^4/128\pi V$,
$\alpha_{\rm string}={\rm e}^{2\phi}\alpha'^3/16\pi V$, where $\phi$ is
the dilaton, $V$ the volume of the compactified space, and $\alpha'$
the Regge slope, thus\footnote{A more careful definition, taking into
account of the renormalization scheme dependence, is given in
ref.~\cite{kaplunovsky}.}
\be
M_{\rm Pl}=\sqrt{\frac{8}{\alpha_{\rm string}}}M_{\rm string}.
\ee
Assuming $V\sim M_{\rm GUT}^{-6}$ (where $M_{\rm
GUT}\simeq 3\times 10^{16}$ GeV), $\alpha_{\rm
string}=\alpha^{}_{\rm GUT}\simeq 1/24$ and ${\rm e}^{2\phi}<1$, one gets
$M_{\rm Pl}\leq \alpha_{\rm GUT}^{-2/3} M_{\rm GUT}\simeq 2\times
10^{17}$ GeV, which is apparently incorrect. Note that this result
relies on the assumptions that the compactified space is symmetric,
i.e., all the dimensions are of similar sizes, and there are no
additional states near the GUT scale. Both assumptions are invalid
in our models.

In heterotic models, it is well known that gauge couplings can
obtain potentially large threshold corrections only if the models
contain ${\mathcal N}=2$ sub-sectors \cite{gauge,KL}. Our $\mz_6$
orbifold models have exactly this kind of property, since both the
${\mathbb Z}_2$ and ${\mathbb Z}_3$ sub-orbifold twists leave
exactly one complex plane unrotated. It is, however, a very
complicate matter to compute gauge threshold corrections in string
theory in the presence of discrete Wilson lines. (The simplest way
to compute these corrections is using the target space duality
symmetry \cite{GPR}, however in the presence of discrete Wilson
lines the duality groups are broken to their discrete subgroups
\cite{td}. Even for the much simpler $\mz_3$ models, threshold
corrections are only known numerically
\cite{MNS}.\footnote{Threshold corrections in models with continuous
Wilson lines \cite{CLM} have been computed in
refs.~\cite{stieberger}. Our models with discrete Wilson lines,
however, are not the limiting cases of those with continuous
lines.})

Fortunately we can use a much simpler field theoretical method to compute gauge threshold corrections in the
orbifold GUT limit, i.e., when one of the $\lgp{SO}{4}$ dimensions is much larger than the string length scale
and other compactified dimensions.\footnote{Admittedly the field theoretical calculation suffers from the usual
UV divergence. The result is sensitive to the cutoff scale and needs be dealt with caution. However, we do not
expect the RG evolution of the difference of gauge couplings to be affected much by our field theoretical
treatment.} In this limit, all the winding modes and KK modes in small dimensions have string scale mass, and
according to refs.~\cite{gauge,KL} one would expect the dominant contributions to the threshold corrections come
from the KK modes of the large dimension. From now on, we shall assume this is correct and neglect all the
contributions from states with string scale mass. The gauge coupling at the string scale, $\alpha_{\rm string}$,
imposes a boundary condition for the renormalization group equations at the cut-off scale, which is taken to be
the string scale, $M_{\rm string}$.

We follow the field theoretical analysis in ref.~\cite{DDG} (see
also \cite{gh}). It has been shown there the correction to a generic
gauge coupling due to a tower of KK states with masses $M_{\rm
KK}=m/R$ is
\be
\alpha^{-1}(\Lambda)=\alpha^{-1}(\mu_0)-\frac{b}{4\pi}
\int_{r\Lambda^{-2}}^{r\mu_0^{-2}}\frac{{\rm d}t}{t}\,
\theta_3\left(\frac{{\rm i}t}{\pi R^2}\right)\,,
\ee
where the integration is over the Schwinger parameter $t$,
$\mu_0$ and $\Lambda$ are the IR and UV cut-offs, and $r=\pi/4$ is a
numerical factor. $\theta_3$ is the Jacobi theta function,
$\theta_3(t)=\sum_{m=-\infty}^\infty {\rm e}^{{\rm i}\pi m^2 t}$, representing
the summation over KK states.

In our models, there are several modifications in the calculation. Firstly there are four sets of KK towers, with
mass $M_{\rm KK}=m/R$ (for $P=P'=+$), $(m+1)/R$ (for $P=P'=-$) and $(m+1/2)/R$ (for $P=+$, $P'=-$ and $P=-$,
$P'=+$), where $m\geq 0$. The summations over KK states give respectively $\frac{1}{2}\left(\theta_3({\rm i}t/\pi
R^2)-1\right)$ for the first two cases and $\frac{1}{2}\theta_2({\rm i}t/\pi R^2)$ for the last two (where
$\theta_2(t)= \sum_{m=-\infty}^\infty {\rm e}^{{\rm i}\pi (m+1/2)^2 t}$), and we have separated out the zero
modes in the $P=P'=+$ case. Secondly, the PS symmetry in our models must be further broken down to that of the
SM. This breaking, in principle, can be induced by brane or bulk states. However, in sect.~\ref{sec:yukawa} we
have shown this breaking is more likely due to non-renormalizable couplings of the states in the $T_{2,4}$
sectors. Since these states are identified with bulk states in the orbifold GUT limit (c.f. sect.~\ref{ogut}), we
shall assume, in what follows, the breaking of PS to the SM is bulk breaking. We further assume the breaking
scale, $M_{\rm PS}$, is smaller than or equal to
the compactification scale, $M_c$, so that we can neglect mass
corrections to the massive KK states.

Tracing the renormalization group evolution from low energy scales,
we are first in the realm of the MSSM, and the beta function
coefficients are $b_i^{\rm MSSM}=(\frac{33}{5},1,-3)$. When we pass
the PS breaking scale, $M_{\rm PS}$, the beta function coefficients
become $(b^{\rm PS}_{++}+b_{\rm brane})_i$, where the two terms
represent contributions from the bulk and brane states. These
coefficients can contain contributions from additional states
besides those of the MSSM, for example, from a vector-like pair
$({\bf 4,1,1})+({\bf\overline 4,1,1})$. The next energy threshold is
the compactification scale $M_c$. From this scale to the string
scale, we have the four sets of KK states. Since we have assumed
$M_{\rm PS}\ll M_c$, the beta function coefficients are those of the
complete PS representations.

Collecting these facts, and using $\theta_2({\rm i}t/\pi R^2)\simeq
\theta_3({\rm i}t/\pi R^2)\simeq\sqrt{\frac{\pi}{t}} R$ for $t/R^2\ll 1$,
we find the GQW equations,
\bea
\frac{2\pi}{\alpha_i(\mu)}&\simeq& \frac{2\pi}{\alpha_{\rm string}}
+b^{\rm MSSM}_i\log\frac{M_{\rm PS}}{\mu} +(b^{\rm PS}_{++}+b_{\rm
brane})_i\log\frac{M_{\rm string}}{M_{\rm PS}}\nn\\
&-& \frac{1}{2}(b^{\rm PS}_{++}+b^{\rm PS}_{--})_i\log \frac{M_{\rm
string}}{M_c}+b^{\mathcal G}\left(\frac{M_{\rm string}}{M_c}
-1\right)\,,\label{rg}
\eea
for $i=1,2,3$, where we have taken the cut-off scales, $\mu_0=M_c$
and $\Lambda=M_{\rm string}$. $b^{\mathcal G}=\sum_{P=\pm, P'=\pm}
b^{\rm PS}_{PP'}$, so in fact it is the beta function coefficient of
the orbifold GUT gauge group, ${\mathcal G}$. The beta function
coefficients in the last two terms have an ${\mathcal N}=2$ nature,
since the massive KK states enjoy a large supersymmetry. For
${\mathcal G}=\lgp{SO}{10}$ and $\lgp{E}{6}$, $b^{\mathcal
G}=-16+4n_{\bf 16}+2n_{\bf 10}$ and $-24+6n_{\bf 27}$, where $n_{\bf
10}$, $n_{\bf 16}$ and $n_{\bf 27}$ are numbers of bulk
hypermultiplets in the respective representations. To accomplish
bulk breaking, we need $n_{\bf 27}=4$ in models A1/B and $n_{\bf
16}=4$, $2\leq n_{\bf 10}\leq 6$ in model A2, therefore $b^{\rm
E_6}=0$ and $b^{\rm SO(10)}=2n_{\bf 10}$.  Coincidentally there is
no power-law running in the $\lgp{E}{6}$ model.


\end{document}